\documentclass[final,3p,times]{elsarticle}

\usepackage{amssymb}
\usepackage{amsmath}
\usepackage{hyperref}
\usepackage{graphbox}
\usepackage{orcidlink}
\usepackage{circuitikz}
\usepackage{calc}
\usepackage{subcaption}
\usetikzlibrary{calc}
\usetikzlibrary{patterns}
\usetikzlibrary{arrows.meta}
\usepackage[export]{adjustbox}
\usepackage[nameinlink,noabbrev,capitalize]{cleveref}
\usepackage{subcaption}
\usepackage{multirow}
\usepackage{booktabs}
\usepackage{tabularx}
\hypersetup{colorlinks=true, linkcolor=blue,urlcolor=blue,filecolor=blue,citecolor=red,allcolors=blue}
\usepackage{bm}
\usepackage{soul}
\usepackage{listings}
\usepackage{xcolor}
\usepackage{diagbox}

\definecolor{codegreen}{rgb}{0,0.6,0}
\definecolor{codegray}{rgb}{0.5,0.5,0.5}
\definecolor{codepurple}{rgb}{0.58,0,0.82}
\definecolor{backcolour}{rgb}{0.95,0.95,0.92}

\lstdefinestyle{mystyle}{
    backgroundcolor=\color{backcolour},   
    commentstyle=\color{codegreen},
    keywordstyle=\color{magenta},
    numberstyle=\tiny\color{codegray},
    stringstyle=\color{codepurple},
    basicstyle=\ttfamily\footnotesize,
    breakatwhitespace=false,         
    breaklines=true,                 
    captionpos=b,                    
    keepspaces=true,                 
    numbers=left,                    
    numbersep=5pt,                  
    showspaces=false,                
    showstringspaces=false,
    showtabs=false,                  
    tabsize=2
}
\usepackage{lineno}

\journal{\href{https://www.sciencedirect.com/journal/aerospace-science-and-technology}{Aerospace Science \& Technology}}

\begin{document}

\begin{frontmatter}


\title{Multiple input tangential interpolation-driven damage detection of a jet trainer aircraft\tnoteref{label5}}
\tnotetext[label5]{\hl{This is an author-accepted manuscript version of the following article: G. Dessena, M. Civera, A. Marcos, B. Chiaia, and O. E. Bonilla-Manrique, ‘Multiple input tangential interpolation-driven damage detection of a jet trainer aircraft’, \textit{Aerospace Science and Technology}, vol. 168, p. 111032, Jan. 2026, doi: 10.1016/j.ast.2025.111032. Available online at} [\url{https://doi.org/10.1016/j.ast.2025.111032}]}

\author[label1]{Gabriele Dessena \orcidlink{0000-0001-7394-9303}}
\affiliation[label1]{organization={Department of Aerospace Engineering, Universidad Carlos III de Madrid},
             addressline={Av.da de la Universidad 30},
             postcode={28911},
             city={Leganés},
             state={Madrid},
             country={Spain}}

\author[label2]{Marco Civera \orcidlink{0000-0003-0414-7440}}

\affiliation[label2]{organization={Department of Structural, Building and Geotechnical Engineering, Politecnico di Torino},
            addressline={Corso Duca degli Abruzzi 24}, 
            city={Turin},
            postcode={10129}, 
            state={Piedmont},
            country={Italy}}

\author[label1]{Andrés Marcos \orcidlink{0000-0002-0116-6681}}
\author[label2]{Bernardino Chiaia \orcidlink{0000-0002-5469-2271}}
\author[label3]{Oscar E. Bonilla-Manrique \orcidlink{0000-0003-0541-8310}}
\affiliation[label3]{organization={Electronic Technology Department, Universidad Carlos III de Madrid},
             addressline={Av.da de la Universidad 30},
             postcode={28911},
             city={Leganés},
             state={Madrid},
             country={Spain}}

\begin{abstract}
{The problem of damage detection and identification is of interest {for} many aerospace and aeronautical engineering {systems}. However, relevant literature mostly focuses on subsystems and parts, rather than full airframes.} In structural dynamics, modal parameters, such as natural frequencies and mode shapes, from any structure are the main building blocks of vibration-based damage detection. However, traditional comparisons of these parameters are often ambiguous in complex systems, complicating damage detection and assessment. The modified total modal assurance criterion (MTMAC), an index well-known in the field of finite element model updating, is extended to address this challenge and is proposed as an index for damage identification and severity assessment. To support the requirement for precise and robust modal identification of Structural Health Monitoring (SHM), the improved Loewner Framework (iLF), known for its reliability and computational performance, is pioneeringly employed within SHM. Since the MTMAC is proposed solely as a damage identification and severity assessment index, the coordinate modal assurance criterion (COMAC), also a well-established tool, but for damage localisation using mode shapes, is used for completeness. The iLF SHM capabilities are validated through comparisons with traditional methods, including least-squares complex exponential (LSCE) and stochastic subspace identification with canonical variate analysis (SSI-CVA) on a numerical case study of a cantilever beam. Furthermore, the MTMAC is validated against the traditional vibration-based approach, which involves directly comparing natural frequencies and mode shapes. Finally, an experimental dataset from a BAE Systems Hawk T1A jet trainer ground vibration test is used to demonstrate the iLF and MTMAC capabilities on a real-life, real-size SHM problem, showing their effectiveness in detecting and assessing damage.
\end{abstract}




\end{frontmatter}
\section{Introduction}\label{sec:int}
{\textit{Damage} can be defined as any structural change in a system that compromises or affects its operational capacity} \cite{Farrar2007}. {The process of implementing a damage detection strategy for aerospace} \cite{Gelman2020}, civil \cite{Huang2024}{, or mechanical engineering} \cite{Zhang2024} {systems is commonly referred to as Structural Health Monitoring (SHM).}

{In the case of vibration-based SHM, modal parameters, such as natural frequencies ($\omega_n$), damping ratios ($\zeta_n$), and mode shapes ($\bm{\phi}_n$), are frequently selected as damage indicators due to their ease of identification from vibrational data }\cite{Rytter1993}{ and their direct relationship with the mass and stiffness of the structure. These parameters can serve various purposes, such as assessing damage severity and/or localisation. $\omega_n$ are particularly effective for evaluating damage severity, but less reliable for localisation since small frequency changes are often masked by environmental and operational variations }\cite{Fan2011}.{ Nevertheless, they were successfully applied in }\cite{Civera2021a}{ for the SHM of a masonry bridge structure using single-input and multiple-output (SIMO) data.
Conversely, $\bm{\phi}_n$ are better suited for damage localisation, typically achieved by detecting deviations between baseline and damaged values. Still, they can also be helpful for damage severity assessment. For example, }\cite{Modesti2024}{ proposed a two-step procedure for damage detection on beam-like structures using $\bm{\phi}_n$.
 $\zeta_n$ are hardly ever used as a standalone damage indicator due to their strong dependence on non-structural factors, which makes it difficult to distinguish changes in damping caused by damage from those due to other influences }\cite{Reynders2008}.

{Beyond these traditional strategies, more advanced approaches have emerged.
Recent advances in vibration-based SHM have focused on civil infrastructure, particularly bridges }\cite{Zhang2025}{ and wind turbines }\cite{Vettori2024}{, with a growing interest in addressing challenges such as uncertainty quantification, noise robustness, and operational variability }\cite{OConnell2024}{. Furthermore, clustering algorithms, such as the Density-based Spatial Clustering of Applications with Noise (DBSCAN), have gained attention for identifying aeronautically relevant structures. Notable applications are the output-only modal identification of a helicopter blade in }\cite{Sibille2023}{ and an aircraft model in }\cite{Wu2025}. 
{Advanced methodologies, such as the Cepstrum-Informed Attention-Based Network, have been introduced to model structural dynamics efficiently while maintaining physical interpretability, using compact damage-sensitive features derived from cepstral coefficients }\cite{Li2025}. {Bayesian modal identification approaches have been developed to incorporate ambient and seismic response data, enabling robust modal parameter estimation with reduced computational complexity }\cite{Ni2025}.{ Additionally, digital twin frameworks have established themselves as effective tools for predictive maintenance, offering unambiguous correlations between dynamic behaviours and structural conditions, such as under-foundation scour in bridges }\cite{Sanchez-Haro2025}. {For wind turbine rotors, operational modal analysis techniques have been adapted to identify localised stiffness reductions and anisotropic effects using time-periodic approximations and Gaussian residual tests }\cite{Cadoret2025}.{ Furthermore, machine learning methods, including convolutional neural networks, have been successfully applied to classify damage scenarios in lightweight bridge structures, addressing the complexities of coupled dynamic responses from moving loads }\cite{Dadoulis2025}. {Moreover, as shown in }\cite{Ren2026}{, traditional guided wave SHM can be enhanced for operational damage detection in aeronautical structures using Gaussian mixture models and their probability function for damage detection and localisation. Lastly, population-based SHM has demonstrated the potential to transfer structural behaviour knowledge from homogeneous building populations to individual targets, offering a scalable solution for damage detection across nominally identical structures }\cite{Astorga2025}.

{Nevertheless, an unequivocal solution for SHM does not yet exist, as most methods and strategies are either underdeveloped or not fully codified in national or international standards }\cite{Cawley2018}.
 {Furthermore, in aerospace and aeronautical engineering, most applications focus on a deep analysis of individual subsystems, like fuel pumps }\cite{Verhulst2022} and tanks \cite{Gao2019}, or parts, like blades \cite{Liu2024}, attachment lugs \cite{Chen2024}, {aircraft engine brackets} \cite{WentaoZhang1, WentaoZhang2, WentaoZhang3}, and spars \cite{Fang2019}, rather than full or partial airframes, despite very early efforts \cite{Pappa1998}.

 {In addition, developing new methods – or refining existing ones - is a challenge due to the lack of diverse datasets from the system experimental setup perspective. In fact, most case studies and benchmarks for vibration-based SHM focus on SIMO }\cite{Figueiredo2009} {or output-only data }\cite{Maeck2001} {rather than multiple input and multiple output (MIMO). This is critical as MIMO testing can offer further insight into modes not fully identified from SIMO or OMA campaigns, such as out-of-plane dominant motion in flexible wings }\cite{Dessena2022} {or civil infrastructure. A well-known example is the Z24 bridge case study, where the weakly excited lateral modes are the most difficult to identify }\cite{Reynders2012a}. {Despite this difficulty, }\cite{Peeters2015} {showed that the first lateral mode was the most indicative of damage occurrence, which can be extrapolated as an argument to use the identification of small amplitude vibration modes to enhance the precision of vibration-based SHM.}
 
{While some efforts were made at the beginning of the 2000s for MIMO-based bridge monitoring }\cite{Catbas1998, Catbas2004}{, these were not followed by further advances or testing campaigns. Additionally, there is a general lack of aeronautically relevant examples for SHM, with the notable exception of }\cite{Dessena2022g, Dessena2024b}. {However, this issue was recently addressed by the Laboratory for Verification and Validation (LVV) at the University of Sheffield, which conducted a thorough vibration testing campaign for a full-scale jet trainer aircraft }\cite{Wilson2024}{ and has released the dataset }\cite{Wilson2024a}. {This includes vibration data from different input signals, including simulated and real damaged cases in a MIMO configuration, offering a thorough standard for future identification and damage detection methods.
 
Finally, coming back to the basics of vibration-based SHM, there is a third, and fundamental, critical point: the reliability of the identified modal parameters.

In this regard, system identification (SI) offers critical approaches to extract information and/or obtain models from simulation, experimental, or real data }\cite{Lju1987}.{ For example, in structural dynamics, SI methods are used for the extraction of modal parameters }\cite{VanOverschee1996}{, such as natural frequencies ($\omega_n$), damping ratios ($\zeta_n$), and mode shapes ($\bm{\phi}_n$), from linear time invariant systems (LTIs)} \cite{Dessena2022}. {Modal parameters have always been the backbone of vibration-based SHM }\cite{Farrar2025}{. Hence, there is a continuous search for novel and efficient methods for extracting modal parameters, such as those presented in }\cite{Civera2021}{, alongside more refined indices }\cite{Allemang2003}{ and analysis methods }\cite{Bull2021}{. The former are necessary to move away from direct comparison of modal parameters – a notoriously challenging task for higher-dimensional systems. For this reason, indices such as the modal assurance criterion (MAC) }\cite{Gres2021}{, which helps condense information to make it usable for SHM, have achieved considerable success over the years. On the other hand, the latter is not restricted to direct methods and can include indirect methods, such as model updating-based approaches }\cite{Dessena2022d, Dessena2022c}{, as well as data-driven techniques }\cite{Huang2024}. 

{Given all these challenges - lack of definitive damage indices, MIMO experimental validation, and efficient SI algorithms -}
We propose a two-step modal \& damage identification framework. The first step uses the improved Loewner Framework (iLF) \cite{Dessena2024} to identify the modal parameters from a MIMO system, while the second step uses the modified total modal assurance criterion (MTMAC) \cite{Perera2006} as a damage assessment index based on the identified modal parameters.
The iLF, a recently developed computationally efficient MIMO modal parameter identification technique, is applied in this article to the jet trainer aircraft dataset mentioned above and, beforehand, validated on a numerical case study. {The work reported here represents the first instance in the literature where iLF is applied for modal parameter-based SHM purposes.} To highlight the capabilities of the iLF, the results are compared to those from other industrially relevant methods, such as least-squares complex exponential (LSCE) \cite{Zimmerman2017} and stochastic subspace identification with canonical variate analysis (SSI-CVA) \cite{VanOverschee1996}. The former is a frequency domain method (as is the iLF), while the latter is an output-only technique in the time domain.
In addition, the MTMAC (modified total modal assurance criterion) is shown to be a valid index for damage assessment. It is highlighted that this criterion was used before, together with MACFLEX (modal flexibility assurance criterion) \cite{Georgioudakis2018}, but is here proposed as a standalone damage assessment index. Further, the well-known Coordinate Modal Assurance Criterion (COMAC) \cite{Lieven1988} is used for damage localisation to ensure the completeness of the damage assessment procedure. The rationale is that the COMAC identifies which degrees-of-freedom (DoFs) measurements contribute to a low value of MAC \cite{Fayyadh2011}; hence, a low value of COMAC corresponds to a local $\bm{\phi}_n$ trajectory deviation, which is a typical sign of damage. This is supported, among many other examples, by the relevant literature on damage localisation on composite laminates \cite{Perez2014}, reinforced concrete beams \cite{Kianfar2024}, and retrofitted unreinforced masonry walls \cite{Romero-Carrasco2024}. 

{Consequently, the main aim of this study is the introduction of a condensed index for damage detection using the MTMAC, which leads to the following objectives:}
\begin{enumerate}
    \item Validate the iLF precision and accuracy for modal parameter-based damage detection via direct comparison on a numerical model;
    \item Show that the MTMAC can be a valid and accurate damage index on a numerical model;
    \item Validate the proposed combined iLF and MTMAC approach on the numerical data;
    \item Apply the damage detection strategy on a complex aeronautical structure: A BAE Systems Hawk T1A airframe
    \end{enumerate}

{Being able to use the proposed combined approach is extremely important for complex systems, such as a full trainer jet, because modal parameters are difficult to monitor directly, and a merged index, like the proposed MTMAC, makes differentiating between healthy and damaged cases a more straightforward task. Moreover, it is noted that the proposed method is formulated for linear time-invariant dynamics, as is customary in modal-based SHM. This is consistent with the ground vibration test dataset, which was acquired under low-amplitude excitations ensuring quasi-linear structural behaviour. Nevertheless, although outside the scope of this work, the interested reader is referred to }\cite{Shaw1991} {for the theoretical background that allows the nonlinear modal analysis of a full aircraft, a Morane-Saulnier Paris, shown in} \cite{Kerschen2013}.

Thus, the main contributions of this work are the following:
\begin{enumerate}
    \item[i.] A two-step modal \& damage identification framework is proposed:
    \begin{enumerate}
    \item[a.] (First step) The iLF is shown to be valid for SHM purposes;
    \item[b.] (Second step) The MTMAC is shown to serve as a damage assessment quantification index.
    \end{enumerate}
    \item[ii.] The framework (iLF and MTMAC) is applied to a comprehensive dataset obtained from a full-scale jet trainer aircraft.
\end{enumerate}

The remaining sections of the paper are organised as follows:
\begin{enumerate}
    \item[2.] \hyperref[sec:met]{\emph{Methodology}}: The Loewner Framework (LF) theory, the backbone of the iLF;
    \item[3.] \hyperref[sec:num]{\emph{Numerical case study}}: A MIMO numerical case study of a cantilever beam with three-step changes in its section profile is analysed. Three damage (element stiffness reduction) and a mass addition (on a single element) cases are used to validate the use of the iLF and the MTMAC;
    \item[4.] \hyperref[sec:exp]{\emph{Experimental case study}}: Modal parameters from twelve cases (one healthy and 11 damage scenarios) of the BAE Hawk T1A vibration dataset are extracted via the iLF and used for damage assessment via the MTMAC and damage localisation via the COMAC;
    \item[5.] \hyperref[sec:con]{\emph{Conclusions}}: Final remarks and a summary of the results are presented.
\end{enumerate}

\section{Methodology}\label{sec:met}
The following subsections outline the iLF theoretical background (\cref{sec:lf}) and the SHM tools considered (\cref{sec:dam}). Then, in \cref{sec:fra}, the proposed approach of using the iLF as a valid tool for SHM purposes is discussed.

\subsection{The Loewner Framework}\label{sec:lf}
The theoretical origin of the LF can be traced back to model order reduction methods \cite{Mayo2007,Antoulas2017}. While its initial application was for modelling multi-port circuits in Electrical Engineering \cite{Lefteriu2009,Lefteriu2010b}, since then, it has been extended to investigate unsteady aerodynamics in Aeronautical Engineering \cite{Quero2019} and recently applied to the extraction of modal parameters from SIMO mechanical systems for SHM \cite{Dessena2022}. The LF computational efficiency for SI via modal parameters extraction has been recently examined in \cite{Dessena2022f}. The basic approach was then extended to the ground vibration testing of wings \cite{Dessena2022e}. More recent work has focused on expanding the LF to output-only systems \cite{Dessena2024f} and to MIMO systems, while also increasing its computational efficiency \cite{Dessena2024}; thus, introducing the improved LF, or iLF, which is the variant considered in this work. 
In essence, the LF approach fits a suitable transfer function to a set of complex-valued data, such as Frequency Response Functions (FRFs), by splitting the data into two subsets (or directions), from which rational tangential interpolants are built by exploiting the Loewner matrix $\mathbb{L}$, first introduced by Karl Loewner in the 1930s \cite{Lowner1934}.

Before moving forward, a brief technical presentation of the LF, on which the iLF computational implementation is based, is given, starting with the description of the Loewner matrix $\mathbb{L}$.  \emph{For a row array of pairs of complex numbers ($\mu_j$,${v}_j$), $j=1$,...,$q$, and a column array of pairs of complex numbers ($\lambda_i$,${w}_i$), $i=1$,...,$\rho$, such that $\lambda_i$, $\mu_j$ distinct, the associated $\boldsymbol{\mathbb{L}}$ can be defined:}
\begin{equation}
\label{eq:LM}
\boldsymbol{\mathbb{L}}=\begin{bmatrix}
\frac{\mathbf{v}_1-\mathbf{w}_1}{\mu_1-\lambda_1} & \cdots & \frac{\mathbf{v}_1-\mathbf{w}_\rho}{\mu_1-\lambda_\rho}\\
\vdots & \ddots & \vdots\\
\frac{\mathbf{v}_q-\mathbf{w}_1}{\mu_q-\lambda_1} & \cdots & \frac{\mathbf{v}_q-\mathbf{w}_\rho}{\mu_q-\lambda_\rho}\\
\end{bmatrix}\:\in \mathbb{C}^{q\times \rho}
\end{equation}
\emph{This is valid if a known underlying function $\pmb{\phi}$ exists, such that $\mathbf{w}_i=\pmb{\phi}(\lambda_i)$ and $\mathbf{v}_j=\pmb{\phi}(\mu_j).$}

Thus, considering a LTI system $\mathbf{\Sigma}$ with $m$ inputs and $n$ outputs, and $k$ internal variables, described by:
\begin{equation}
\begin{split}
  \mathbf{\Sigma}:\;\mathbf{E}\frac{d}{dt}\mathbf{x}(t)=\mathbf{A}\mathbf{x}(t)+\mathbf{B}\mathbf{u}(t)\\
\mathbf{y}(t)=\mathbf{C}\mathbf{x}(t)+\mathbf{D}\mathbf{u}(t)  
\label{eq:LTI}
\end{split}
\end{equation}

\noindent where $\mathbf{x}(t)\:\in\: \mathbb{R}^{k}$ is the state, $\mathbf{u}(t)\:\in\:\mathbb{R}^{m}$ is the input and $\mathbf{y}(t)\:\in\:\mathbb{R}^{p}$ is the output. The system matrices are $\mathbf{E}$, $\mathbf{A}\in \mathbb{R}^{k\times k}$, $\mathbf{B}\in \mathbb{R}^{k\times m}$, $\mathbf{C}\in \mathbb{R}^{p\times k}$, and $\mathbf{D}\in \mathbb{R}^{p\times m}$. Its transfer function $\mathbf{H}(s)$ can be defined as:
\begin{equation}
    \mathbf{H}(s)=\mathbf{C}(s\mathbf{E}-\mathbf{A})^{-1}\mathbf{B}+\mathbf{D}
    \label{eq:trans}
\end{equation}

\noindent Next, the structure of tangential interpolation, often referred to as the rational interpolation in tangential directions, can be exploited to yield the following associated right (\cref{eq:RID}) and left (\cref{eq:LID}) interpolation data:
\begin{equation}
\begin{gathered}
   (\lambda_i;\mathbf{r}_i,\mathbf{w}_i),\: i = 1,\dots,\rho
    \quad \quad
    \begin{matrix}
        \mathbf{\Lambda}=\text{diag}[\lambda_1,\dotsc,\lambda_k]\in \mathbb{C}^{\rho\times \rho}\\
        \mathbf{R}=[\mathbf{r}_1\;\dotsc \mathbf{r}_k]\in \mathbb{C}^{m\times \rho}\\
        \mathbf{W} = [\mathbf{w}_1\;\dotsc\;\mathbf{w}_k]\in \mathbb{C}^{\rho\times \rho}
        \end{matrix}\Bigg\}
\end{gathered}
\label{eq:RID}
\end{equation}
 \noindent Similarly, the left interpolation data:
\begin{equation}
    \begin{gathered} 
        (\mu_j,\mathbf{l}_j,\mathbf{v}_j),\: j = 1,\dots,q
        \quad \quad
        \begin{matrix}
            \mathbf{M}=\mathrm{diag}[\mu_1,\dotsc,\mu_q]\in \mathbb{C}^{q\times q}\\
            \mathbf{L}^T=[\mathbf{l}_1\;\dotsc \mathbf{l}_q]\in \mathbb{C}^{p\times q}\\
            \mathbf{V}^T = [\mathbf{v}_1\;\dotsc\;\mathbf{v}_q]\in \mathbb{C}^{m\times q}
        \end{matrix}\Bigg\}
    \end{gathered}
\label{eq:LID}
\end{equation}

\noindent The indices $\lambda_i$ and $\mu_j$ represent, respectively, the values at which the right ($\mathbf{R}$) and left ($\mathbf{L}$) subsets are sampled to produce the approximation $\mathbf{H}(s)$, while $\rho$ and $q$ correspond to their dimension. The vectors $\mathbf{w}_i$ and $\mathbf{v}_j$ denote the right and left tangential general directions, typically chosen randomly in practice \cite{Quero2019}, while the values $\mathbf{r}_i$ and $\mathbf{l}_j$  represent the right and left tangential data. 

The rational interpolation problem is resolved when the transfer function $\mathbf{H}(s)$, associated with realisation $\mathbf{\Sigma}$ in \cref{eq:LTI}, is linked to $\mathbf{w}_i$ and $\mathbf{v}_j$ in such a way that the Loewner pencil satisfies \cref{eq:LS2}:
\begin{equation}
 \begin{split}
\mathbf{H}(\lambda_i)\mathbf{r}_i=\mathbf{w}_i,\:i=1.\dots,\rho\\
\mathbf{l}_j\mathbf{H}(\mu_j)=\mathbf{v}_j,\:j=1,\dots,q
 \end{split}
 \label{eq:LS2}
 \end{equation}
\noindent such that the Loewner pencil satisfies \cref{eq:LS2}.

Now, given a set of points $Z=\{z_1,\dots,z_N\}$ in the complex plane, the corresponding value of $\mathbf{H}(s)$ can be separated into left $\lambda_{i=1...\rho}$ and right $\mu_{j=1...q}$ data with $N=q+\rho$:

\begin{equation}
\begin{split}
	Z=\{\lambda_1,\dots,\lambda_\rho\} \cup \{\mu_1,\dots,\mu_q\}\\
	Y=\{\mathbf{w}_1,\dots,\mathbf{w}_\rho\} \cup \{\mathbf{v}_1,\dots,\mathbf{v}_q\}
\end{split}
\label{eq:ZY}
\end{equation}
\noindent Hence, the matrix $\boldsymbol{\mathbb{L}}$ becomes:
\begin{equation}
\label{eq:LM2}
\boldsymbol{\mathbb{L}}=\begin{bmatrix}
\frac{\mathbf{v}_1\mathbf{r}_1-\mathbf{l}_1\mathbf{w}_1}{\mu_1-\lambda_1} & \cdots & \frac{\mathbf{v}_1\mathbf{r}\rho-\mathbf{l}_1\mathbf{w}\rho}{\mu_1-\lambda\rho}\\
\vdots & \ddots & \vdots\\
\frac{\mathbf{v}_q\mathbf{r}_1-\mathbf{l}_q\mathbf{w}_1}{\mu_q-\lambda_1}& \cdots & \frac{\mathbf{v}_q\mathbf{r}\rho-\mathbf{l}_q\mathbf{w}\rho}{\mu_q-\lambda\rho}\\
\end{bmatrix}\:\in \mathbb{C}^{q\times \rho}
\end{equation}

\noindent Since $\mathbf{v}_q\mathbf{r}_\rho$ and $\mathbf{l}_q\mathbf{w}_\rho$ are scalars, the Sylvester equation is satisfied by $\boldsymbol{\mathbb{L}}$ in such a fashion:
\begin{equation}
    \boldsymbol{\mathbb{L}}\mathbf{\Lambda}-\mathbf{M}\boldsymbol{\mathbb{L}}=\mathbf{L}\mathbf{W}-\mathbf{V}\mathbf{R}
    \label{eq:syl}
\end{equation}

\noindent Thus, the \emph{shifted Loewner matrix}, $\boldsymbol{\mathbb{L}}_s$, that is the $\boldsymbol{\mathbb{L}}$ corresponding to $s\mathbf{H}(s)$, can be defined as:

\begin{equation}
\label{eq:LS}
\boldsymbol{\mathbb{L}}_s=\begin{bmatrix}
\frac{\mu_1\mathbf{v}_1\mathbf{r}_1-\lambda_1\mathbf{l}_1\mathbf{w}_1}{\mu_1-\lambda_1} & \cdots & \frac{\mu_1\mathbf{v}_1\mathbf{r}_\rho-\lambda_\rho\mathbf{l}_1\mathbf{w}_\rho}{\mu_1-\lambda_\rho}\\
\vdots & \ddots & \vdots\\
\frac{\mu_q\mathbf{v}_v\mathbf{r}_1-\lambda_1\mathbf{l}_q\mathbf{w}_1}{\mu_q-\lambda_1}& \cdots & \frac{\mathbf{v}_q\mathbf{r}_\rho-\mathbf{l}_q\mathbf{w}_\rho}{\mu_q-\lambda_\rho}\\
\end{bmatrix}\:\in \mathbb{C}^{q\times \rho}
\end{equation}

\noindent Similarly, the Sylvester equation can be fulfilled now as follows:
\begin{equation}
    \boldsymbol{\mathbb{L}}_s\Lambda-\mathbf{M}\boldsymbol{\mathbb{L}}_s=\mathbf{L}\mathbf{W}\mathbf{\Lambda}-\mathbf{M}\mathbf{V}\mathbf{R}
    \label{eq:syl2}
\end{equation}

\noindent Without loss of generality, $\mathbf{D}$ can be considered 0 since its contribution does not affect the tangential interpolation of LF~\cite{Mayo2007}. Thus, for ease of presentation, the remainder of the presentation will focus on:
\begin{equation}
\mathbf{H}(s)=\mathbf{C}(s\mathbf{E}-\mathbf{A})^{-1}\mathbf{B}
\label{eq:fin}
\end{equation}

\noindent A realisation of minimal dimension is achievable only when the system is fully controllable and observable. Thus, if the data is assumed to be sampled from a system where the transfer function is represented by \cref{eq:fin}, the generalised tangential observability, $\mathcal{O}_q$, and generalised tangential controllability matrices  $\mathcal{R}_\rho$ can be derived from \cref{eq:syl2,eq:fin}~\cite{Lefteriu2010b}: as follows:
\begin{equation}
    \mathcal{O}_q=\begin{bmatrix}
    \mathbf{l}_1\mathbf{C}(\mu_1\mathbf{E}-\mathbf{A})^{-1}\\
    \vdots\\
   \mathbf{l}_q\mathbf{C}(\mu_q\mathbf{E}-\mathbf{A})^{-1}
    \end{bmatrix}\:\in \mathbb{R}^{q\times k}
    \label{eq:Oq}
\end{equation}
\begin{equation}
    \mathcal{R}_\rho=\begin{bmatrix}
    (\lambda_1\mathbf{E}-\mathbf{A})^{-1}\mathbf{Br}_1 &
    \cdots &
    (\lambda_\rho\mathbf{E}-\mathbf{A})^{-1}\mathbf{Br}_\rho 
    \end{bmatrix}\:\in \mathbb{R}^{k\times \rho}
    \label{eq:Rk}
\end{equation}

\noindent Now, combining \cref{eq:Oq,eq:Rk} with, respectively, \cref{eq:LM2,eq:LS} yields:

\begin{equation}
\begin{split}
    \boldsymbol{\mathbb{L}}_{j,i} = \dfrac{\mathbf{v}_j\mathbf{r}_i-\mathbf{l}_j\mathbf{w}_i}{\mu_j-\lambda_i}=\dfrac{\mathbf{l}_j\mathbf{H}(\mu_i)\mathbf{r}_i-\mathbf{l}_j\mathbf{H}(\lambda_i)\mathbf{r}_i}{\mu_j-\lambda_i}=\\=
    -\mathbf{l}_j\mathbf{C}(\mu_j\mathbf{E-\mathbf{A}})^{-1}\mathbf{E}(\lambda_i\mathbf{E}-\mathbf{A})^{-1}\mathbf{Br}_i
\end{split}
\label{eq:LH}
\end{equation}

\begin{equation}
\begin{split}
    (\boldsymbol{\mathbb{L}}_s)_{j,i} = \dfrac{\mu_j\mathbf{v}_j-\lambda_i\mathbf{l}\mathbf{w}_i}{\mu_j-\lambda_i}
    =\dfrac{\mu_j\mathbf{l}_j\mathbf{H}(\mu_i)\mathbf{r}_i-\lambda_i\mathbf{l}_j\mathbf{H}(\lambda_i)\mathbf{r}_i}{\mu_j-\lambda_i}=\\= -\mathbf{l}_j\mathbf{C}(\mu_j\mathbf{E}-\mathbf{A})^{-1}\mathbf{A}(\lambda_i\mathbf{E}-\mathbf{A})^{-1}\mathbf{Br}_i
\end{split}
\label{eq:lLH}
\end{equation}

\noindent Thus, for minimal data, where the right and left data have equal length ($p = v$), the following is required:
\begin{align}
    \boldsymbol{\mathbb{L}}=-\mathcal{O}_q\mathbf{E}\mathcal{R}_\rho &&
    \boldsymbol{\mathbb{L}}_s=-\mathcal{O}_q\mathbf{A}\mathcal{R}_\rho
    \label{eq:LLs}
\end{align}

\noindent This leads to the following system matrix assignment:
\begin{align}
    \mathbf{E}=-\boldsymbol{\mathbb{L}},&&\mathbf{A}=-\boldsymbol{\mathbb{L}}_s,&&\mathbf{B}=\mathbf{V},&&\mathbf{C}=\mathbf{W}
\end{align}
\noindent Consequently, the interpolating rational function can be expressed now as:
\begin{equation}
    \mathbf{H}(s)=\mathbf{W}(\boldsymbol{\mathbb{L}}_s-s\boldsymbol{\mathbb{L}})^{-1}\mathbf{V}
\end{equation}
\noindent The previous derivation applies to the minimal data case, which is hardly ever the case when working with real-world data. However, the LF approach can be extended to redundant data. To begin, assume: 

\begin{equation}
\begin{split}
    \mathrm{rank}[\zeta\boldsymbol{\mathbb{L}}-\boldsymbol{\mathbb{L}}_s]=\mathrm{rank}[\boldsymbol{\mathbb{L}}\:\boldsymbol{\mathbb{L}}_s]=\\
   =\mathrm{rank}
    \begin{bmatrix}
    \boldsymbol{\mathbb{L}}\\ \boldsymbol{\mathbb{L}}_s
    \end{bmatrix}=k,\; 
    \forall\; \zeta \in \{\lambda_j\}\cup\{\mu_i\}
    \end{split}
    \label{eq:cond1}
\end{equation}
\noindent Then, the short Singular Value Decomposition (SVD) of $\zeta\boldsymbol{\mathbb{L}}-\boldsymbol{\mathbb{L}}_s$ is computed:
\begin{equation}
    \textrm{svd}(\zeta\boldsymbol{\mathbb{L}}-\boldsymbol{\mathbb{L}}_s)=\mathbf{Y}\mathbf{\Sigma}_l\mathbf{X}
\label{eq:cond2}
\end{equation}
\noindent where $\mathrm{rank}(\zeta\boldsymbol{\mathbb{L}}-\boldsymbol{\mathbb{L}}_s)=\mathrm{rank}(\mathbf{\Sigma}_l)=\mathrm{size}(\mathbf{\Sigma}_l)=k,\mathbf{Y}\in\mathbb{C}^{q \times k}$ and $\mathbf{X}\in\mathbb{C}^{k\times \rho}$.
Further, note that:
\begin{equation}
\begin{split}
-\mathbf{A}\mathbf{X}+\mathbf{E}\mathbf{X}\mathbf{\Sigma}_l = \mathbf{Y}^*\boldsymbol{\mathbb{L}}_s\mathbf{X}^*\mathbf{X}-\mathbf{Y}^*\boldsymbol{\mathbb{L}}\mathbf{X}^*\mathbf{X}\mathbf{\Sigma}_l=\\=\mathbf{Y}^*(\boldsymbol{\mathbb{L}}_s-\boldsymbol{\mathbb{L}}\mathbf{\Sigma}_l )=\mathbf{Y}^*\mathbf{V}\mathbf{R}=\mathbf{BR}
\end{split}
\label{eq:cond3}
\end{equation}
and likewise, $-\mathbf{Y}\mathbf{A}+\mathbf{M}\mathbf{Y}\mathbf{E}=\mathbf{L}\mathbf{C}$ such that $\mathbf{X}$ and $\mathbf{Y}$ are the generalised controllability and observability matrices for the system $\mathbf{\Sigma}$ with $\mathbf{D}=0$. After verifying the right and left interpolation conditions, the Loewner realisation for redundant data is as follows:
\begin{equation}
\begin{split}
\mathbf{E}=-\mathbf{Y}^*\boldsymbol{\mathbb{L}}\mathbf{X},\quad \quad
\mathbf{A}=-\mathbf{Y}^*\boldsymbol{\mathbb{L}}_s\mathbf{X,}\\
\mathbf{B}=\mathbf{Y}^*\mathbf{V},\quad \quad
\mathbf{C}=\mathbf{W}\mathbf{X}
\end{split}
\label{eq:real}
\end{equation}

Finally, the system modal parameters can be extracted through an eigen-analysis of the system matrices $\mathbf{A}$ and $\mathbf{C}$ in \cref{eq:real}. This was the main contribution of \cite{Dessena2022}. The reader is encouraged to consult \cite{Mayo2007,Ionita2014} for a more comprehensive discussion of each step of the LF, \cite{Dessena2024} for insights on the iLF computational implementation, and \cite{Dessena2024e} for its MATLAB implementation. 

\subsection{Damage assessment, quantification, and localisation}\label{sec:dam}
As hinted in \cref{sec:int}, damage can be assessed by identifying a difference in $\omega_n$ between the baseline and a successive structural state. Furthermore, quantifying this difference can pinpoint the severity level of the damage, e.g., higher differences might be related to a more severe damage state. On the other hand, localisation is achieved by comparing the baseline and damaged $\bm{\phi}_n$, where deviations from the baseline shape indicate the damage location. These methodologies are well-established in the literature. Nonetheless, these deviations can also be used for assessment and quantification since a more significant deviation corresponds to a more severe damage scenario. Nevertheless, these comparisons become challenging to visualise and interpret for complex structures. Hence, in this work, the complement of the well-known MTMAC \cite{Perera2006} is proposed for damage assessment, to pair the $\omega_n$ and $\bm{\phi}_n$ contributions to damage assessment and quantification: 
\begin{equation}
    \text{MTMAC}= 1-\prod^n_{i=1}\dfrac{\text{MAC}(\bm{\phi}_n^E,\bm{\phi}_n^N)}{1+\left|\frac{\omega_i^N-\omega_i^E}{\omega_i^N+\omega_i^E}\right|}
    \label{eq:mtamac}
\end{equation}
Note that the subscript $i$ is the mode of interest, and the superscripts $E$ and $N$ are, respectively, the baseline and damaged condition. The MAC is defined as per the classical work of Allemang \& Brown \cite{Allemang1982}:
\begin{equation}
    \text{MAC}(i,j)= \dfrac{\left|(\bm{\phi}_i^E)^T(\bm{\phi}_j^N)\right|^2}{\left((\bm{\phi}_i^E)^T(\bm{\phi}_i^E)\right)\left((\bm{\phi}_j^N)^T(\bm{\phi}_j^N)\right)}
    \label{eq:mac}
\end{equation}
where subscripts $i$ and $j$ are the two modes of interest. Since the MTMAC is a function of both $\omega_n$ and $\bm{\phi}_n$, its value diminishes proportionally with the deviation between the baseline and damaged modal parameters, due to the local variations in mass or stiffness. On the other hand, damage localisation suffers from the same issues as damage quantification. Therefore, in this work, the COMAC \cite{Lieven1988} is used for damage localisation since it has been historically used for determining the local sensitivity to system changes between two sets of $\bm{\phi}_n$. The COMAC is defined as:
\begin{equation}
    \text{COMAC}(p)= \dfrac{\sum_{i=1}^n\left|(\bm{\phi}_{p,i}^E)(\bm{\phi}_{p,i}^N)\right|^2}{\sum_{i=1}^n(\bm{\phi}_{p,i}^E)^2\sum_{i=1}^n(\bm{\phi}_{p,i}^N)^2}
    \label{eq:comac}
\end{equation}
where $p$ represents the corresponding DoF of the system.

To benchmark the use of the MTMAC as a damage assessment and severity index and the effectiveness of the COMAC for damage localisation, two traditional damage quantification and localisation methods will also be applied to the same case study: the absolute difference of $\omega_n$ in percentage and $\bm{\phi}_n$ trajectory deviations. 
\subsection{A novel two-step modal \& damage identification method}\label{sec:fra}




{As many SHM approaches rely on identified modal parameters, a more accurate and efficient method can be developed by combining the two tools introduced earlier: the iLF and the MTMAC.

The proposed SHM workflow consists of:}
\begin{enumerate}
    \item {Modal identification of the system;}

    \item {Using the identified parameters to validate the MTMAC for damage assessment and quantification;}

    \item {Applying the COMAC, for completeness, to determine the location of the damage.}
\end{enumerate}

{To benchmark this method, the conventional LSCE and SSI-CVA techniques are employed, and their results are compared with those obtained using the iLF. In addition, a traditional vibration-based damage assessment, quantification, and localisation approach, based on comparing $\omega_n$ and $\bm{\phi}_n$, is applied to further validate the MTMAC as a damage index.}

\section{Numerical case study}\label{sec:num}
To assess the suitability of the iLF and MTMAC for damage detection, a numerical case study is presented using a 2D hollow rectangular cross-section cantilever beam. 
{The rationale here is to isolate the performance of the modal extraction algorithm (iLF) and damage index (MTMAC) in an intentionally simple benchmark, fully controlled, without unmodeled nonlinearities (e.g., joints and contacts) that can introduce confounding effects when interpreting damage indices. The paired full-scale aircraft experiment, which will be discussed in the next Section, will provide the necessary complexity for real-life validity.}

\begin{figure}[!htb]
\centering
		{\includegraphics[align=c,width=.45\textwidth,keepaspectratio]{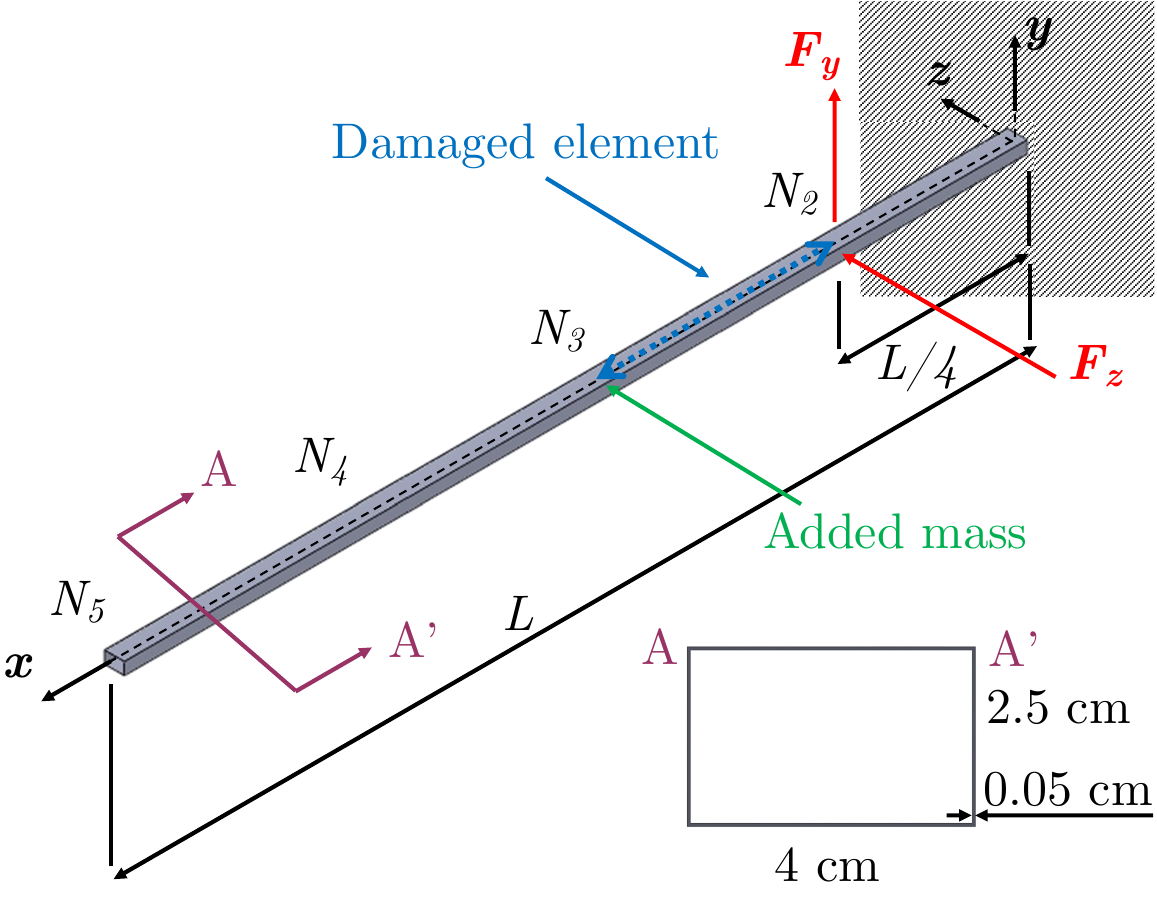}}
	\caption{Numerical case study: schematic of the 2D beam with the dimensions {of its hollow rectangular} cross-section. Not in scale.}
	\label{fig:fig1}
\end{figure}

The beam {hollow rectangular} cross-section is shown in \cref{fig:fig1}. It is made of aluminium, with a Young’s modulus (E) of 69 GPa and density ($\rho$) of 2700 kgm\textsuperscript{-3}, and a length (L) of 1.8 m. The beam is discretised, in MATLAB, into four 2D (y and z displacements) Euler-Bernoulli beam elements of equal length. Sample beam element mass and stiffness matrices are available in \hyperref[sec:apa]{Appendix A}. {The beam is fully constrained in all DoFs at the wall, }$N_1$.

\begin{figure}[!htb]
\centering
	\begin{subfigure}[t]{.49\textwidth}
	\centering
		{\includegraphics[width=\textwidth,keepaspectratio]{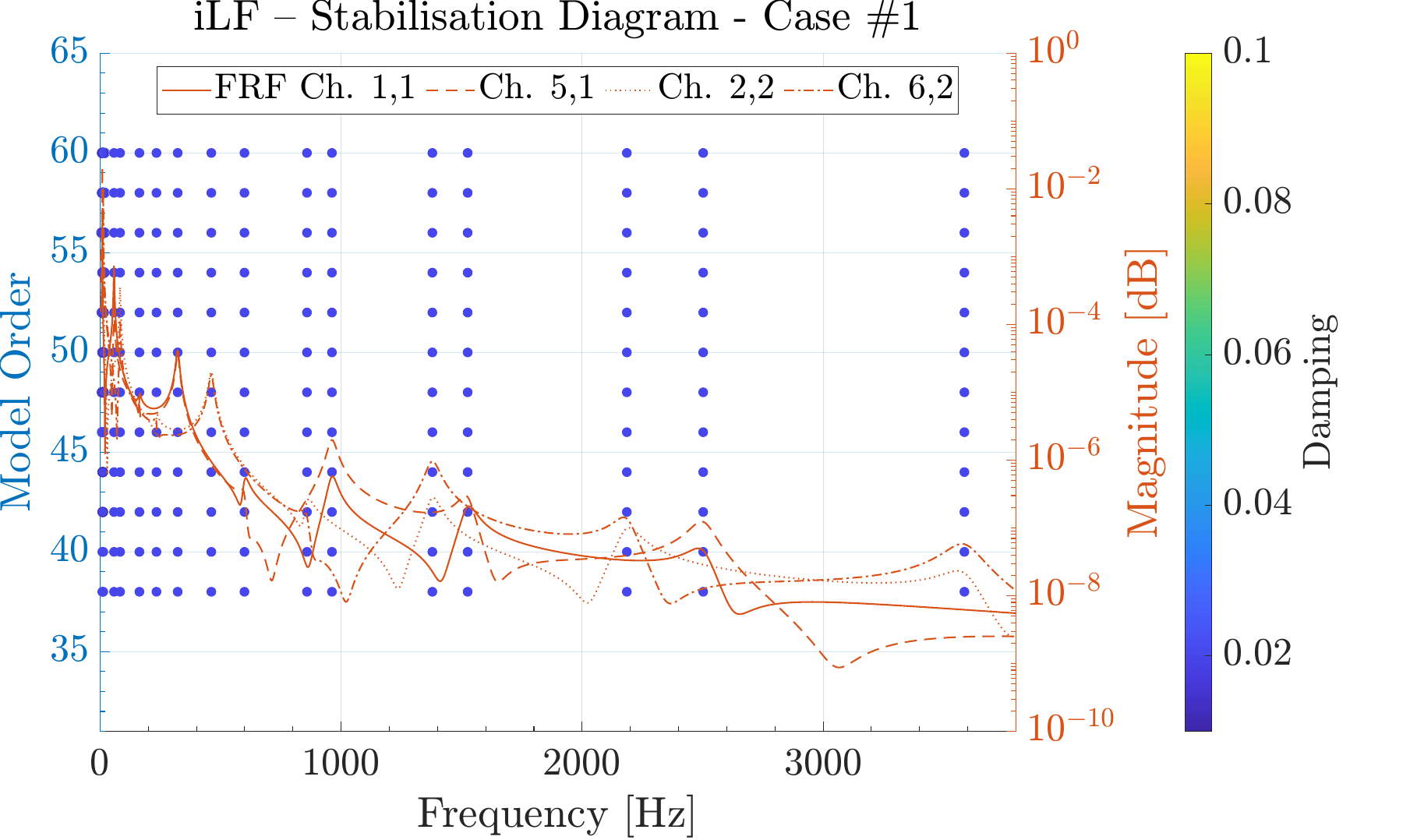}}
		\captionsetup{font={it},justification=centering}
		\subcaption{\label{fig:fig2a}}	
	\end{subfigure}
    \begin{subfigure}[t]{.49\textwidth}
	\centering
		\includegraphics[width=\textwidth,keepaspectratio]{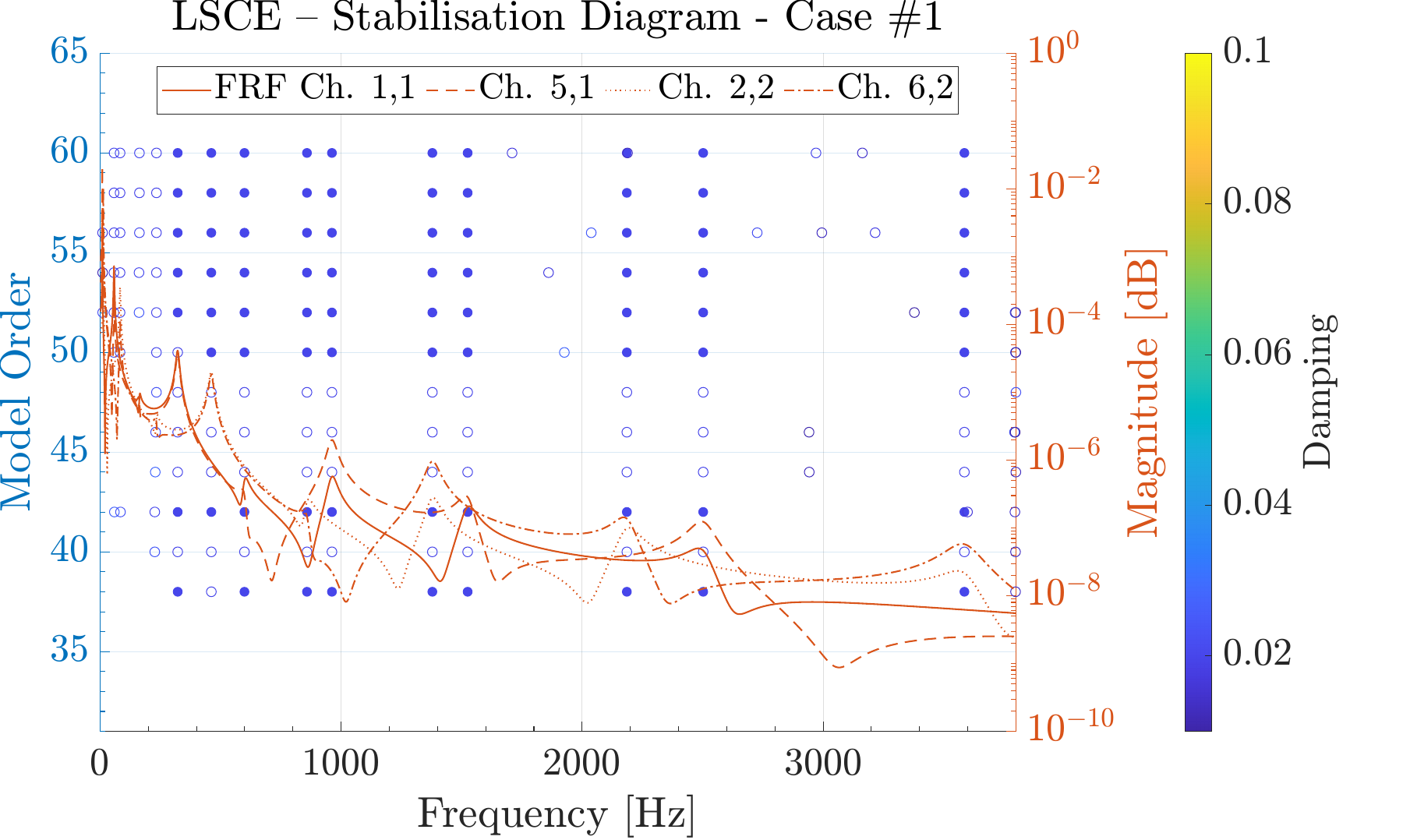}
		\captionsetup{font={it},justification=centering}
		\subcaption{\label{fig:fig2b}}	
	\end{subfigure}
    \begin{subfigure}[t]{.49\textwidth}
	\centering
		{\includegraphics[width=\textwidth,keepaspectratio]{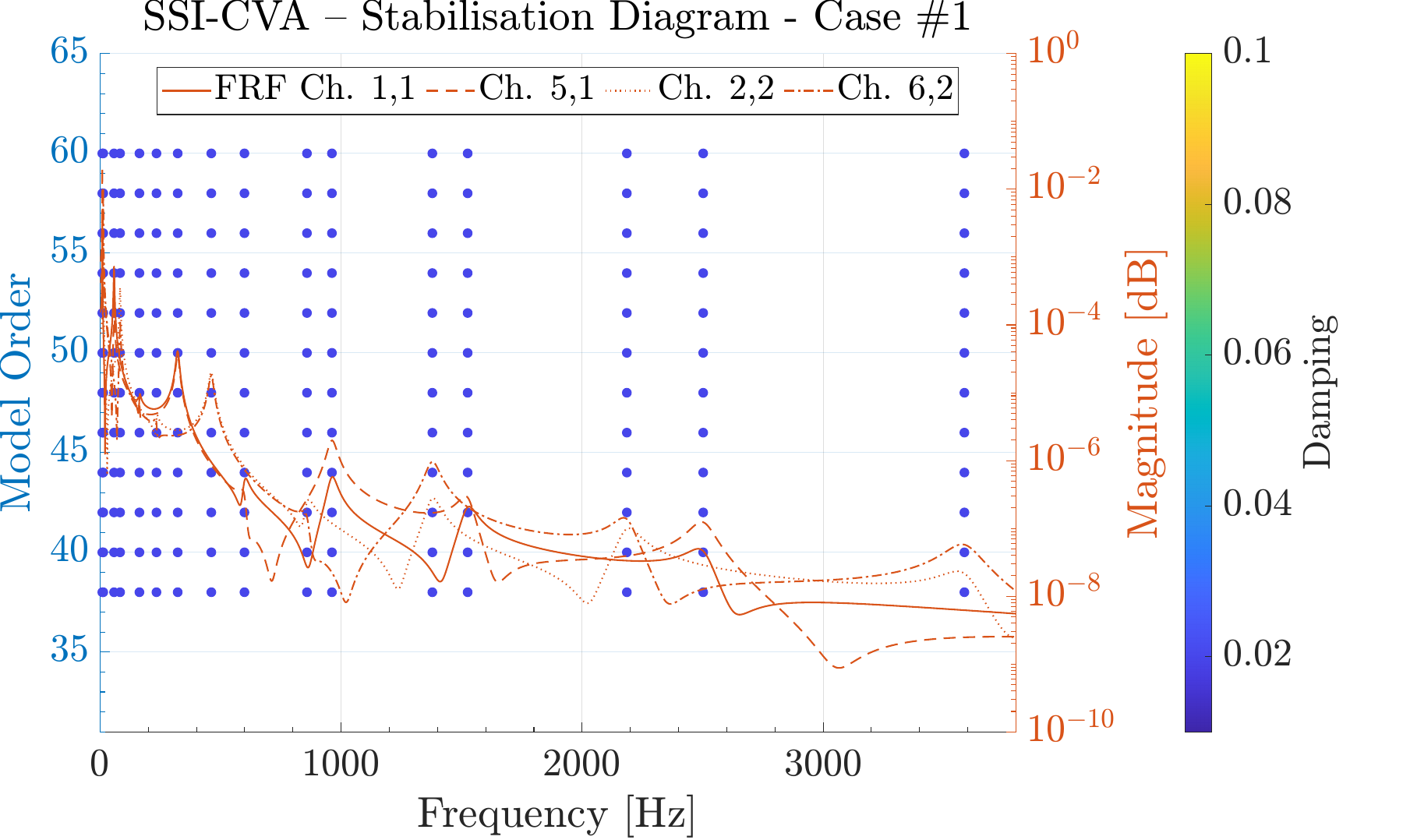}}
		\captionsetup{font={it},justification=centering}
		\subcaption{\label{fig:fig2c}}	
	\end{subfigure}
	\caption{Numerical case study: Stabilisation diagrams for the modes identified via iLF (\cref{fig:fig2a}), LSCE (\cref{fig:fig2a}), and SSI-CVA (\cref{fig:fig2c}) for the baseline scenario.}
	\label{fig:fig2}
\end{figure}

For the numerical MIMO experiment, two 1 N step input forces are applied in the y and z directions at node \#2 ($N_2$ -- $F_y$ and $F_z$ act at L/4 from the origin) {at time zero of a 60 s recording}. These and the displacements in the two directions are recorded at a sampling frequency of fs = 16384 Hz\footnote{This frequency was required to simulate the resulting mass-spring-damper system using the \texttt{lsim} function (\url{https://uk.mathworks.com/help/control/ref/dynamicsystem.lsim.html})  in MATLAB}, thus allowing the inspection of all modes up to 8192 Hz as per the Nyquist criterion. This enables the identification of all system modes. For all the modes, a 2\% modal damping ratio is superimposed{, using the uncoupled damping assumption.}
In addition to the baseline case (case \#1) described above, the stiffness of the second element (see \cref{fig:fig1}) is reduced by 5\% (case \#2), 10\% (case \#3), and 20\% (case \#4) to simulate different damage scenarios on the beam. In addition, a fifth case (case \#5) is considered, consisting of a mass addition of 100 g (just under 20\% of the beam mass) to node 3, simulating an underwing payload or an engine pod mount assembly. This case allows us to examine the method sensitivity in distinguishing between damage and mass changes.
{The analytical results are obtained by simple eigenanalysis of the system mass and stiffness matrices. The modal damping is considered when computing the damped natural frequencies.} 
For the identification process, the output displacement time series resulting from the excitation above are recorded and used to obtain the compliance FRFs in the usual fashion: Elementwise division of the output spectra by the input spectra. This process is repeated for the two inputs, thus resulting, in MATLAB notation, in an \texttt{m:n:p} matrix, where \texttt{m} is the number of outputs, \texttt{n} is the number of inputs, and \texttt{p} is the number of frequency bins. In this case, there are eight displacement output channels, one per direction (y and z) at each node, and two force input channels. LSCE and iLF require the FRFs, while the output time series are used with SSI-CVA. \Cref{tab:tab1} shows the $\omega_{n}$ and $\zeta_{n}$ identified analytically and by applying the iLF and two traditional methods, SSI-CVA \cite{VanOverschee1996} and LSCE \cite{Zimmerman2017}, to the baseline case. Notably, only the displacement DoFs are considered in the analysis. The percentage differences with respect to the analytical values are also reported (in brackets). The modal parameters are obtained with the help of stabilisation diagrams, as shown in \cref{fig:fig2}, for model orders ranging from 24 to 50. {All numerical system analytical, numerical, and identified results are obtained in MATLAB 2024b.}

\begin{table*}[!htb]
\centering
\caption{Numerical case study: Analytical and identified $\omega_{n}$ and $\zeta_{n}$ from the Baseline case}\label{tab:tab1}
{\footnotesize
\begin{tabular}{@{}ccccccccc@{}}
\toprule
\multirow{3}{*}{Mode \#} &
\multicolumn{4}{c}{\textbf{Natural Frequency [Hz]} – (difference [\%])} &
\multicolumn{4}{c}{\textbf{Damping Ratio [-]} – (difference [\%])} \\
\cmidrule(lr){2-5} \cmidrule(lr){6-9}
& Analytical & iLF & SSI-CVA & LSCE & Analytical & iLF & SSI-CVA & LSCE \\
\midrule
1  & 9.23 & 9.23 & 9.23 & -& 0.02 & 0.02 & 0.02 & -\\
   & (0.00) & (0.00) & (0.00) & (NA) & (0.00) & (0.00) & (0.00) & (NA) \\
2  & 13.23 & 13.23 & 13.23 & -& 0.02 & 0.02 & 0.02 & -\\
   & (0.00) & (0.00) & (0.00) & (NA) & (0.00) & (0.00) & (0.00) & (NA) \\
3  & 57.92 & 57.92 & 57.92 & -& 0.02 & 0.02 & 0.02 & -\\
   & (0.00) & (0.00) & (0.00) & (NA) & (0.00) & (0.00) & (0.00) & (NA) \\
4  & 83.01 & 83.01 & 83.01 & -& 0.02 & 0.02 & 0.02 & -\\
   & (0.00) & (0.00) & (0.00) & (NA) & (0.00) & (0.00) & (0.00) & (NA) \\
5  & 163.25 & 163.25 & 163.25 & -& 0.02 & 0.02 & 0.02 & -\\
   & (0.00) & (0.00) & (0.00) & (NA) & (0.00) & (0.00) & (0.00) & (NA) \\
6  & 233.96 & 233.96 & 233.96 & -& 0.02 & 0.02 & 0.02 & -\\
   & (0.00) & (0.00) & (0.00) & (NA) & (0.00) & (0.00) & (0.00) & (NA) \\
7  & 322.06 & 322.06 & 322.06 & 322.06 & 0.02 & 0.02 & 0.02 & 0.02 \\
   & (0.00) & (0.00) & (0.00) & (0.00) & (0.00) & (0.00) & (0.00) & (0.00) \\
8  & 461.54 & 461.54 & 461.54 & 461.54 & 0.02 & 0.02 & 0.02 & 0.02 \\
   & (0.00) & (0.00) & (0.00) & (0.00) & (0.00) & (0.00) & (0.00) & (0.00) \\
9  & 599.02 & 599.02 & 599.02 & 599.02 & 0.02 & 0.02 & 0.02 & 0.02 \\
   & (0.00) & (0.00) & (0.00) & (0.00) & (0.00) & (0.00) & (0.00) & (0.00) \\
10 & 858.45 & 858.45 & 858.45 & 858.45 & 0.02 & 0.02 & 0.02 & 0.02 \\
   & (0.00) & (0.00) & (0.00) & (0.00) & (0.00) & (0.00) & (0.00) & (0.00) \\
11 & 962.02 & 962.02 & 962.02 & 962.02 & 0.02 & 0.02 & 0.02 & 0.02 \\
   & (0.00) & (0.00) & (0.00) & (0.00) & (0.00) & (0.00) & (0.00) & (0.00) \\
12 & 1378.67 & 1378.67 & 1378.67 & 1378.67 & 0.02 & 0.02 & 0.02 & 0.02 \\
   & (0.00) & (0.00) & (0.00) & (0.00) & (0.00) & (0.00) & (0.00) & (0.00) \\
13 & 1525.13 & 1525.13 & 1525.13 & 1525.13 & 0.02 & 0.02 & 0.02 & 0.02 \\
   & (0.00) & (0.00) & (0.00) & (0.00) & (0.00) & (0.00) & (0.00) & (0.00) \\
14 & 2185.65 & 2185.65 & 2185.65 & 2185.65 & 0.02 & 0.02 & 0.02 & 0.02 \\
   & (0.00) & (0.00) & (0.00) & (0.00) & (0.00) & (0.00) & (0.00) & (0.00) \\
15 & 2502.41 & 2502.41 & 2502.41 & 2502.41 & 0.02 & 0.02 & 0.02 & 0.02 \\
   & (0.00) & (0.00) & (0.00) & (0.00) & (0.00) & (0.00) & (0.00) & (0.00) \\
16 & 3586.19 & 3586.19 & 3586.19 & 3586.19 & 0.02 & 0.02 & 0.02 & 0.02 \\
   & (0.00) & (0.00) & (0.00) & (0.00) & (0.00) & (0.00) & (0.00) & (0.00) \\
\bottomrule
\end{tabular}}
\end{table*}

\begin{figure*}[!htb]
\centering
{\includegraphics[align=c,width=.6\textwidth,keepaspectratio]{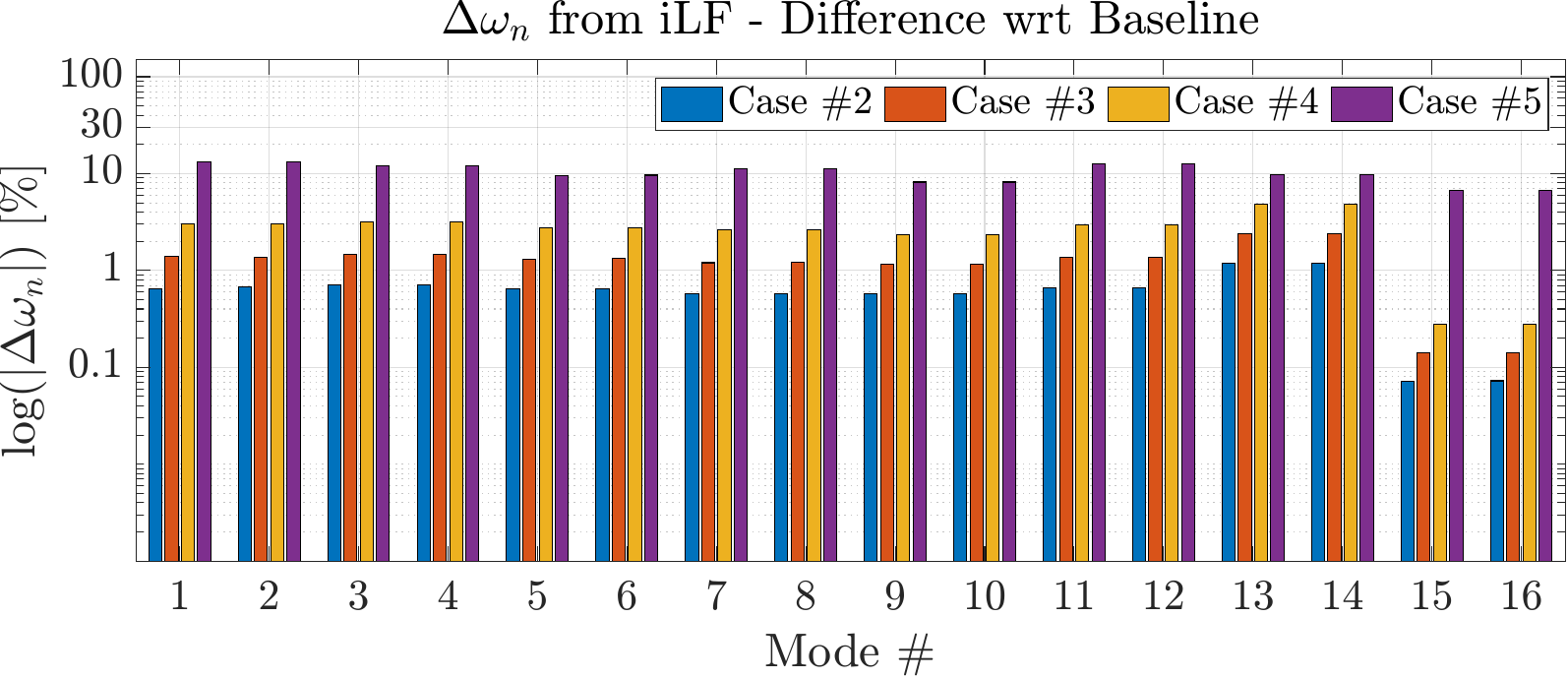}}
	\caption{{Numerical case study: Difference between the iLF-identified $\omega_n$ of the baseline scenario and Cases \#2-5.}}
	\label{fig:fig3}
\end{figure*}

The maximum error of the identified parameters ($\omega_{n}$ and $\zeta_{n}$) reported in \cref{tab:tab1} is 0\%. However, LSCE fails to identify lower modes in higher-dimensional systems with closely spaced modes, such as modes 1-2, as well as modes 3-6. These results align with expectations, as this issue is well-documented in the literature \cite{Dessena2024,Dessena2022}. Conversely, \cref{fig:fig2} clearly shows that the stabilisation diagrams from iLF (\cref{fig:fig2a}) and SSI-CVA (\cref{fig:fig2c}) can accurately identify the stable modes, while for LSCE (\cref{fig:fig2b}) this is troublesome, resulting in only ten (out of sixteen) modes identified as stable. 
{The stabilisation diagrams in question are constructed by varying the system order 
from $k_{\min}=32$ to $k_{\max}=60$ with an increment $\Delta k = 2$, yielding the extraction set $\mathbf{k}=\{\,k_{\min}:\Delta k:k_{\max}\,\}$. 
Hard criteria are imposed through the admissible damping range $\zeta_n \in [\zeta_{\min}, \zeta_{\max}] = [0.005,\,0.03]$ and the natural frequency range 
$\omega_n \in [\omega_{\min}, \omega_{\max}] = [0,\,3800]$. Soft criteria include frequency stabilisation $\Delta \omega_{\mathrm{stab}} = 0.01$, damping stabilisation $\Delta \zeta_{\mathrm{stab}} = 0.05$, and MAC stabilisation $\mathrm{MAC}_{\mathrm{stab}} = 0.95$. Additionally, the number of MAC comparisons $N_{\mathrm{MAC}} = 10$ are considered in the selection process.}
It should be noted, although not included in \cref{tab:tab1}, that the MAC between the analytical $\bm{\phi}_n$ and those identified via iLF and SSI-CVA is equal to, or very close to, 1. On the other hand, the LSCE identified $\bm{\phi}_n$ MAC, when compared to the analytical values, never exceeds 0.58.
 
These results demonstrate that the iLF is, for MIMO experimental modal analysis testing, (1) at least as accurate as SSI-CVA and (2) significantly more accurate than LSCE. In this instance, cases with artificially added white Gaussian noise are not considered, as it has already been demonstrated in \cite{Dessena2024} that measurement noise has a very limited effect on the accuracy of the iLF identification.

\subsection{Investigation of damage effects}
Given the good accuracy and precision shown for the baseline case by the iLF when compared to the SSI-CVA and LSCE methods, only the iLF-derived modal parameters are used for damage assessment in the numerical case study. {It must be noted that performing these using the analytical modal parameters yields the same results as those presented below, since the iLF identified modal parameters perfectly match their analytical counterparts.} The iLF capability to support damage detection will be assessed by applying it to the four other cases (cases \#2-5) and determining if a relationship exists between the absolute difference of the damaged and baseline $\omega_n$. In addition, the $\bm{\phi}_n$ change across the different cases will be used for damage localisation.

\begin{figure*}[!htb]
\centering
		{\includegraphics[align=c,width=.85\textwidth,keepaspectratio]{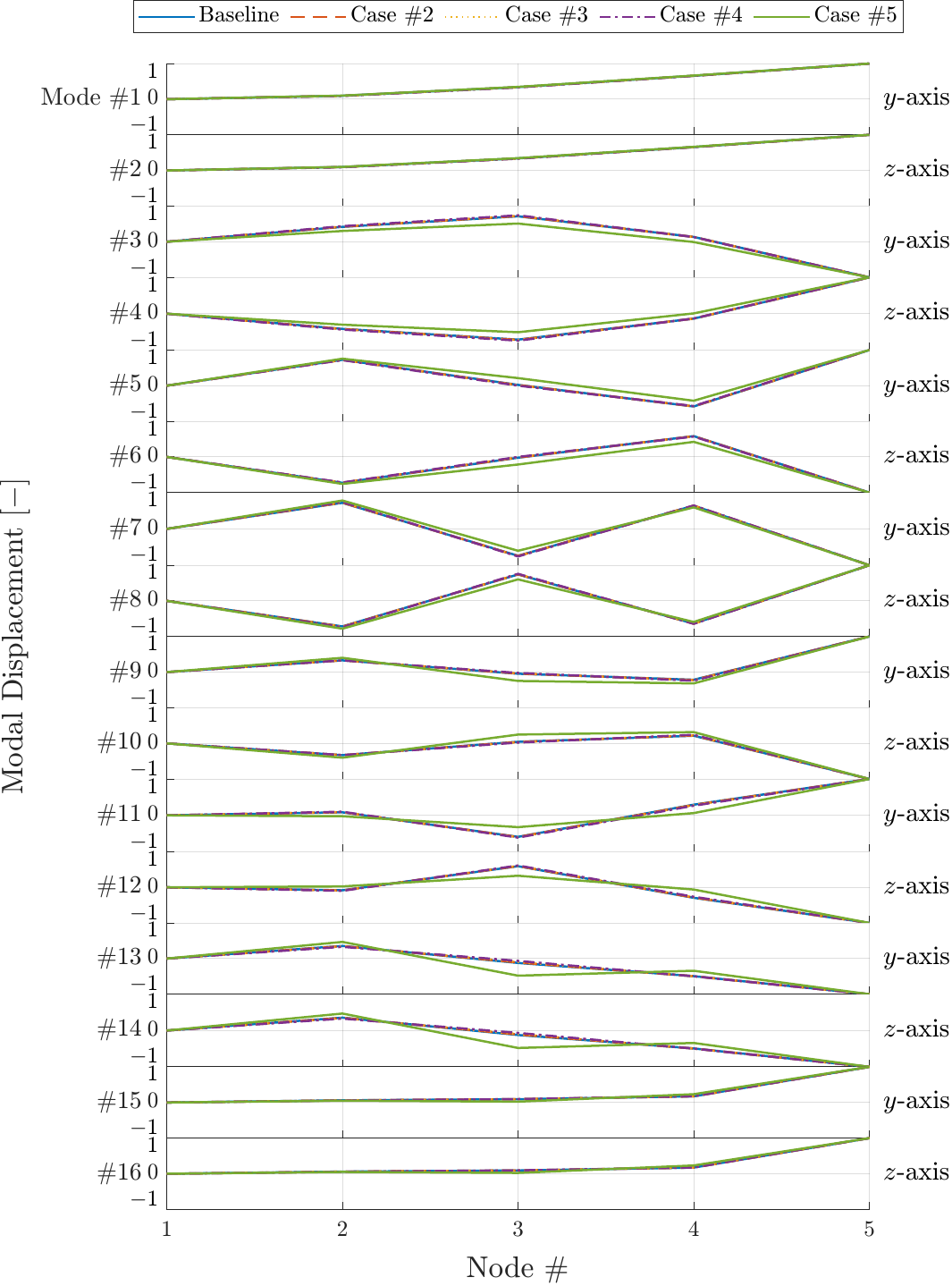}}
	\caption{Numerical case study: $\bm{\phi}_n$ of the modes identified by iLF. Note that the secondary y-axis labels represent the dominant displacement shown.}
	\label{fig:fig4}
\end{figure*}

\Cref{fig:fig3} shows the aforementioned absolute difference of natural frequencies ($\text{log}|\Delta\omega_n|$) between the damaged and baseline cases, while \Cref{fig:fig4} displays the dominant axis displacement (as marked in the plot secondary y-axis) for all $\bm{\phi}_n$.

From \cref{fig:fig3}, it is clear that a relationship between damage intensity and identified frequency is confirmed, as the difference increases with increased damage. It should be noted that despite the absolute value plot, all changes are negative, as expected for stiffness reductions and mass additions. On the other hand, a more notable change is observed for the case with the added mass, which, similarly, exhibits a clear distinction from the baseline case. In addition, the iLF captured the slight differences in the identified $\omega_n$, as those shown for modes \#15-16, which are well under 0.5\% in absolute value. Hence, it can be said that the iLF can perform $\omega_n$-based damage detection and damage severity assessment on a numerical system.

\begin{figure}[!htb]
\centering
		{\includegraphics[align=c,width=.55\textwidth,keepaspectratio]{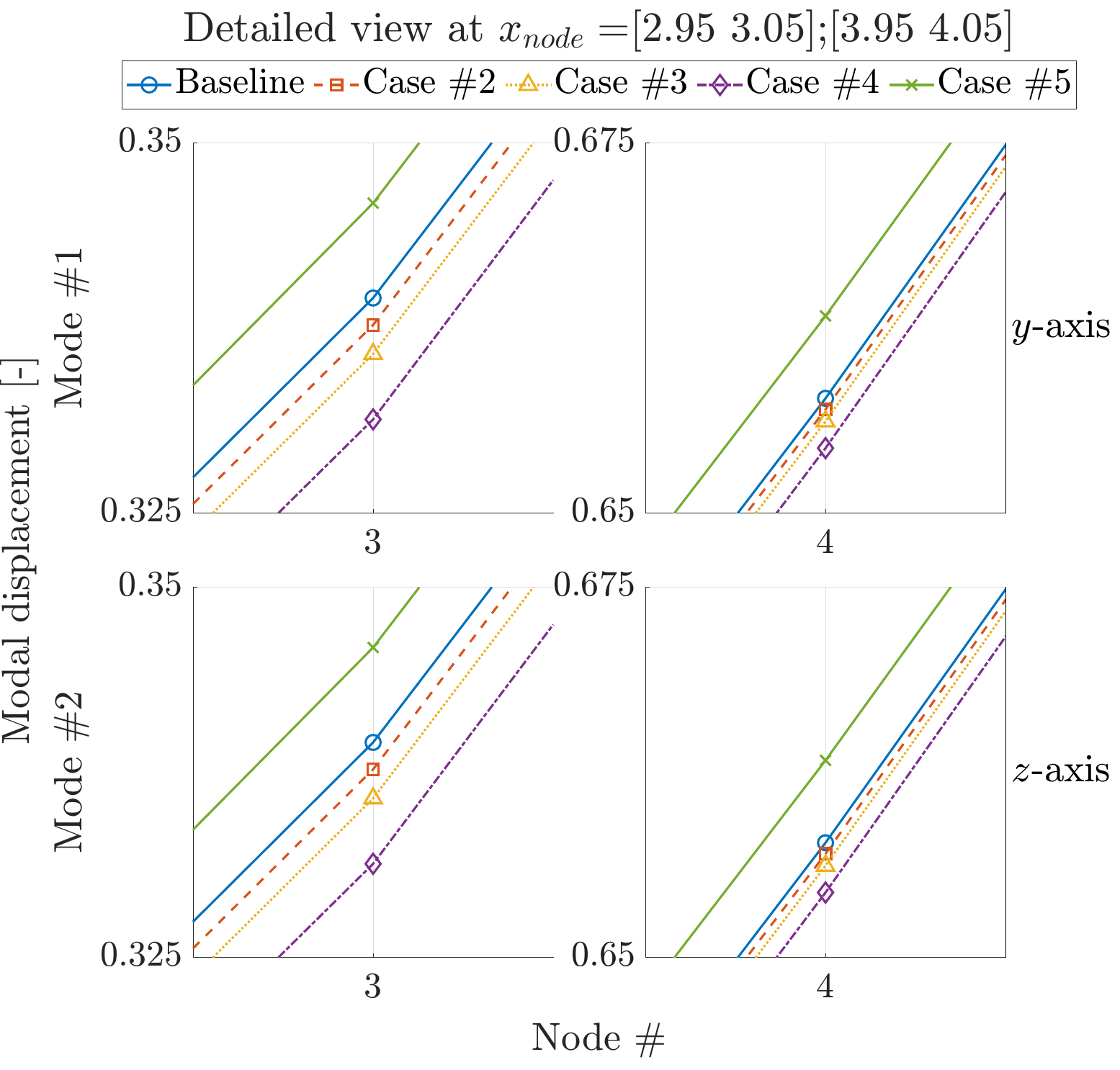}}
	\caption{Numerical case study: Detailed view of $\bm{\phi}_{1,2}$ identified by iLF, showing the trajectory change according to the case number. Note that the secondary y-axis labels represent the dominant displacement shown.}
	\label{fig:fig5}
\end{figure}

\Cref{fig:fig4} shows the $\bm{\phi}_n$ identified via iLF for the baseline and damaged cases. Recalling that the damage has been simulated in the second element between nodes \#2 and 3, it can be seen that the deviation increases in that area with increasing damage. However, a more precise assessment can be carried out in the zoomed-in view of $\bm{\phi}_{1,2}$ in \cref{fig:fig5}, where it is even more apparent that both the damage intensity and position are correctly identified via the modal parameter extraction with iLF.

\begin{figure}[!htb]
\centering
		{\includegraphics[align=c,width=.45\textwidth,keepaspectratio]{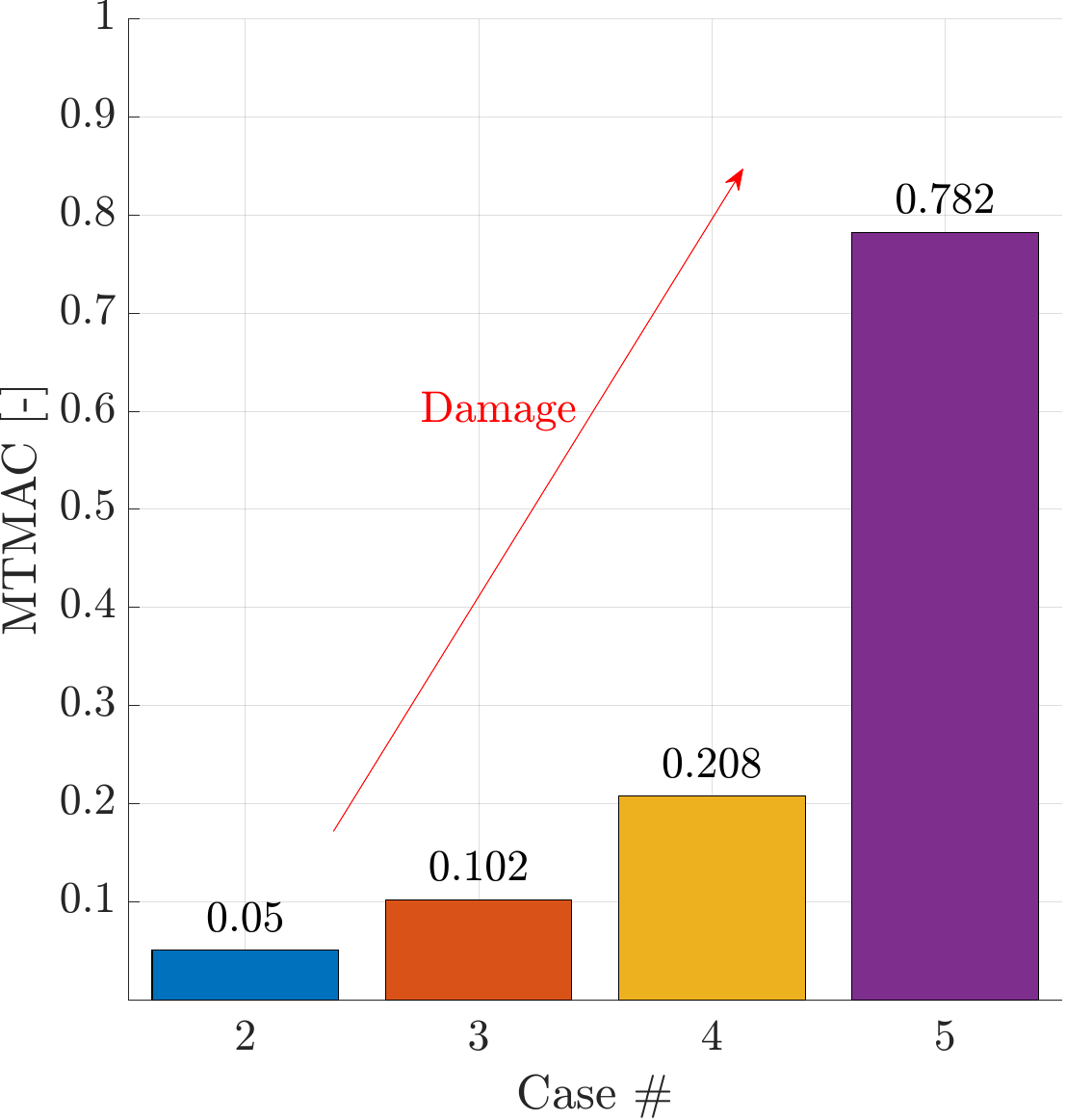}}
	\caption{Numerical case study: MTMAC between the baseline and cases \#2-5 (damaged) modal parameters identified with iLF.}
	\label{fig:fig6}
\end{figure}

Now that it has been assessed that the iLF can serve as a suitable tool for damage detection, the MTMAC capabilities should be verified. The MTMAC values between the damaged and added mass cases versus the baseline case are compared in \cref{fig:fig6}. These are computed considering all modes identified via iLF.

It is observed that the MTMAC increases with damage, exhibiting sufficient granularity to differentiate between 5\% increments (a minimal change). This demonstrates that the MTMAC can be effectively used to assess and quantify damage (at least in the ideal conditions provided by numerical cases). In addition, the larger MTMAC value for case \#5 indicates that this index can distinguish between damage cases and added mass cases that alter the structure behaviour.

{In this sense, a needed remark concerns thresholding. The current study is designed as a proof-of-concept validation of the iLF and MTMAC framework, focusing on sensitivity and consistency across different (numerical and experimental) datasets. However, decision thresholds are natural in applied SHM. Thus, while this work is exploratory/validation-oriented, and the Authors do not want to overclaim, it is still possible to define some quantitative MTMAC values that could serve as illustrative thresholds (with the caveat that they are case-study dependent). More specifically, based on the outcomes shown in }\cref{fig:fig6} {and discussed previously:}

\begin{itemize}
    \item {Case \#2 (5\% stiffness reduction): MTMAC $\approx$ 0.05;}
    \item {Case \#3 (10\% stiffness reduction): MTMAC $\approx$ 0.102;}
    \item {Case \#4 (20\% stiffness reduction): MTMAC $\approx$ 0.208;}
    \item {Case \#5 (mass addition equal to 2\% of the beam whole mass, no stiffness damage): MTMAC $\approx$ 0.782.}
\end{itemize}

\noindent {It is possible to say, at least for the beam example:}

\begin{itemize}
    \item {MTMAC $\leq 0.1-0.15$ may correspond to mild/local damage;}
    \item {MTMAC $\geq 0.15-0.2$ may correspond to clear/severe stiffness reduction ($\geq 20\%$);}
    \item {MTMAC $\gg 0.5$ may indicate very severe damage and/or a substantial structural change (e.g., a large added mass).}
\end{itemize}

\begin{figure}[!htb]
\centering
	\begin{subfigure}[t]{.45\textwidth}
	\centering
		{\includegraphics[width=\textwidth,keepaspectratio]{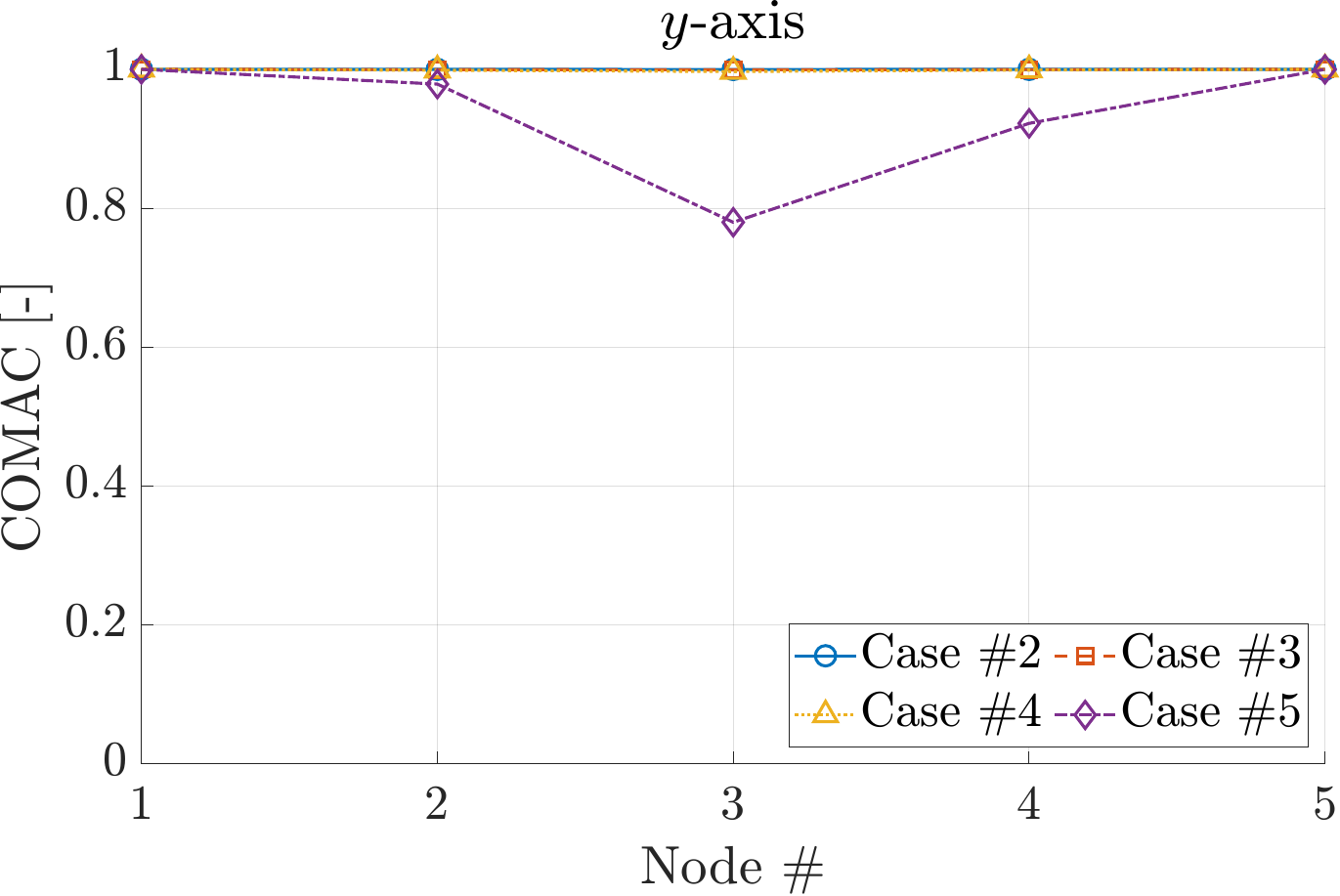}}
		\captionsetup{font={it},justification=centering}
		\subcaption{\label{fig:fig7a}}	
	\end{subfigure}
    \begin{subfigure}[t]{.47\textwidth}
	\centering
		\includegraphics[width=\textwidth,keepaspectratio]{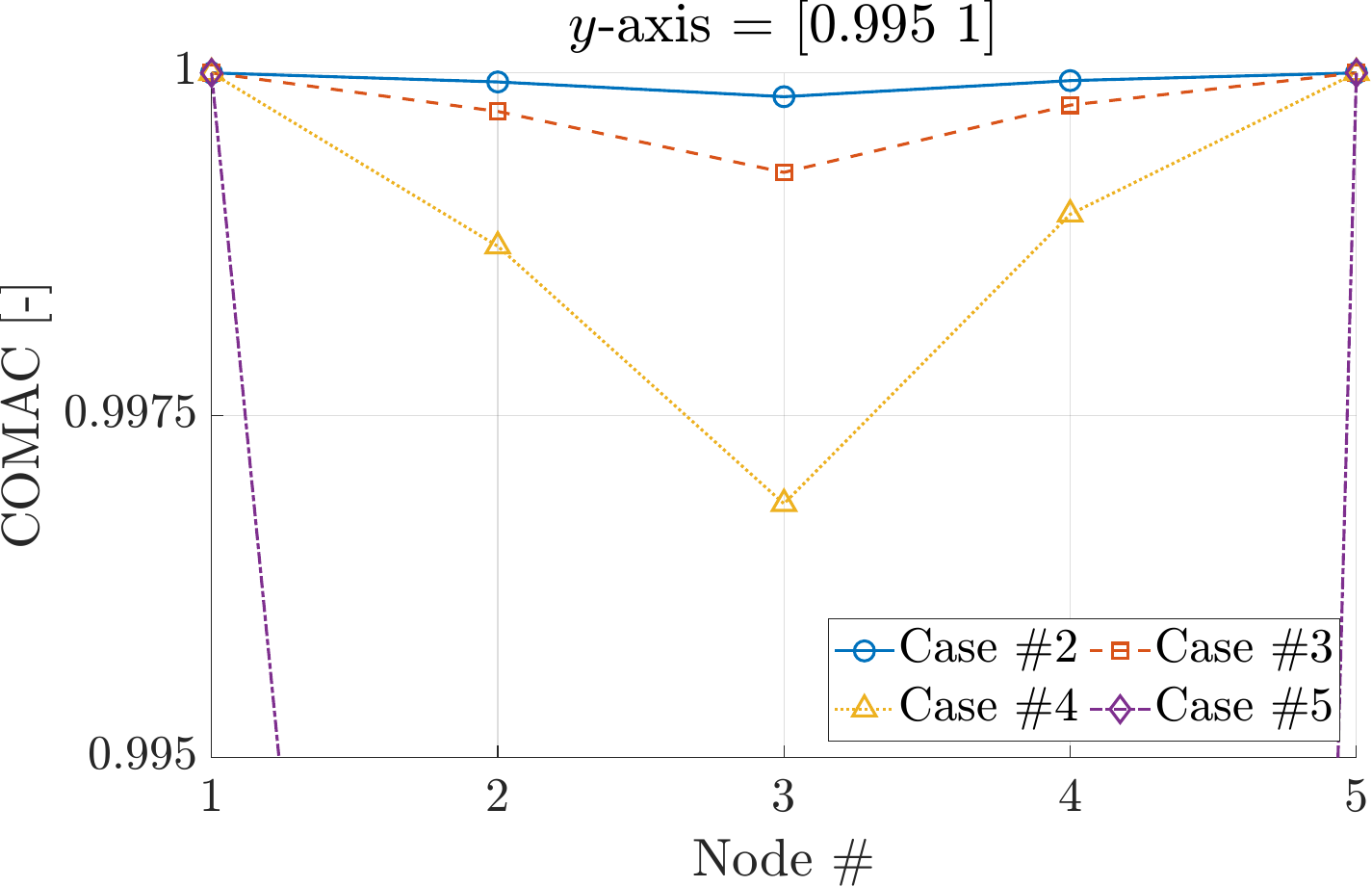}
		\captionsetup{font={it},justification=centering}
		\subcaption{\label{fig:fig7b}}	
	\end{subfigure}
	\caption{Numerical case study: COMAC values between the iLF identified $\bm{\phi}_n$ of the baseline case and cases \#2-5; (\hyperref[fig:fig7a]{a}) {across the y-axis and }(\hyperref[fig:fig7b]{b}) {across the y-axis – Detail for COMAC }$\in$ [0.995, 1].}
	\label{fig:fig7}
\end{figure}

Finally, the COMAC is employed in \cref{fig:fig7} to validate its applicability to damage localisation. A similar situation to that described in \cref{fig:fig4,fig:fig5} is observed: the COMAC values decrease significantly around nodes \#2 and 4, i.e., the damaged area. In particular, for case \#5, the COMAC deviation precisely detects the added mass location at node \#3. Note that only the results for y-direction modal displacements are presented, for the sake of brevity, since a similar result is found for the z-direction.

\section{Experimental case study}\label{sec:exp}
\begin{figure}[!htb]
\centering
		{\includegraphics[align=c,width=.5\textwidth,keepaspectratio]{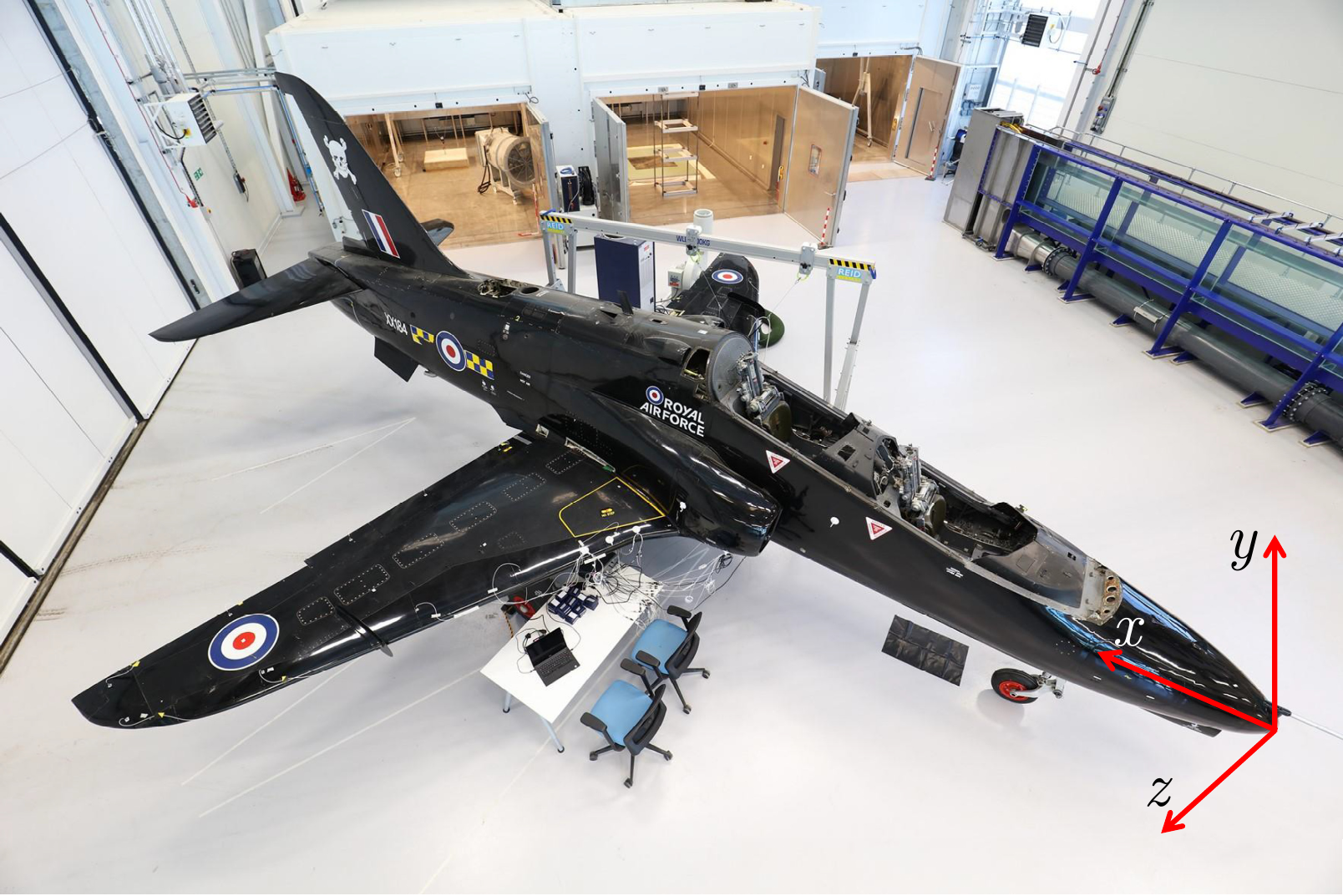}}
	\caption{BAE Systems Hawk T1A aircraft experimental setup at the LVV at the University of Sheffield{, showing the reference frame and its origin (adapted }from \cite{Wilson2024}).}
	\label{fig:fig8}
\end{figure}

The BAE Systems Hawk T1A experimental dataset has recently been made available by the Dynamics Research Group at the University of Sheffield \cite{Wilson2024}. This dataset extends a previous database \cite{Haywood-Alexander2024} that focused on the starboard wing of the same aircraft. The decommissioned full jet trainer aircraft, once in service with the British Royal Air Force, is shown in \cref{fig:fig8} at the LVV facilities at the University of Sheffield{, along with the global coordinate reference frame and its origin.} The data used in this study refer to the entire Hawk airframe structure, retrieved from \cite{Wilson2024a}. The same dataset was also used previously to validate the MIMO extension of the LF for linear SI \cite{Dessena2024} and in \cite{Berghout2025} for the preliminary validation of a Convolutional Neural Network with a Reversed Mapping approach to SHM. Here in this work, the dataset is used to validate the proposed damage assessment procedure. For the sake of brevity, only the key aspects of the experimental setup, testing procedures, and recorded data used in this research work will be detailed. A more detailed discussion can be found in \cite{Wilson2024,Haywood-Alexander2024}.

{Importantly, this laboratory dataset includes multiple substructures, connections, and operational complexities typical of aeronautical assemblies, complementing the controlled numerical benchmark and addressing realism and complexity}.
{In this context, it must be recalled that the System Identification approach used here (iLF) is intended for LTI systems; thus, structural nonlinearities cannot be explicitly modelled in this analytical formulation. However, this limitation is inherent in any modal-based SHM approach, as the Fourier Transform itself is a linear operator. }
{Hence, nonlinearities in the mechanical system will manifest as noise-like distortions in the Fourier spectra of the signal and therefore in the Frequency Response Functions (FRFs) from which the modal parameters are estimated. For an overview of nonlinear modal analysis approaches to full-scale aeronautical systems, the interested reader can refer to} \cite{Noel2013,Dossogne2015}{ for an application to a General Dynamics F-16.}
{Nevertheless, the specific signals employed here correspond to the system vibrational response under low-amplitude broadband excitations, where the aircraft is proven to behave mostly in the linear regime (as a standard practice in ground vibration tests).} 

\subsection{Experimental setup}
\begin{figure}[!htb]
\centering
	\begin{subfigure}[t]{.35\textwidth}
	\centering
		{\includegraphics[width=.8\textwidth,keepaspectratio]{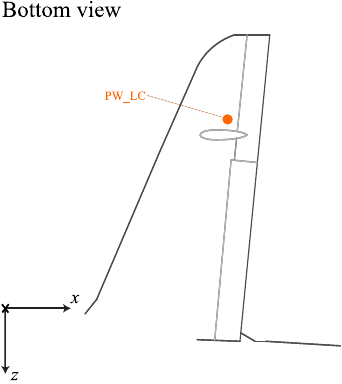}}
		\captionsetup{font={it},justification=centering}
		\subcaption{\label{fig:fig9a}}	
	\end{subfigure}
    \begin{subfigure}[t]{.35\textwidth}
	\centering
		\includegraphics[width=.8\textwidth,keepaspectratio]{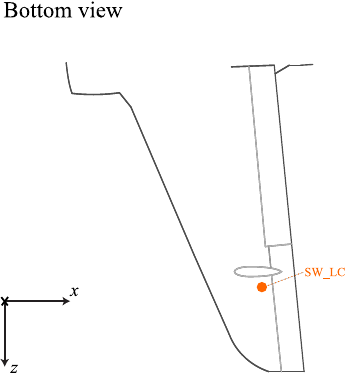}
		\captionsetup{font={it},justification=centering}
		\subcaption{\label{fig:fig9b}}	
	\end{subfigure}
    \begin{subfigure}[t]{.35\textwidth}
	\centering
		{\includegraphics[width=.8\textwidth,keepaspectratio]{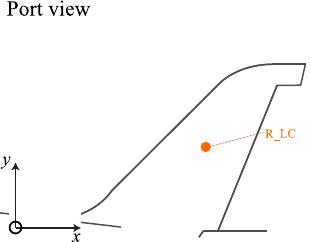}}
		\captionsetup{font={it},justification=centering}
		\subcaption{\label{fig:fig9c}}	
	\end{subfigure}
    \begin{subfigure}[t]{.35\textwidth}
	\centering
		\includegraphics[width=.5\textwidth,keepaspectratio]{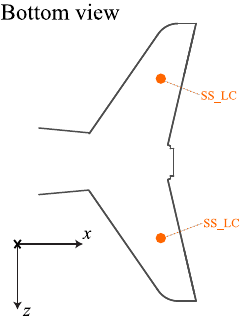}
		\captionsetup{font={it},justification=centering}
		\subcaption{\label{fig:fig9d}}	
	\end{subfigure}
	\caption{BAE Systems Hawk T1A aircraft: Shaker positions for the MIMO tests on the (\hyperref[fig:fig9a]{a}) starboard and the (\hyperref[fig:fig9b]{b}) port wing, the (\hyperref[fig:fig9c]{c}) horizontal stabiliser, and the (\hyperref[fig:fig9d]{d}) vertical stabiliser (adapted from \cite{Wilson2024}).}
	\label{fig:fig9}
\end{figure}

The airframe was equipped with 85 PCB Piezotronics (uniaxial and triaxial) accelerometers, featuring different nominal sensitivities ranging from 10 mVg\textsuperscript{-1} to 100 mVg\textsuperscript{-1}, along with other sensors, totalling 139 recording channels. The sensor layout was designed to cover specific target components: the two wings, the two horizontal stabilisers, the vertical stabiliser, key points on the fuselage, and the three landing gears. For reasons described later, only the accelerometers located on the port wing have been considered in this study. None of the other sensors -– fibre-Bragg grating strain gauges, temperature sensors, etc. -– were considered. For MIMO linear SI, five Tira\textsuperscript{TM} TV 51140-MOSP modal shakers\footnote{\url{https://www.tira-gmbh.de/fileadmin/inhalte/download/schwingprueftechnik/schwingpruefanlagen/modal/EN/100N_bis_2700N/data_sheet_system_tv_51140-MOSP_eng_V05.pdf}} were used, attached to PCB Piezotronics\textsuperscript{TM} 208C02 force transducers (sensitivity: 11.11 mVN\textsuperscript{-1})\footnote{\url{https://www.pcb.com/products?m=208c02}}, applying input forces to the respective locations as indicated in \cref{fig:fig9}. 

\subsubsection{Testing procedures}
The dataset contains white Gaussian noise (WGN) inputs, with five amplitude levels increasing from 0.1 V to 0.5 V. Pink Gaussian noise excitations are also available for linear SI. In contrast, odd random-phase multi-sine excitations can be used to investigate the nonlinear response of the system (which is out of the scope of this work). SIMO and MIMO tests were performed, varying the input source location for SIMO tests. However, these have not been considered here either since the LF was already proven viable for SIMO vibration-based dynamic monitoring in previous works \cite{Dessena2024}. {The excitation amplitudes used here (0.5 V) ensure that the structure remains in its linear regime, as is standard in ground vibration testing.}

\subsubsection{Recorded data and selected recordings}
The acceleration time series were recorded at all output channels with a sampling frequency of $f_s=$ 2048 Hz. The data used here were retrieved directly as pre-processed transfer functions, defined over 8192 spectral lines. Further, the dataset contains (as of the time it was last accessed, 14\textsuperscript{th} of June 2024) 216 different tests, accounting for different types of excitations, input amplitudes, number and location(s) of the input source(s), and kind of damage (simulated –- mass addition –- or actual -– panel removal).

\begin{figure}[!htb]
\centering
		{\includegraphics[align=c,width=.3\textwidth,keepaspectratio]{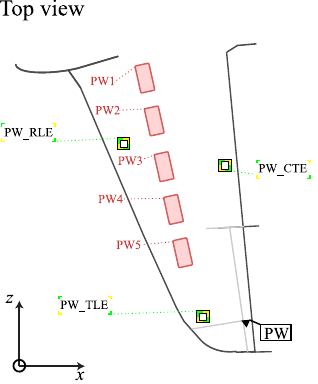}}
	\caption{BAE Systems Hawk T1A aircraft: Simulated damage (PW\_xxE) and removed panels (PW$n$) positions on the port wing (PW) (adapted from \cite{Wilson2024}).}
	\label{fig:fig10}
\end{figure}

The criteria followed in this work for the selection of the cases were:
\begin{enumerate}
    \item For MIMO testing, only tests using the five available input sources were considered;
    \item To replicate what was done in \cite{Dessena2024} and for consistency with \cite{Haywood-Alexander2024}, WGN inputs were selected. These were defined by an excitation bandwidth in the 5–256 Hz range;
    \item The maximum input amplitude voltage, 0.5 V, was considered for all cases;
    \item Six cases of added mass were considered. Specifically, these arise from adding different weights at three locations: the tip of the leading edge (PW\_TLE), the root of the same (PW\_RLE), and the centre of the trailing edge (PW\_CTE), with the mass always added on the upper side, as shown in \cref{fig:fig10}. The added weights considered are M1 (254.3 g), M2 (616.8 g), and M3 (916.8 g); 
    \item The interest was focused on damage cases simulated on the port wing, which covered several types of damage. These include removed panels (at different locations) and added masses (located at different sites and with different weights to simulate growing damage severity);
    \item All the relevant damage scenarios were separately tested;
    \item Five damage cases were used, each one consisting of a single removed panel, from PW1 to PW5, moving from the fuselage to the wing tip (see \cref{fig:fig10}). Taken together, they account for different damage locations, severity, and typologies.
\end{enumerate}

Thus, a total of twelve cases were considered. These are summarised in \cref{tab:tab2}.

\begin{table}[!htb]
\centering
\caption{BAE Systems Hawk T1A aircraft: Test scenarios considered.}
{\footnotesize	\begin{tabular}{ccl}
\hline
\textbf{Case \#} & \textbf{Scenario} & \textbf{Description} \\
\hline
1 & HS\_WN05 & Healthy state with 0.5 V input \\
2 & DS\_WN10 & M1 mass addition at PW\_CTE \\
3 & DS\_WN12 & M3 mass addition at PW\_CTE \\
4 & DS\_WN13 & M1 mass addition at PW\_RLE \\
5 & DS\_WN15 & M3 mass addition at PW\_RLE \\
6 & DS\_WN17 & M2 mass addition at PW\_TLE \\
7 & DS\_WN18 & M3 mass addition at PW\_TLE \\
8 & PR\_WN03 & PW1 panel removal \\
9 & PR\_WN06 & PW2 panel removal \\
10 & PR\_WN09 & PW3 panel removal \\
11 & PR\_WN12 & PW4 panel removal \\
12 & PR\_WN15 & PW5 panel removal \\
\hline
\end{tabular}}
\label{tab:tab2}
\end{table}

\begin{figure}[!htb]
\centering
		{\includegraphics[align=c,width=.65\textwidth,keepaspectratio]{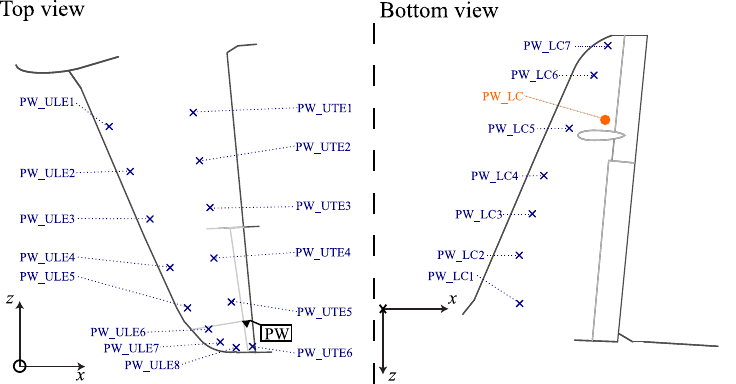}}
	\caption{BAE Systems Hawk T1A aircraft: Sensors position (\textcolor{blue}{\textbf{X}}) on the port wing (adapted from \protect\cite{Wilson2024}). Note: In the original image of the bottom view \protect\cite{Wilson2024}, the sensors were ordered from the wing tip to the root. However, the numbering in the dataset metadata \cite{Wilson2024a} reflects the modified figure presented here.}
	\label{fig:fig11}
\end{figure}

Moreover, aiming at MIMO SI and vibration-based SHM, these signals have all been compared to the health baseline for the same input amplitude (HS\_WN5), using the 21 uniaxial 10 mVg\textsuperscript{-1} accelerometers included on the wing port. These sensors were recording in the vertical direction only, arranged in a grid of two lines on the top side (running parallel to the leading and trailing edges) and one line on the bottom side (parallel to the leading edge), with respectively 14 and 7 uniaxial accelerometers, for a total of 21 output channels (see \cref{fig:fig11}).

\subsection{Results}
\begin{figure}[!htb]
\centering
	\begin{subfigure}[t]{.49\textwidth}
	\centering
		{\includegraphics[width=\textwidth,keepaspectratio]{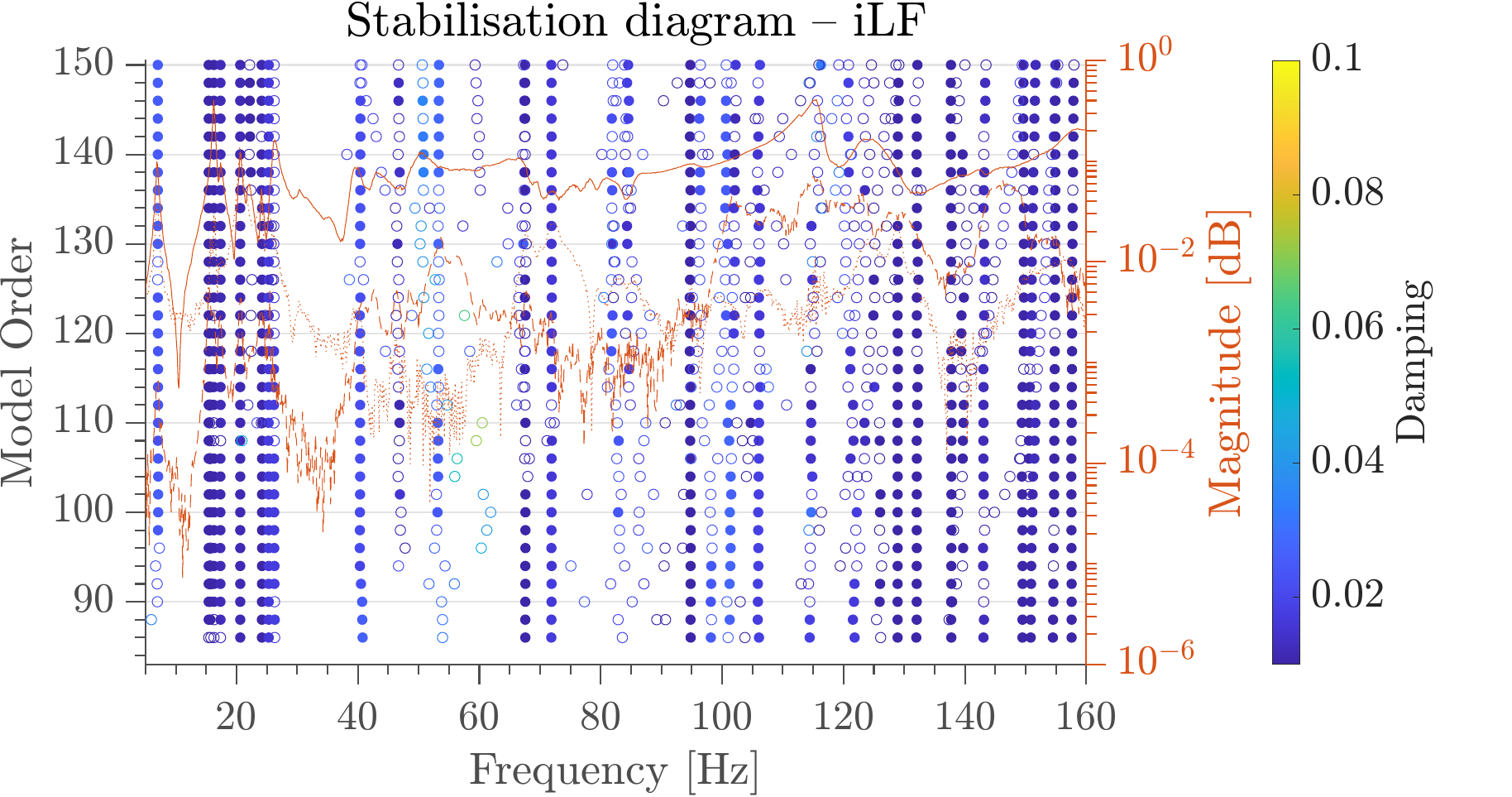}}
		\captionsetup{font={it},justification=centering}
		\subcaption{\label{fig:fig12a}}	
	\end{subfigure}
    \begin{subfigure}[t]{.49\textwidth}
	\centering
		\includegraphics[width=\textwidth,keepaspectratio]{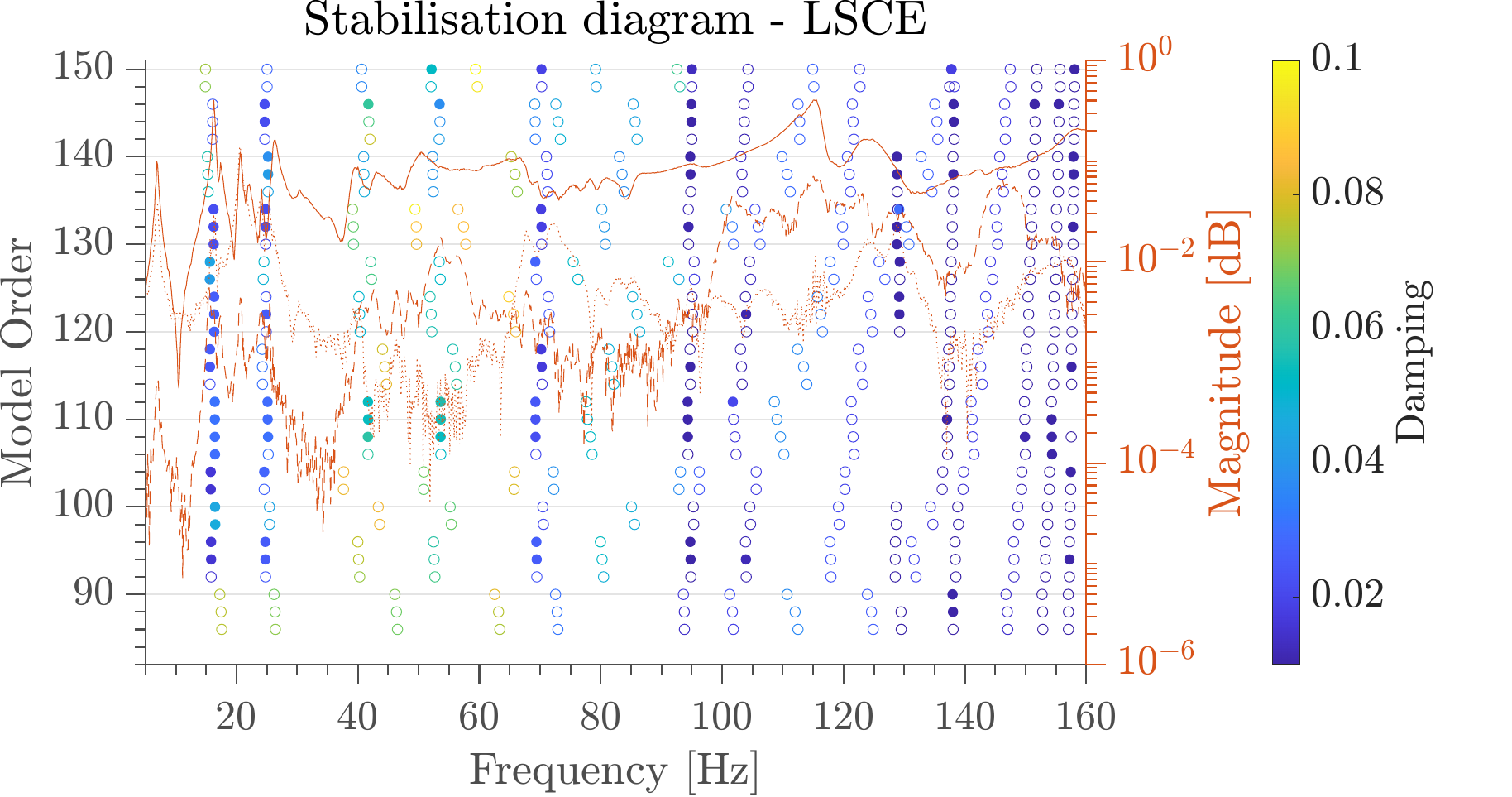}
		\captionsetup{font={it},justification=centering}
		\subcaption{\label{fig:fig12b}}	
	\end{subfigure}
    \begin{subfigure}[t]{.49\textwidth}
	\centering
		{\includegraphics[width=\textwidth,keepaspectratio]{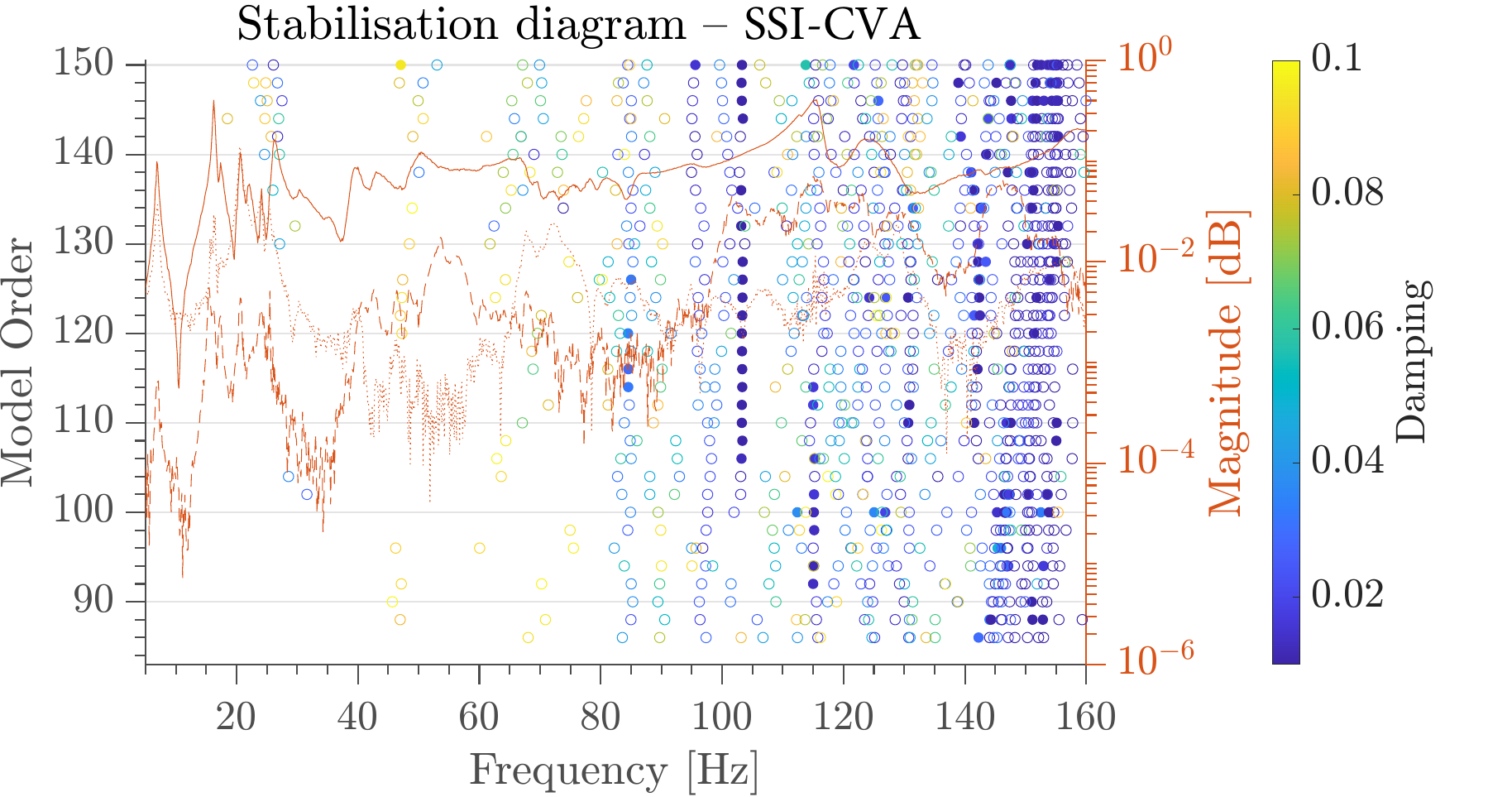}}
		\captionsetup{font={it},justification=centering}
		\subcaption{\label{fig:fig12c}}	
	\end{subfigure}
	\caption{BAE Systems Hawk T1A aircraft: {Stabilisation diagrams for the modes identified via (}\hyperref[fig:fig12a]{a}{) iLF, (}\hyperref[fig:fig12b]{b}{) LSCE, and (}\hyperref[fig:fig12c]{c}) SSI-CVA for the baseline scenario.}
	\label{fig:fig12}
\end{figure}

Before addressing the effectiveness of the proposed methodology for SHM purposes, it is worthwhile to investigate the accuracy of modal identification via iLF compared to that obtained with the benchmark methods SSI-CVA and LSCE on this experimental dataset. A further source of comparison is identified in the literature, i.e. the results reported in \cite{Dessena2024}. However, please consider that those results were obtained for the case with 0.4 V input voltage, while all data in this work refer to the 0.5 V input case.
The first step involves identifying the stable modes for the healthy case, case \#1, via iLF and benchmark methods. In \cref{fig:fig12}, the three stabilisation diagrams are presented. It is clear that the iLF (\cref{fig:fig12a}) identification succeeds in finding the most stable modes, while SSI-CVA (\cref{fig:fig12c}) and especially LSCE (\cref{fig:fig12b}) identifications return fewer and less stable poles. {The stabilisation diagrams are constructed by varying $k$ from $k_{\min}=84$ to $k_{\max}=150$ with $\Delta k = 2$, yielding the extraction set $\mathbf{k}=\{\,k_{\min}:\Delta k:k_{\max}\,\}$. Hard criteria are imposed through the admissible ranges $\zeta_n \in [0.005,\,0.1]$ and $\omega_n \in [5,\,160]$. Soft criteria include $\Delta \omega_{\mathrm{stab}} = 0.05$, $\Delta \zeta_{n,\mathrm{stab}} = 0.3$, and $\mathrm{MAC}_{\mathrm{stab}} = 0.95$. Additionally, $N_{\mathrm{MAC}} = 3$ are considered in the selection process.}

\begin{figure}[!htb]
\centering
	\begin{subfigure}[t]{.43\textwidth}
	\centering
		{\includegraphics[width=\textwidth,keepaspectratio]{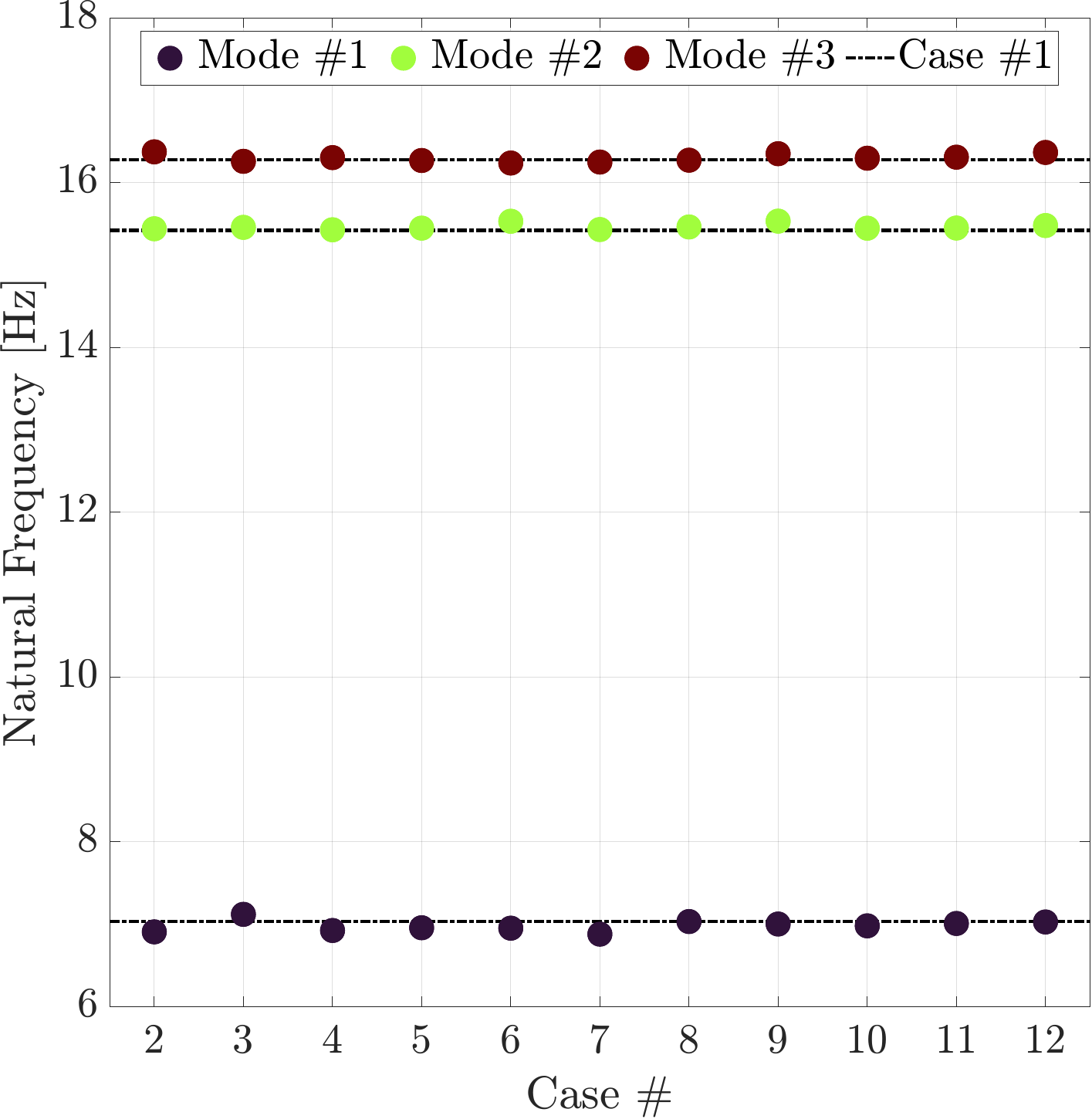}}
		\captionsetup{font={it},justification=centering}
		\subcaption{\label{fig:fig13a}}	
	\end{subfigure}
    \begin{subfigure}[t]{.45\textwidth}
	\centering
		\includegraphics[width=\textwidth,keepaspectratio]{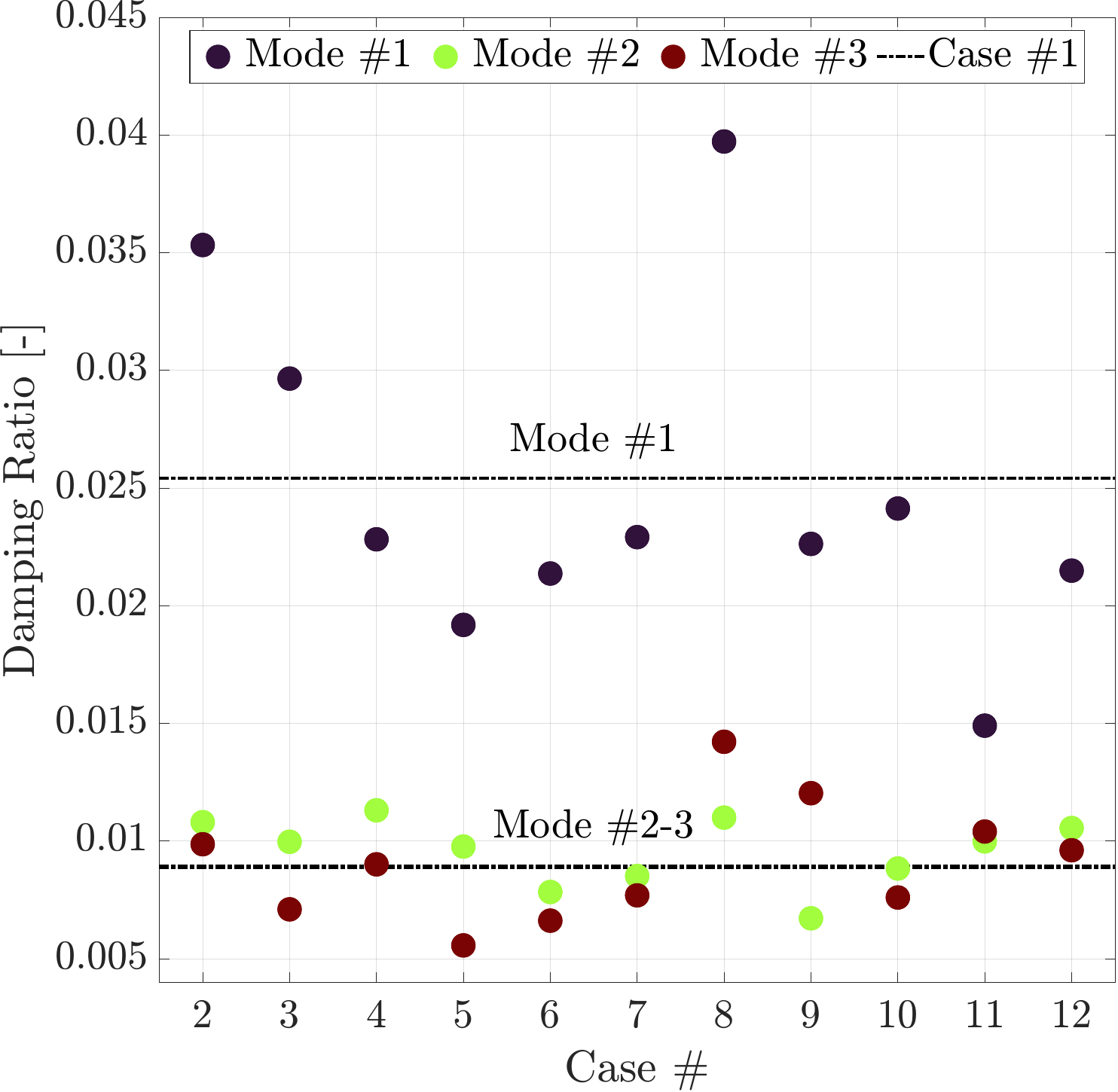}
		\captionsetup{font={it},justification=centering}
		\subcaption{\label{fig:fig13b}}	
	\end{subfigure}
	\caption{BAE Systems Hawk T1A aircraft: (\hyperref[fig:fig13a]{a}) $\omega_{1-3}$ and (\hyperref[fig:fig13b]{b}) $\zeta_{1-3}$ identified via iLF for the 12 cases under scrutiny.}
	\label{fig:fig13}
\end{figure}

From \cite{Dessena2024} and the current iLF identification, 41 modes were identified. However, only the first three are used here for the SHM task. Hence, in \cref{tab:tab3}, the first three pairs of natural frequencies ($\omega_n$) and damping ratios ($\bm{\phi}_n$) identified in this work for the undamaged Hawk T1A Aircraft are presented alongside the aforementioned literature results in \cref{tab:tab3}. {
Notably, only the iLF is able to identify all modes of interest, so SSI-CVA and LSCE results are disregarded from now onwards.} The iLF-identified $\omega_n$ and $\bm{\phi}_n$ for all cases are shown \cref{fig:fig13a,fig:fig13b}. {All the processing carried out in this work for obtaining the modal parameters of the BAE Systems Hawk T1A is done in MATLAB 2024b.}


\begin{table}[!htb]
\centering
\caption{BAE Systems Hawk T1A aircraft: $\omega_{1-3}$ and $\zeta_{1-3}$ identified via iLF, SSI-CVA, and LSCE. 0.4~V input case refers to the literature result in \cite{Dessena2024} for a healthy case with a lower input amplitude.}
\label{tab:tab3}
{\footnotesize\begin{tabular}{ccccccccc}
\toprule
\multirow{2}{*}{\textbf{Mode \#}} &
\multicolumn{4}{c}{\textbf{Natural Frequency} [Hz] -(difference [\%])} &
\multicolumn{4}{c}{\textbf{Damping Ratio} [-] -(difference [\%])} \\
\cline{2-5}
 & 0.4 V input \cite{Dessena2024} & iLF & SSI-CVA & LSCE & 0.4 V input \cite{Dessena2024} & iLF & SSI-CVA & LSCE \\
\midrule
1 & 6.98 & 7.03 & -& -& 0.03 & 0.03 & -& -\\
  &      & (0.7) & (-) & (-) &      & (0) & (-) & (-) \\
2 & 15.43 & 15.42 & -& -& 0.01 & 0.01 & -& -\\
  &       & (-0.1) & (-) & (-) &      & (0) & (-) & (-) \\
3 & 16.32 & 16.28 & -& 16.25 & 0.01 & 0.01 & -& 0.03 \\
  &       & (-0.3) & (-) & (-0.4) &      & (0) & (-) & (200) \\
\bottomrule
\end{tabular}}
\end{table}

As shown in \cref{fig:fig13}, a deviation is observed between the identified $\omega_{1-3}$ of cases \#2-12 and case \#1. {However, linking a frequency trend (decrease or increase) to the damage (mass addition or panel removal) in this complex structure is not a straightforward task.} This is a well-known issue with complex assembled systems. However, this falls in the context of damage (or more correctly, anomaly) classification, which is another task, even more difficult than damage localisation and severity assessment. Damage classification is almost impossible to assess in an unsupervised fashion \cite{Worden2007} and, therefore, is outside of the aim of this research work. Similarly, the situation shown in \cref{fig:fig13} for $\zeta_{1-3}$ does not offer any cues for inferring a clear (linear or nonlinear) relationship between the damage case and the identified damping. Nevertheless, this is particularly expected due to the damping own nature \cite{Civera2021}.

\begin{figure*}[!htb]
\centering
		{\includegraphics[align=c,width=\textwidth,keepaspectratio]{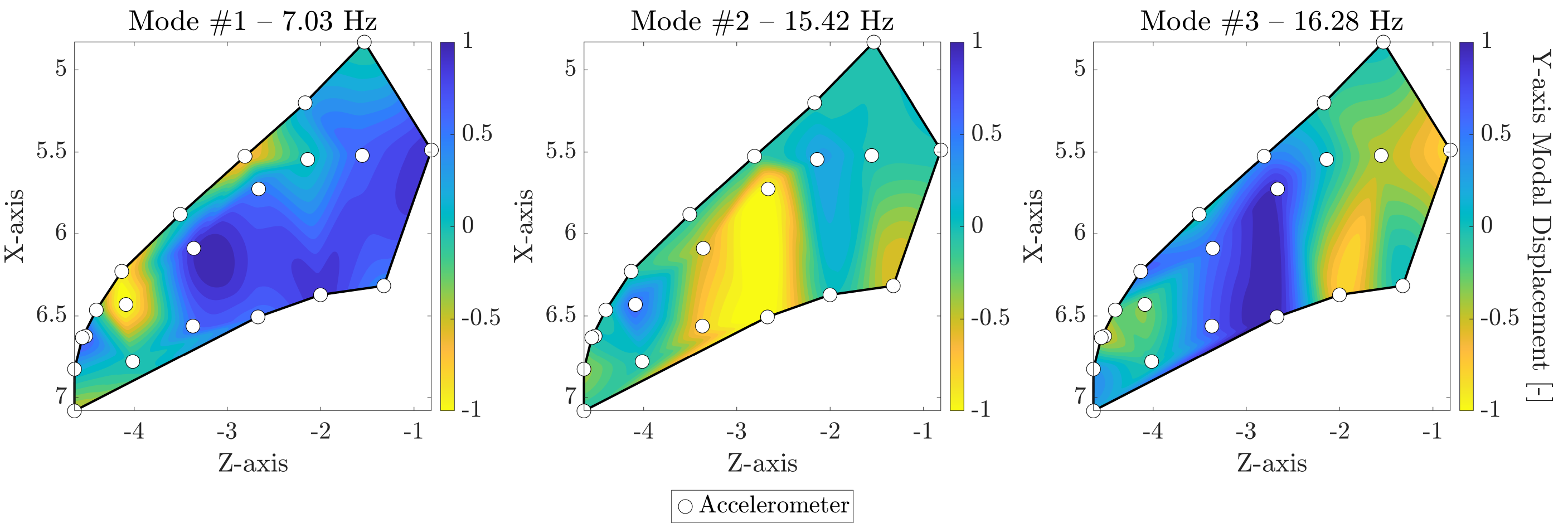}}
	\caption{BAE Systems Hawk T1A aircraft: $\bm{\phi}_{1-3}$ identified via iLF for case \#1.}
	\label{fig:fig14}
\end{figure*}

Finally, the $\bm{\phi}_{1-3}$ identified via iLF for the healthy state are considered in \cref{fig:fig14}. It should be noted that representing these mode shapes is particularly challenging, as the wing has been isolated from the entire aircraft system, to allow for local scaling of the modal displacements. Nevertheless, the dominant displacement motion of the port wing can be recalled from \cite{Dessena2024}:

\begin{enumerate}
    \item[] \textit{Mode \#1}: 1\textsuperscript{st} bending (downwards);
    \item[] \textit{Mode \#2}: 2\textsuperscript{nd} bending (downwards);
    \item[] \textit{Mode \#3}: 1\textsuperscript{st} coupled between bending (downwards) and torsion.
\end{enumerate}

The identified $\bm{\phi}_{1-3}$ for cases \#2-12 can be compared to those of case \#1 via the MAC value. These are shown in \cref{tab:tab4}. Here, it can be seen that as the mass increases (cases \#2-7), the MAC values of one or more modes go down, showing a change in the $\bm{\phi}_{1-3}$ trajectory. These changes become even more pronounced for the panel-removed cases (cases \#8-12), in particular for cases \#8 and 11.

\begin{table}[!htb]
\centering
\caption{BAE Systems Hawk T1A aircraft: MAC matrices diagonal values between the damaged and removed panel cases and case \#1, the healthy state.}
\label{tab:tab4}
{\footnotesize\begin{tabular}{c*{11}{c}}
\hline
\diagbox[width=6em, height=3em]{\textbf{Mode \#}}{\textbf{Case}} & \textbf{2} & \textbf{3} & \textbf{4} & \textbf{5} & \textbf{6} & \textbf{7} & \textbf{8} & \textbf{9} & \textbf{10} & \textbf{11} & \textbf{12} \\\cline{2-12}
\textit{1} & 0.99 & 0.92 & 0.86 & 0.72 & 0.90 & 0.85 & 0.58 & 0.80 & 0.88 & 0.60 & 0.85 \\\hline
\textit{2} & 1.00 & 1.00 & 0.99 & 1.00 & 0.96 & 1.00 & 0.99 & 0.99 & 0.91 & 0.91 & 0.90 \\\hline
\textit{3} & 0.99 & 0.97 & 0.98 & 0.93 & 0.99 & 0.98 & 0.98 & 0.99 & 0.99 & 0.99 & 0.99 \\\hline
\end{tabular}}
\end{table}

Hence, we can assess that the model parameters, specifically $\omega_{1-3}$ and $\bm{\phi}_{1-3}$ of the damaged cases, deviate from those of case \#1. However, assessing such a system by simple comparison is not a straightforward task. Hence, the combined use of MTMAC (for damage severity assessment) and COMAC (for damage localisation), already validated on the numerical system, is here tested for the damage assessment of this experimental case study, as reported in the next sections.

\subsubsection{Experimentally simulated damage}
\begin{figure}[!htb]
\centering
		{\includegraphics[align=c,width=.55\textwidth,keepaspectratio]{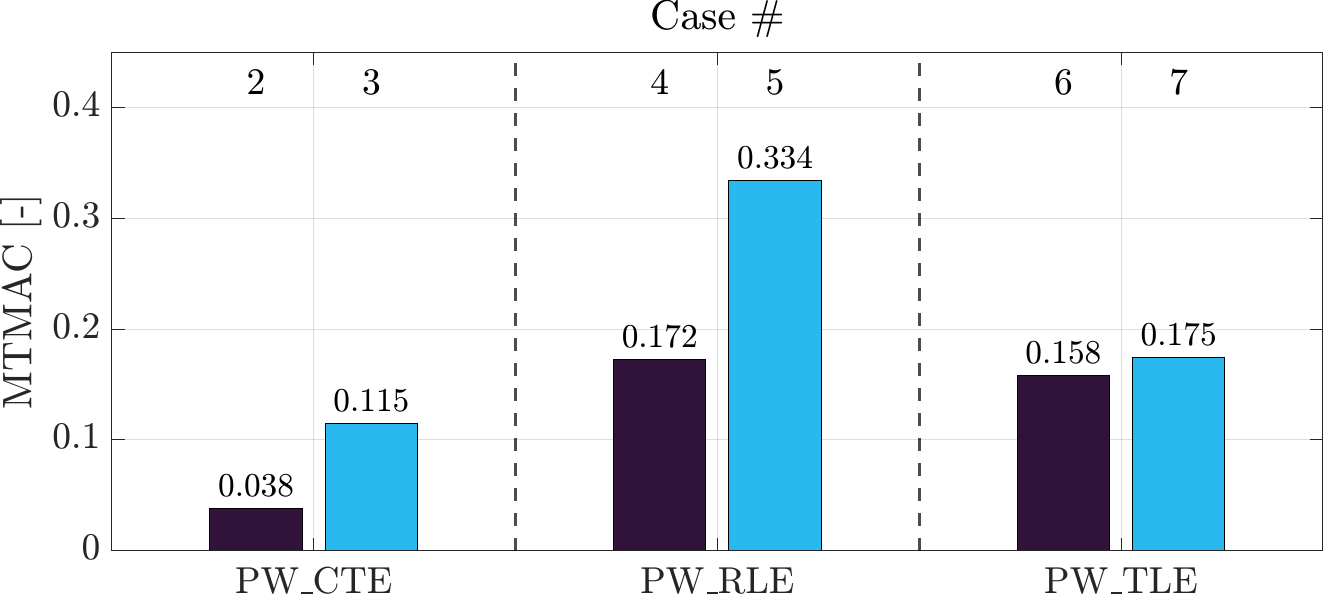}}
	\caption{BAE Systems Hawk T1A aircraft: MTMAC for cases \#2-7 w.r.t. case \#1.}
	\label{fig:fig15}
\end{figure}

In this subsubsection, the simulated damage conditions (via mass addition), corresponding to cases \#2-7, are investigated. In \cref{fig:fig15}, the MTMAC values for these cases are shown. A relationship exists between the different mass quantities and positions. In fact, cases \#4 and 5 show the highest MTMAC values of the six cases. Hence, there is a more pronounced change in modal parameters with the baseline case. Traditionally, this is interpreted as a location within the structure that is more sensitive to damage effects. 

Now, by looking at all cases with the same mass (M3 in cases \#3, 5, and 7), it can be asserted that the location PW\_RLE is the most sensitive to damage and that PW\_CTE is the least, while PW\_TLE lies in the middle. Nevertheless, a deviation from the baseline $\omega_{1-3}$ and $\bm{\phi}_{1-3}$ can be identified in all three locations; most importantly, the magnitude of the added mass influences them by increasing the deviations.

{In this sense, it is possible to compare these findings to the threshold defined before for the numerical case study. As mentioned there, establishing reliable thresholds would require a larger dataset of repeated measurements under varying environmental/operational conditions, and well-defined requirements regarding acceptable margins of stiffness or frequency reduction. However, just to showcase the generalisability of the findings across datasets, one can notice that:}

\begin{itemize}
    \item {across all locations, smaller added masses (M1, $\approx 250$ g) produce MTMAC values in the same order of magnitude ($\approx 0.04–0.17$) and that are, except PW\_RLE, comparable to what was labelled as 'mild' or \textit{local} damage in the beam numerical model;}
    \item {in the same locations, larger masses (M3, $\approx 900$ g) increase the MTMAC values up to $\approx 0.12-0.33$; again, except for a single case (in this instance, PW\_CTE), these values are aligned with the ones caused by severe stiffness reduction in the beam model.}
\end{itemize}

{Summarising, it is clear that no absolute and universally-valid thresholds can be defined a priori, and in a complex system such as an entire aircraft, these depend on the specific location under scrutiny. Nevertheless, these values represent some insight about which orders of magnitude one should expect from this index.}

\begin{figure}[!htb]
\centering
	\begin{subfigure}[t]{.34\textwidth}
	\centering
		{\includegraphics[width=\textwidth,keepaspectratio]{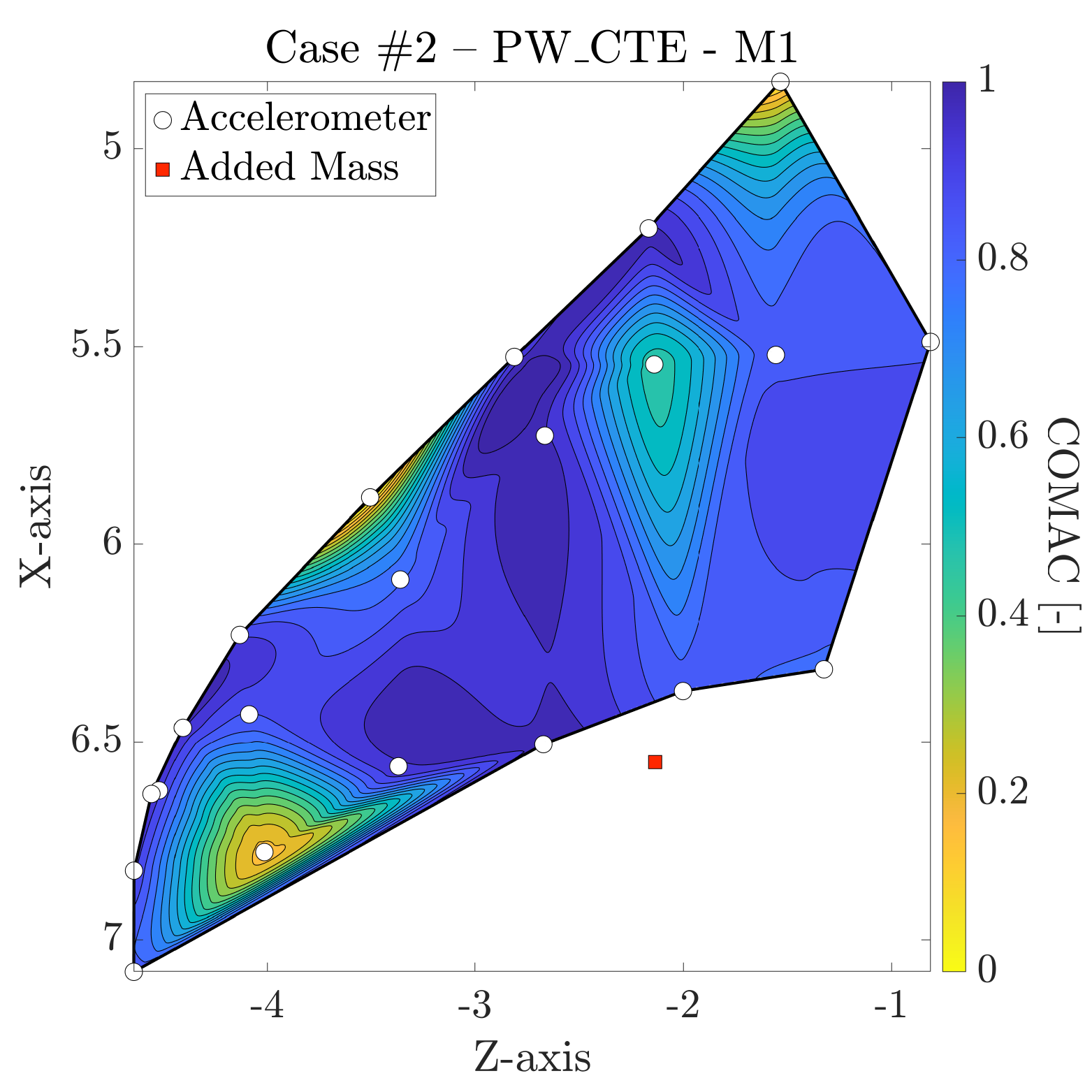}}
		\captionsetup{font={it},justification=centering}
		\subcaption{\label{fig:fig16a}}	
	\end{subfigure}
    \begin{subfigure}[t]{.34\textwidth}
	\centering
		\includegraphics[width=\textwidth,keepaspectratio]{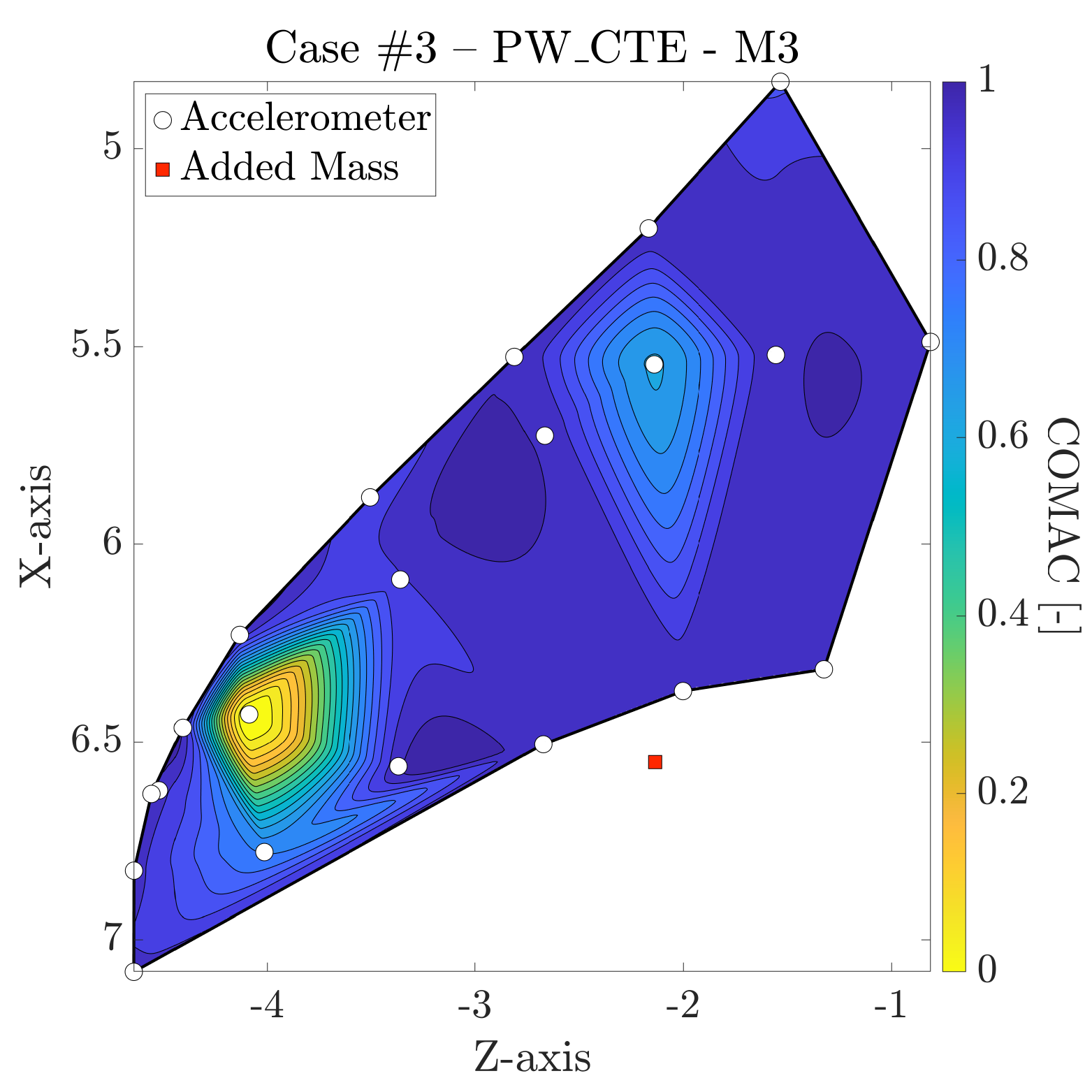}
		\captionsetup{font={it},justification=centering}
		\subcaption{\label{fig:fig16b}}	
	\end{subfigure}
    \begin{subfigure}[t]{.34\textwidth}
	\centering
		{\includegraphics[width=\textwidth,keepaspectratio]{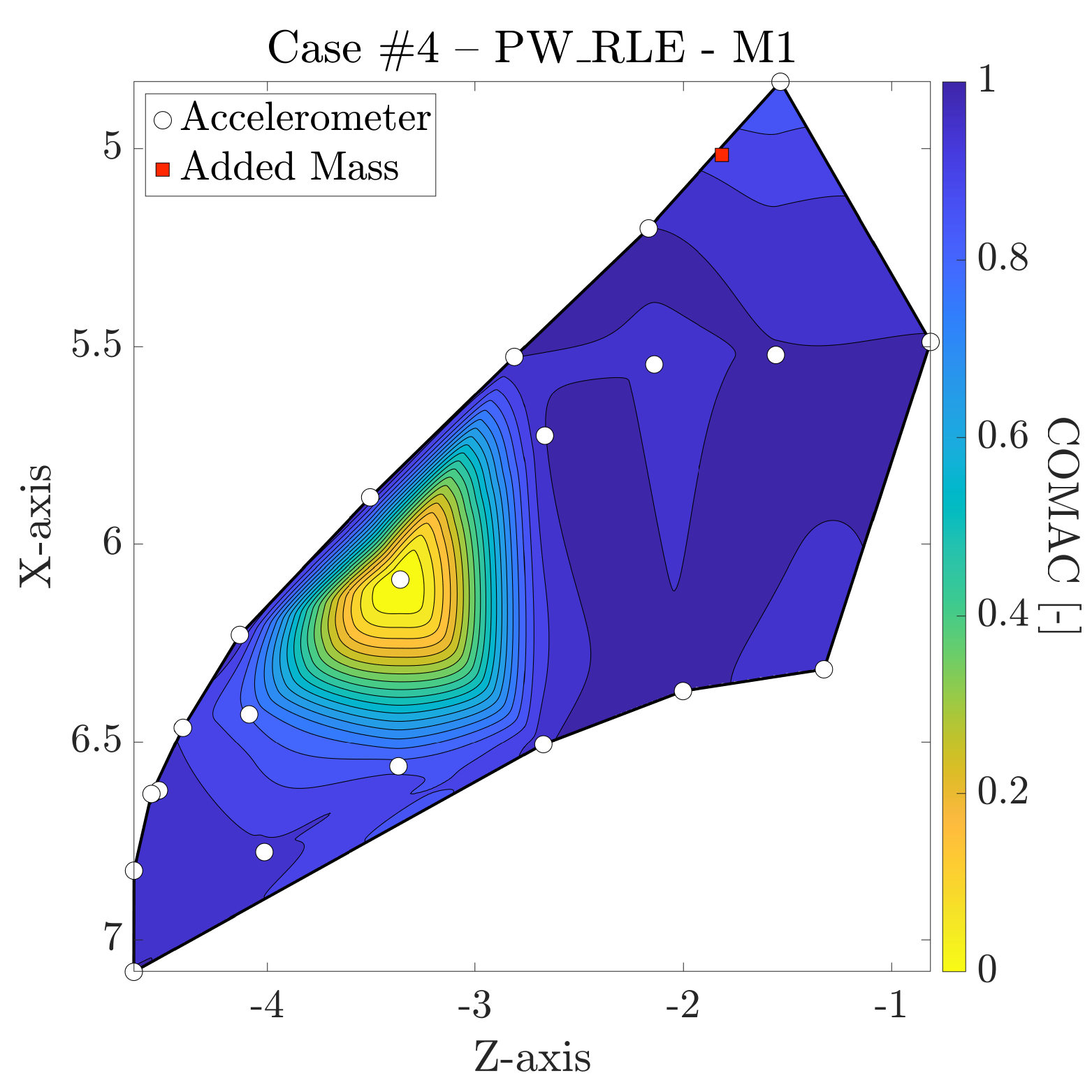}}
		\captionsetup{font={it},justification=centering}
		\subcaption{\label{fig:fig16c}}	
	\end{subfigure}
    \begin{subfigure}[t]{.34\textwidth}
	\centering
		\includegraphics[width=\textwidth,keepaspectratio]{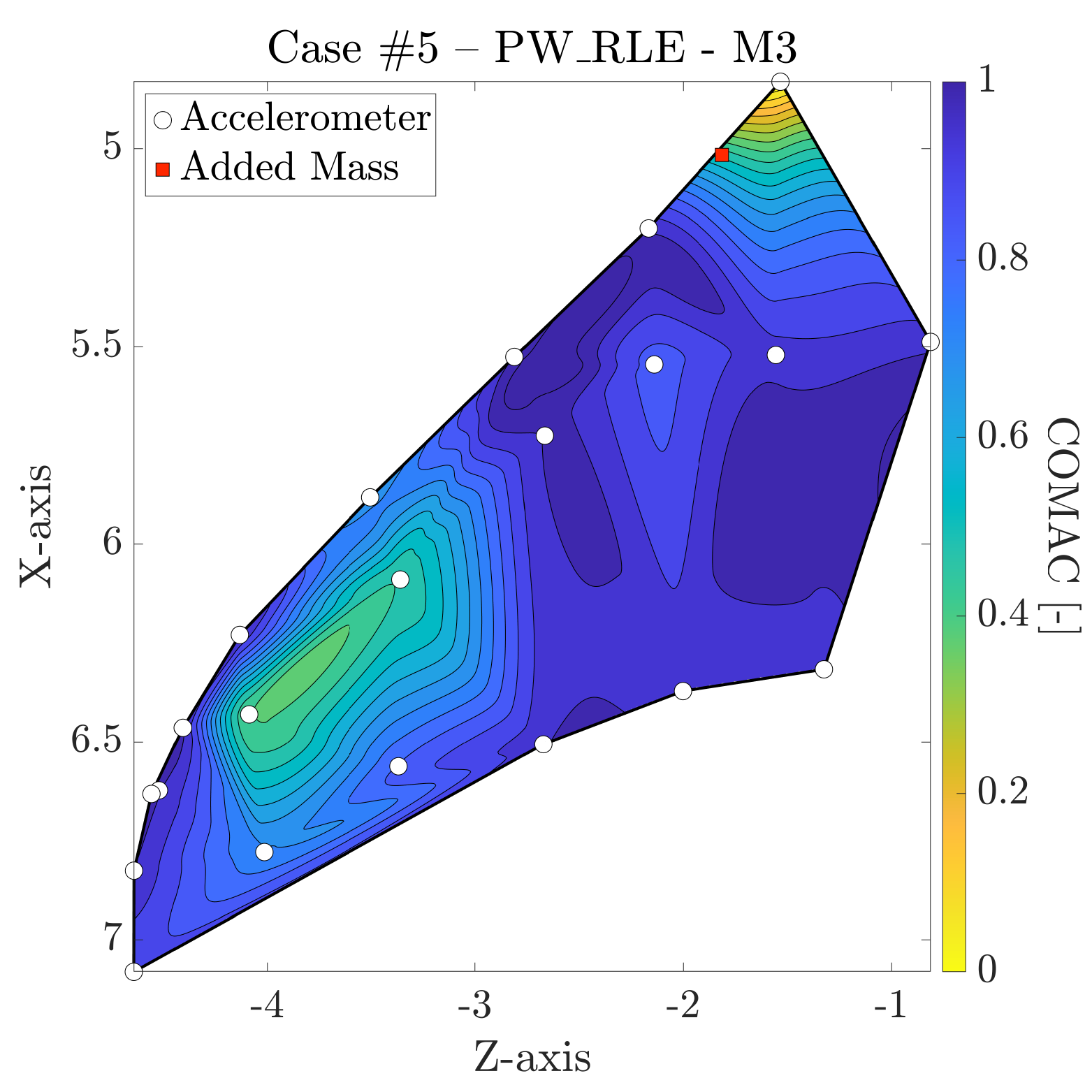}
		\captionsetup{font={it},justification=centering}
		\subcaption{\label{fig:fig16d}}	
	\end{subfigure}
        \begin{subfigure}[t]{.34\textwidth}
	\centering
		{\includegraphics[width=\textwidth,keepaspectratio]{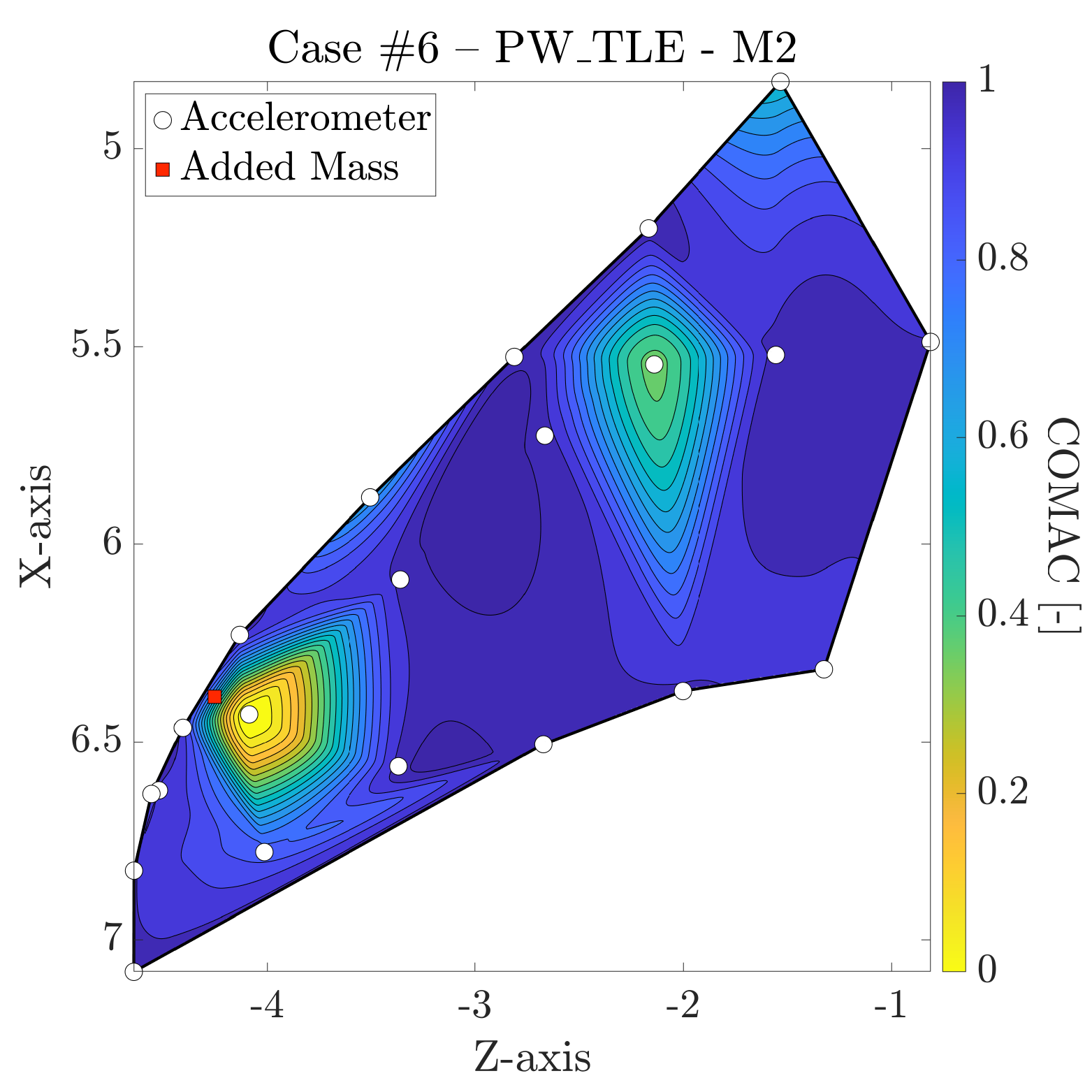}}
		\captionsetup{font={it},justification=centering}
		\subcaption{\label{fig:fig16e}}	
	\end{subfigure}
    \begin{subfigure}[t]{.34\textwidth}
	\centering
		\includegraphics[width=\textwidth,keepaspectratio]{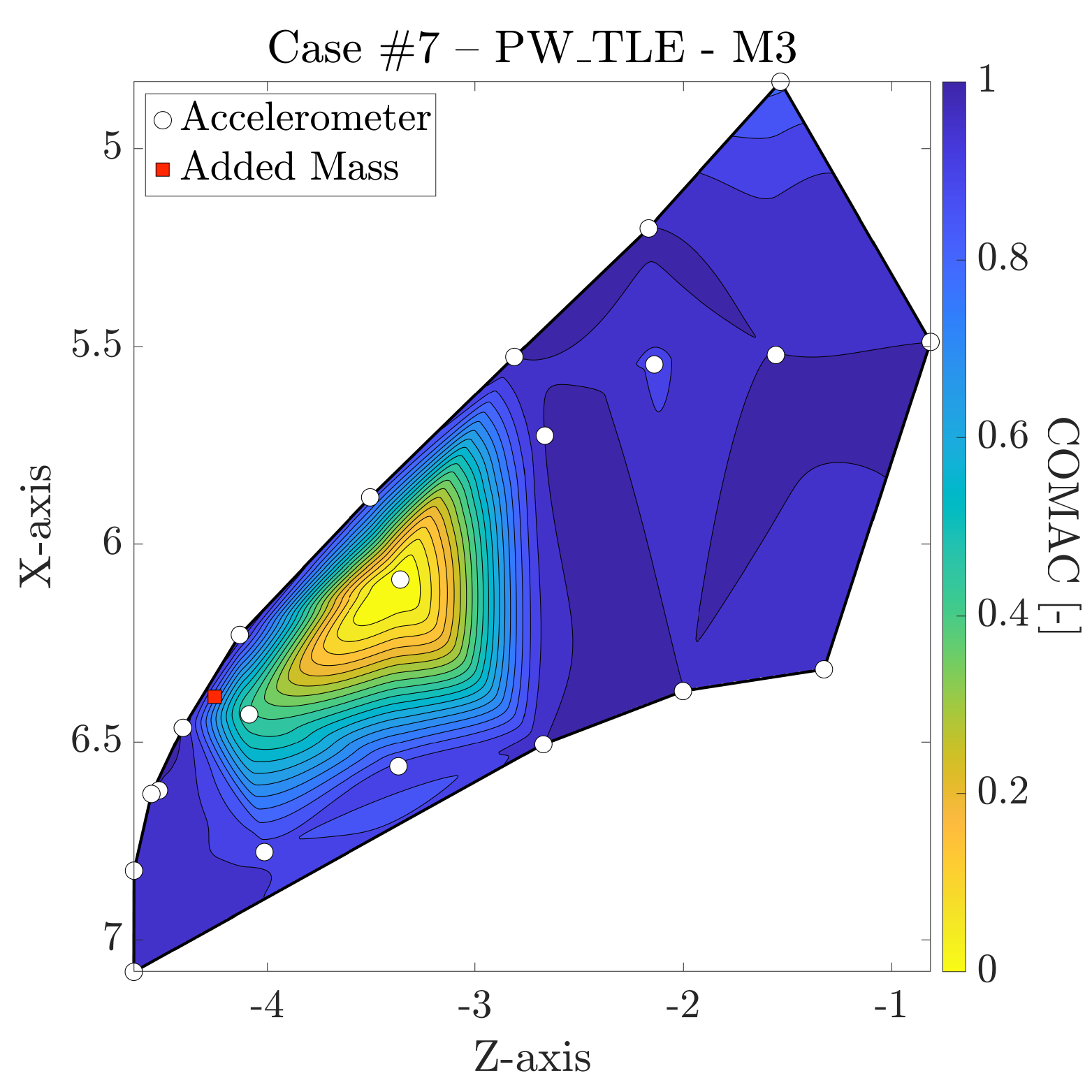}
		\captionsetup{font={it},justification=centering}
		\subcaption{\label{fig:fig16f}}	
	\end{subfigure}
	\caption{BAE Systems Hawk T1A aircraft: Scaled COMAC for the port wing identified via iLF from cases (\hyperref[fig:fig16a]{a}) \#1, (\hyperref[fig:fig16b]{b}) 2, (\hyperref[fig:fig16c]{c}) 3, (\hyperref[fig:fig16d]{d}) 4, (\hyperref[fig:fig16e]{e}) 5, and (\hyperref[fig:fig16f]{f}) 6.}
	\label{fig:fig16}
\end{figure}

Given that the MTMAC can indicate the general state, in terms of damage, of the system for the given cases, the damage localisation task remains. To this aim, the scaled COMAC of cases \#2-7 are plotted in \cref{fig:fig16}. Please note that the COMAC values are normalised between 0 and 1 to better highlight the mode shape trajectory changes, as the damage quantification is already addressed with the MTMAC. The results are first discussed for the case pairs with the same additional mass location, followed by a global discussion.

\textbf{Cases \#2 and 3} (\cref{fig:fig16a,fig:fig16b}): from the two images, it is clear that the areas most influenced by the added mass near the centre of the trailing edge (TE) are near the tip and closer to the root at $z \approx$ -2 m. Notably, as the mass is increased from M1 to M3, the deviation in the mode shapes becomes more dominant near the tip, while retaining some deviation at the same near-root area.

\textbf{Cases \#4 and 5} (\cref{fig:fig16c,fig:fig16d}): In these cases, the most impacted areas are near the leading edge (LE) root and around mid-span. In particular, for the M1 case, the mid-span deviation is more dominant than that at the root. However, this changes for M3, as the root sees the highest deviation and the mid-span change extends, with a lower magnitude, towards the tip. As will be discussed later, this is the only case where the highest deviation exactly matches the damage location.

\textbf{Cases \#6 and 7} (\cref{fig:fig16e,fig:fig16f}): the most prominent areas where a deviation in $\bm{\phi}_{1-3}$ are seen are the near-tip region and the same close to the root section in cases \#2 and 3. Notably, the maximum deviation for both cases is seen near the damaged area. However, when the mass is increased from M2 to M3, the maximum deviation point shifts slightly towards the root (from $x \approx$ -4 m to -3.5 m).

\subsubsection{Real, panel-off, damage}
\begin{figure}[!htb]
\centering
		{\includegraphics[align=c,width=.55\textwidth,keepaspectratio]{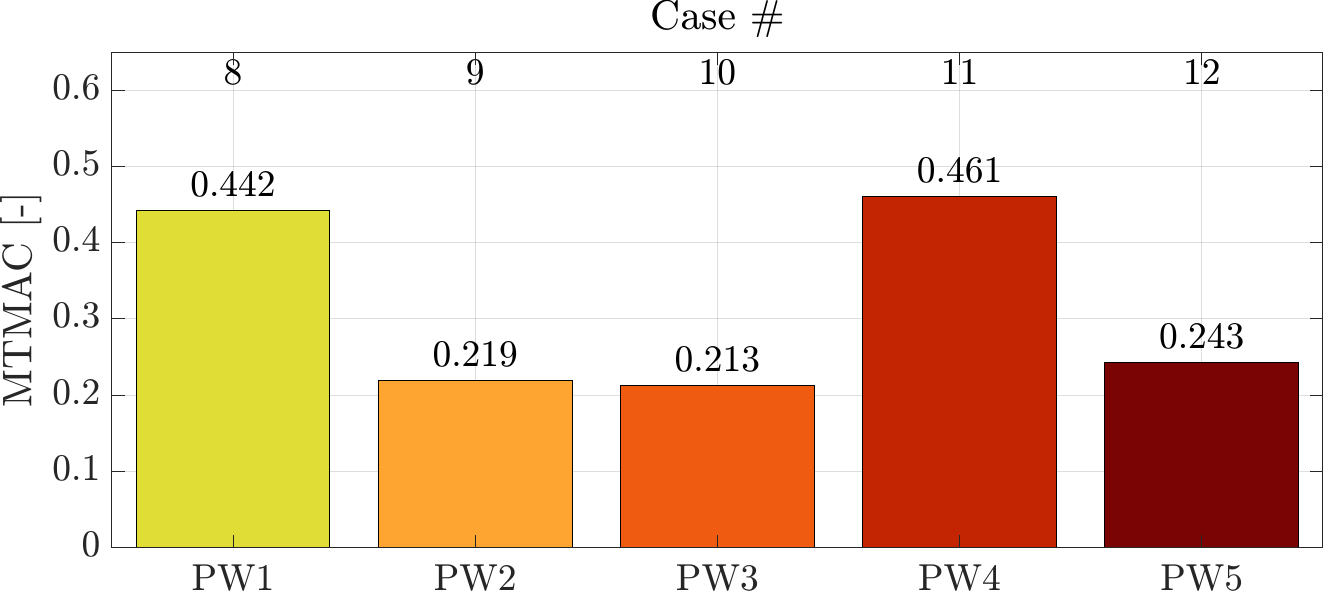}}
	\caption{BAE Systems Hawk T1A aircraft: MTMAC values for cases \#8-12 w.r.t. case \#1.}
	\label{fig:fig17}
\end{figure}

In \cref{fig:fig17}, the MTMAC values for the panel-off cases (\#8-12) are shown. Notably, the values for the panels removed at PW2-3 and 5 have a similar magnitude, while for the PW1 and PW4 cases, the MTMAC values are doubled. These larger deviations in $\omega_{1-3}$ and $\bm{\phi}_{1-3}$ from case \#1 show that the latter locations have a higher sensitivity to damage.

\begin{figure}[!htb]
\centering
	\begin{subfigure}[t]{.34\textwidth}
	\centering
		{\includegraphics[width=\textwidth,keepaspectratio]{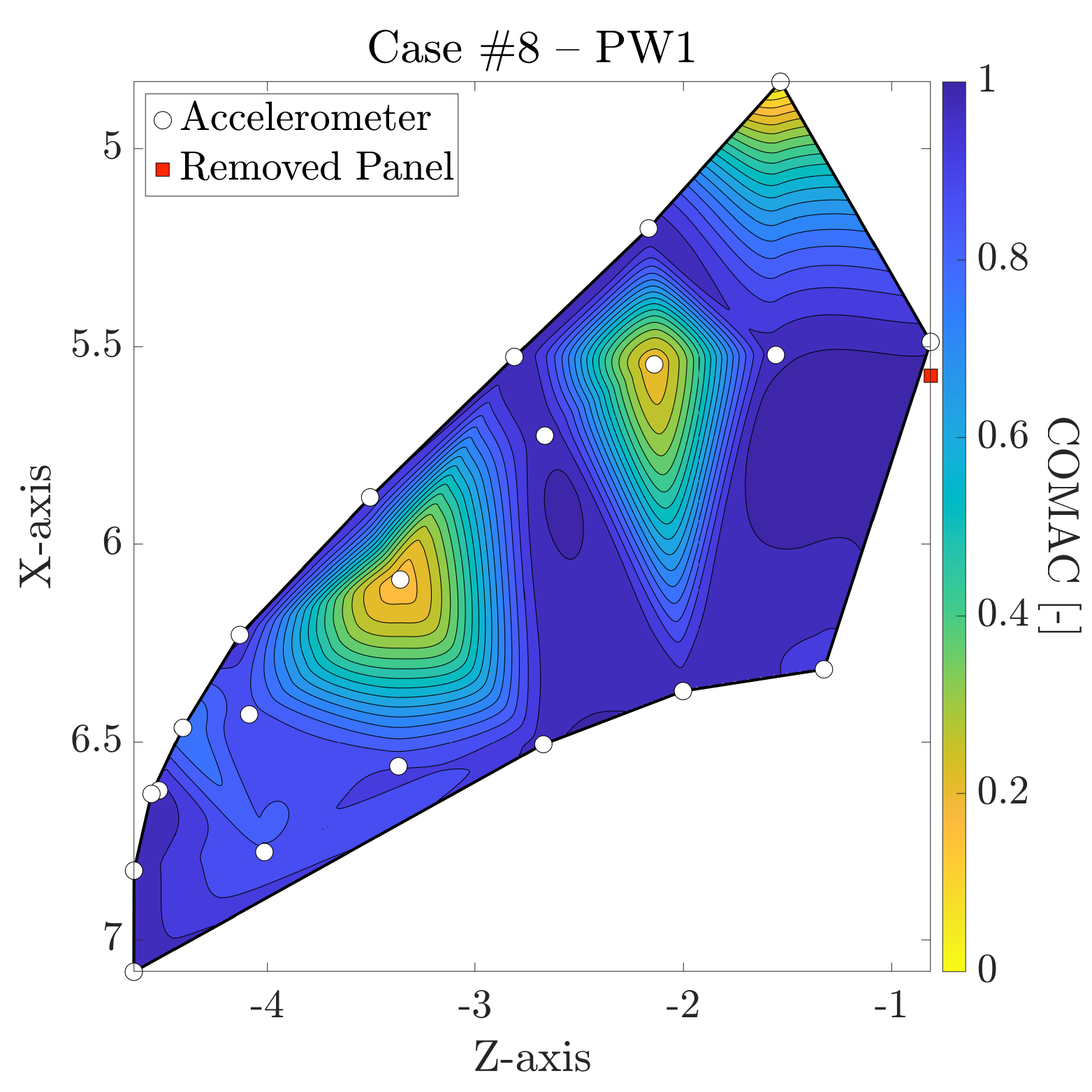}}
		\captionsetup{font={it},justification=centering}
		\subcaption{\label{fig:fig18a}}	
	\end{subfigure}
    \begin{subfigure}[t]{.34\textwidth}
	\centering
		\includegraphics[width=\textwidth,keepaspectratio]{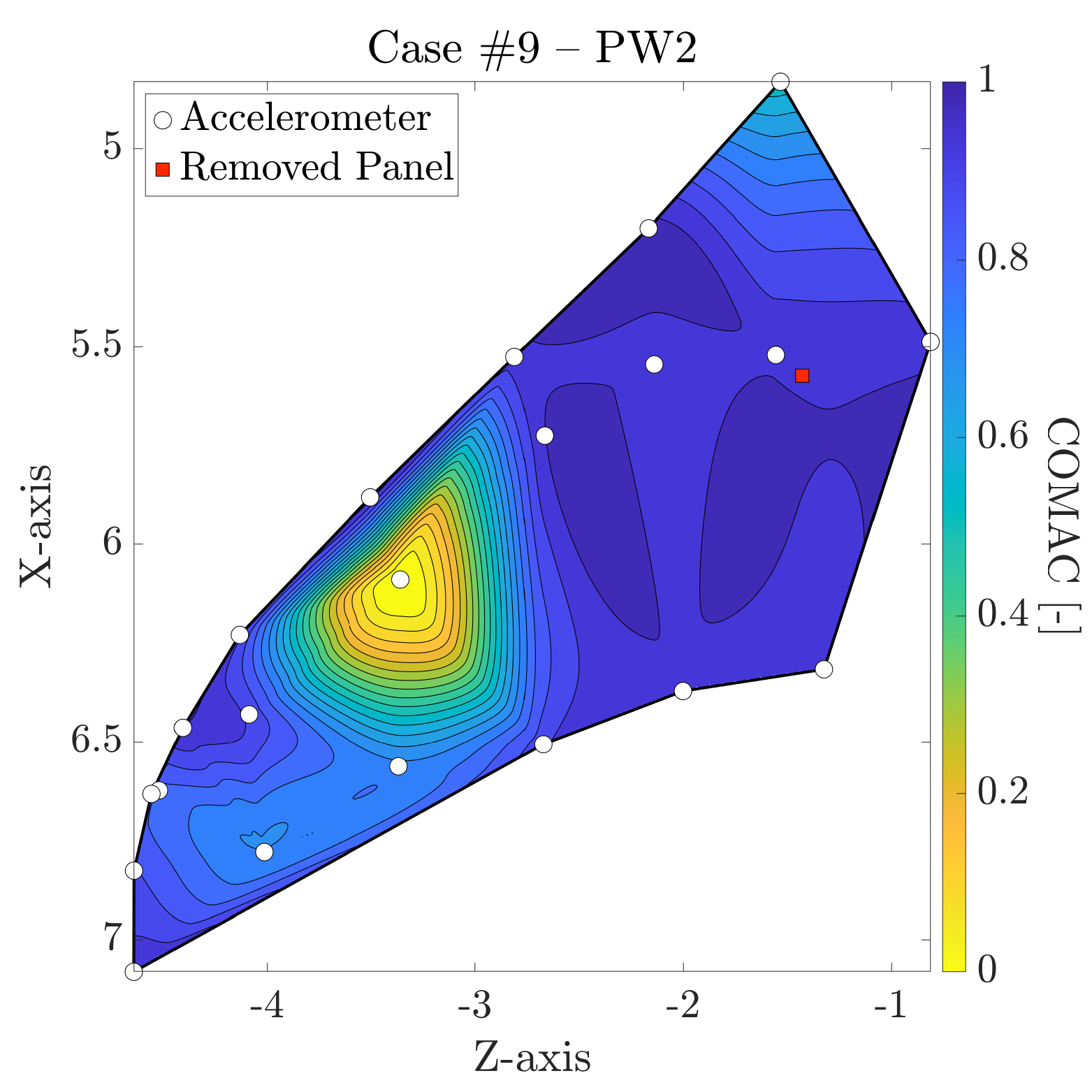}
		\captionsetup{font={it},justification=centering}
		\subcaption{\label{fig:fig18b}}	
	\end{subfigure}
    \begin{subfigure}[t]{.34\textwidth}
	\centering
		{\includegraphics[width=\textwidth,keepaspectratio]{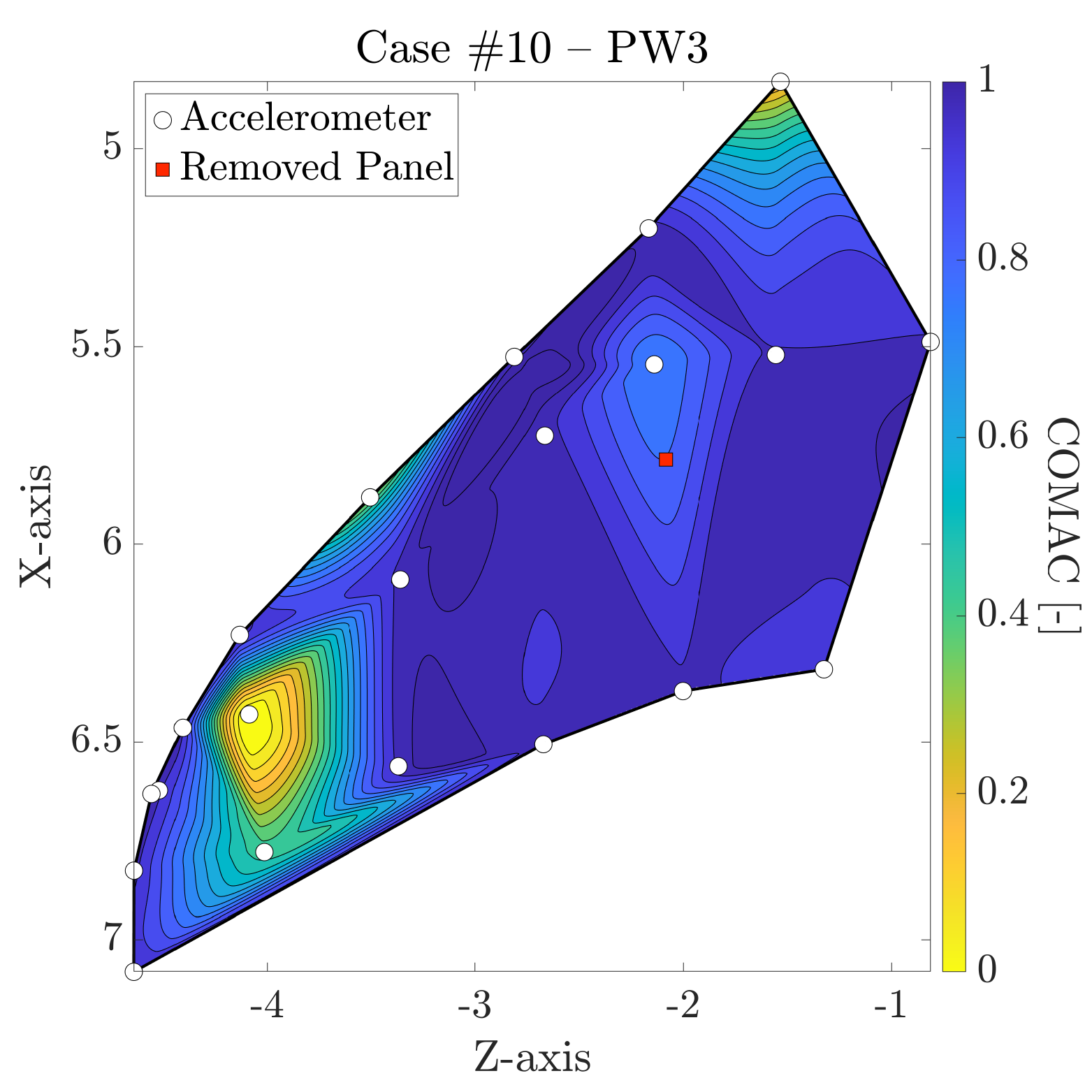}}
		\captionsetup{font={it},justification=centering}
		\subcaption{\label{fig:fig18c}}	
	\end{subfigure}
    \begin{subfigure}[t]{.34\textwidth}
	\centering
		\includegraphics[width=\textwidth,keepaspectratio]{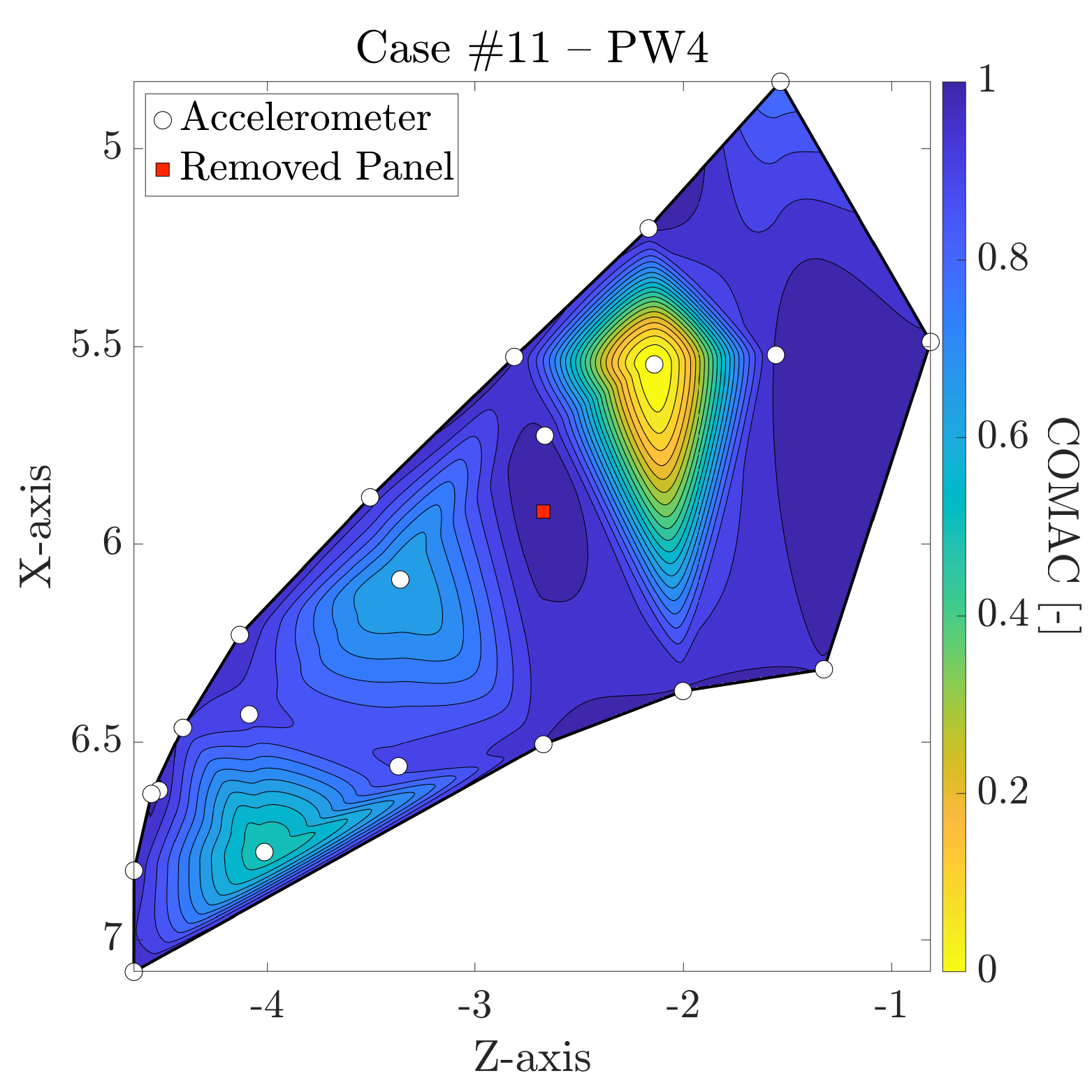}
		\captionsetup{font={it},justification=centering}
		\subcaption{\label{fig:fig18d}}	
	\end{subfigure}
        \begin{subfigure}[t]{.34\textwidth}
	\centering
		{\includegraphics[width=\textwidth,keepaspectratio]{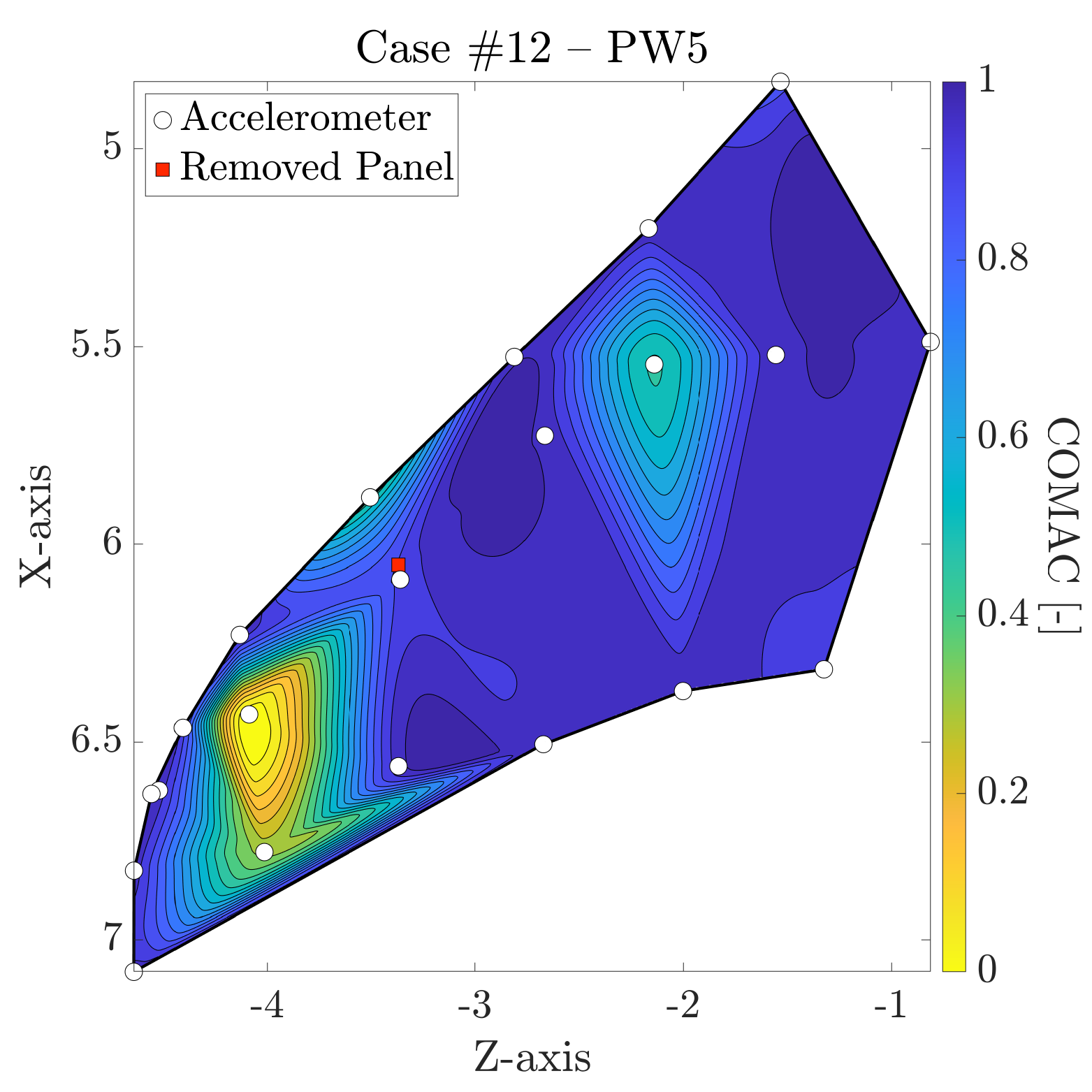}}
		\captionsetup{font={it},justification=centering}
		\subcaption{\label{fig:fig18e}}	
	\end{subfigure}
	\caption{BAE Systems Hawk T1A aircraft: Scaled COMAC for the port wing identified via iLF from cases (\hyperref[fig:fig18a]{a}) \#10, (\hyperref[fig:fig18b]{b}) 11, (\hyperref[fig:fig18c]{c}) 12, (\hyperref[fig:fig18d]{d})  13, and (\hyperref[fig:fig18e]{e}) 14.}
	\label{fig:fig18}
\end{figure}

\Cref{fig:fig18} shows the scaled COMAC for all of the panel-off cases. It is important to see that for all different locations, different deviations are identified with the COMAC; however, the maximum COMAC never exactly matches the damaged area. {This is discussed later in the article.}

\subsection{Discussion}
For all cases, added mass and panel-off, the MTMAC successfully detects a change in the system modal parameters $\omega_{1-3}$ and $\bm{\phi}_{1-3}$. In particular, from the two subsets (added mass and panel-off), it is possible to see which locations are more sensitive to changes in the structure. The highest MTMAC values are recorded for cases \#8 and 11, which, respectively, represent panel-off cases at locations near the root (PW1) and at mid-span (PW4). The other three panel-off cases show a similar MTMAC value, hence showing a similar sensitivity. Notably, all panels are roughly placed between the quarter chord and mid-chord. On the other hand, the added mass locations are close to either the LE or TE. For them, the most sensitive location to damage is case \#5, near the root leading edge. In contrast, the location least sensitive to damage is at TE at mid-span. Nevertheless, in all cases, it is possible to identify the increase in the added mass with an increase in the MTMAC, as shown in \cref{fig:fig15}. {This means that MTMAC achieves unequivocality in the damage assessment, which is what is lacking in carrying out this assessment through modal parameters. This is confirmed by the early identified results in }\cref{fig:fig13a} and \cref{tab:tab4}.

With respect to damage localisation, this is directly achieved only for cases \#5-7. Nevertheless, each couple of added mass cases is somewhat coherent in their COMAC values distribution and, most importantly, it is unequivocally different from the others. The latter also holds for panel-off cases. These findings indicate that the COMAC can be effectively used for damage location in complex structures, particularly when paired with a high-fidelity finite element model of the system under examination or an empirical measure to correlate the COMAC value distribution with the predicted damage location. Unfortunately, neither was available for the wing under scrutiny at the time of writing; this aspect is not addressed in this work, as it lies beyond its scope, i.e., introducing the use of iLF for MTMAC-based SHM in MIMO tests. 

{Before moving to the conclusions, it is important to address some limitations of the newly-proposed method. As mentioned, the identification and damage assessment framework is limited to the linear (or linearised) response of the system under scrutiny; this, however, is conventional in current GVT applications. Regarding the interpretation of modal deformations, it must be said that while COMAC helps, the interpretation of complex mode shape deviations remains challenging, and location accuracy may be reduced in highly complex structures. This is, in fact, a recognised limitation of all mode-shape-based approaches. Finally, for field test validation, only one real aircraft dataset was used, with limited damage scenarios. Broader validation (other structures, other types of damage) would be needed for further generalisation, even if iLF has already proved effective in other Aerospace} \cite{Dessena2024}{, Mechanical} \cite{Dessena2022} {and even Civil Engineering }\cite{Dessena2024f} {applications. Furthermore, no environmental effects have been considered in this work as the data used in this work was collected in a controlled experimental environment.}

{However, iLF proved capable of robustly identifying the (linear) modal parameters of the system, which MTMAC successfully employed for structural assessment. In this regard, mode shapes can be extracted indifferently from displacement, velocities, or acceleration time series; in this work, acceleration time histories were considered only because they are much more common in standard practice.}

\section{Conclusions\label{sec:con}}
{In this work, a novel vibration-based damage detection and localisation method was applied, targeting the linearised dynamics (the regime in which most SHM and modal analysis methods are validated) of a mechanical system. In particular, the proposal has been verified on a numerically simulated hollow beam and a complex, real-life aeronautical case study of a trainer jet aircraft.} 

A two-step procedure has been introduced, including the identification of the modal parameters with the improved Loewner framework (iLF) and the detection and quantification of damage with the modified total modal assurance criterion (MTMAC). This is the first instance that the iLF has been used in structural health monitoring and that the MTMAC has been employed as a standalone damage assessment index. 

{First, the proposed approach has been validated on a numerical, fully controlled case study. In this benchmark, no nonlinear joints or contact effects were modelled; instead, the hollow beam was designed as a purely linear system to isolate the performance of the method.}

Then, for experimental validation, damage detection and localisation have been innovatively carried out on the multi-input multi-output vibration dataset of the BAE Systems Hawk T1A aircraft from the Laboratory for Verification and Validation at the University of Sheffield. {The proposed method has been shown to retain generality within the LTI assumption, as it applies equally to SIMO and MIMO testing, as shown by moving from the numerically simulated beam to the experimental dataset.} 

These two analyses resulted in the following findings:

\begin{itemize}
    \item The iLF has been numerically and experimentally proven to detect small damage-related changes in modal parameters from multi-input multi-output (MIMO) case studies - either numerically-simulated multiple degrees of freedom systems and complex experimental data (the Hawk T1A jet trainer);
    \item {The use of the MTMAC has been validated for damage assessment and quantification. In particular, it has been proved that this modal parameters-based damage-sensitive feature gives a condensed index that has been shown to work well in both numerical and experimental data;}
    \item Damage was successfully identified in 11 cases of the recently introduced Hawk T1A aircraft MIMO dataset, a first in the literature. {Supporting the methodology feasibility for real-world SHM campaigns;}
    \item{The lack of partial and/or full airframe damage detection applications and methods is addressed.}
\end{itemize}

{Concerning future directions, this study focused solely on System Identification, aiming at data-driven damage assessment. However, the procedure presented here can also be paired with calibrated numerical models to compare the results with those obtained by FEA, potentially to identify regions of stress concentration (for damage localisation) and/or quantify damage (for severity assessment). Furthermore, while the framework successfully differentiated between healthy and damaged cases, no absolute, universally valid threshold has been set to define a ‘critical’ or irreversible structural state, because these will necessarily be case-specific. In practice, such thresholds could be determined statistically from repeated baseline measurements or derived from FEA/digital twin models coupled with certification criteria. Integrating these aspects will be an important direction for the practical deployment of the proposed method.}

\appendix
\section{Euler-Bernoulli beam mass and stiffness matrices}\label{sec:apa}

The 2D Euler-Bernoulli beam mass and stiffness matrices are shown in, respectively, \cref{eq:a1,eq:a2}.

{\scriptsize%
\begin{equation}
M_e = \frac{\rho A a}{105}
\begin{bmatrix}
78 & 0 & 0 & 22a & 27 & 0 & 0 & -13a \\
0 & 78 & -22a & 0 & 0 & 27 & 13a & 0 \\
0 & -22a & 8a^2 & 0 & 0 & -13a & -6a^2 & 0 \\
22a & 0 & 0 & 8a^2 & 13a & 0 & 0 & -6a^2 \\
27 & 0 & 0 & 13a & 78 & 0 & 0 & -22a \\
0 & 27 & -13a & 0 & 0 & 78 & 22a & 0 \\
0 & 13a & -6a^2 & 0 & 0 & 22a & 8a^2 & 0 \\
-13a & 0 & 0 & -6a^2 & -22a & 0 & 0 & 8a^2 \\
\end{bmatrix}
\label{eq:a1}
\end{equation}
}
{\scriptsize%
\begin{equation}
K_e =
\begin{bmatrix}
\frac{3EI_z}{2a^3} & 0 & 0 & \frac{3EI_z}{2a^2} & -\frac{3EI_z}{2a^3} & 0 & 0 & \frac{3EI_z}{2a^2}\\
0 & \frac{3EI_y}{2a^3} & -\frac{3EI_y}{2a^3} & 0 & 0 & -\frac{3EI_y}{2a^2} & -\frac{3EI_y}{2a^2} & 0\\
0 & -\frac{3EI_y}{2a^3} & \frac{2EI_y}{a} & 0 & 0 & 0 & \frac{2EI_y}{a} & 0\\
\frac{3EI_z}{2a^2} & 0 & 0 & \frac{2EI_z}{a} & -\frac{3EI_z}{2a^2} & 0 & 0 & \frac{EI_z}{a}\\
-\frac{3EI_z}{2a^3} & 0 & 0 & -\frac{3EI_z}{2a^2} & \frac{3EI_z}{2a^3} & 0 & 0 & -\frac{3EI_z}{2a^2}\\
0 & -\frac{3EI_y}{2a^2} & 0 & 0 & 0 & \frac{3EI_y}{2a^3} & \frac{3EI_y}{2a^2} & 0\\
0 & -\frac{3EI_y}{2a^2} & \frac{2EI_y}{a} & 0 & 0 & \frac{3EI_y}{2a^2} & \frac{3EI_y}{2a^3} & 0\\
\frac{3EI_z}{2a^2} & 0 & 0 & \frac{EI_z}{a} & -\frac{3EI_z}{2a^2} & 0 & 0 & \frac{2EI_z}{a}
\end{bmatrix}
\label{eq:a2}
\end{equation}
}

{\small
\section*{{\small Corresponding author}}
\noindent Gabriele Dessena \orcidlink{0000-0001-7394-9303} - Primary corresponding author\\
E-mail address: \href{mailto:gdessena@ing.uc3m.es}{gdessena@ing.uc3m.es}\\
\noindent Marco Civera \orcidlink{0000-0003-0414-7440} - Corresponding author\\
E-mail address: \href{mailto:marco.civera@polito.it}{marco.civera@polito.it}

\section*{{\small Author contributions}}
\noindent {\small Conceptualisation, G.D., and M.C.; methodology, G.D., and M.C.; software, G.D.; validation, G.D. and M.C.; formal analysis, G.D.; investigation, G.D. and M.C.; resources, G.D, M.C., B.C., A.M., and OE.BM.; data curation, G.D. and M.C.; writing---original draft preparation, G.D. and M.C.; writing---review and editing, G.D, M.C., A.M., B.C., and OE.BM.; visualisation, G.D., M.C, and A.M.; funding acquisition, G.D, M.C., B.C., A.M., and OE.BM..}

\section*{{\small Declaration of conflicting interests}}
\noindent {\small The author(s) declared no potential conflicts of interest with respect to the research, authorship, and/or publication of this article.}

\section*{{\small Funding}}
\noindent {\small
The first author is supported by the grants for research activity of young PhD holders, part of the Universidad Carlos III de Madrid (UC3M) Own Research Program (\textit{Ayudas para la Actividad Investigadora de los Jóvenes Doctores, del Programa Propio de Investigación de la UC3M}). Furthermore, the authors from UC3M disclosed receipt of the following financial support for the research, authorship, and/or publication of this article: This work has been supported by the Madrid Government (\textit{Comunidad de Madrid}– Spain) under the Multiannual Agreement with the UC3M (IA\_aCTRl-CM-UC3M). The second and fourth authors are supported by the \textit{Centro Nazionale per la Mobilità Sostenibile} (MOST – Sustainable Mobility Center), Spoke 7 (Cooperative Connected and Automated Mobility and Smart Infrastructures), Work Package 4 (Resilience of Networks, Structural Health Monitoring and Asset Management). The third author gladly acknowledges funding as a Beatriz Galindo Distinguished Senior Researcher by the Spanish Government.}

\section*{{\small Acknowledgements}}
\noindent {\small
The \href{https://orda.shef.ac.uk/articles/dataset/BAE_T1A_Hawk_Full_Structure_Modal_Test/24948549}{BAE T1A Hawk Full Structure Modal Test} has been retrieved from the University of Sheffield data repository ORDA and is cited in this work. The authors thank the dataset authors and the LVV at the University of Sheffield for making them openly available.

\section*{{\small Data availability statement}\label{sec:6_data}}
{\small 
The experimental and numerical results (Data file 1) supporting this work are openly available from the Zenodo repository at [\url{https://zenodo.org/records/13866467}]. In addition, this study used third-party experimental data (Data file 2) made available at [\url{orda.shef.ac.uk/articles/dataset/BAE_T1A_Hawk_Full_Structure_Modal_Test/24948549}] under a licence that the authors do not have permission to share:
\begin{itemize}
    \item Data file 1. Data supporting: Multiple input tangential interpolation-driven damage detection of a jet trainer aircraft;
    \item Data file 2. BAE T1A Hawk Full Structure Modal Test.
\end{itemize}
Data file 1 is available under the terms of the [GNU General Public License v3.0 (GPL 3.0)].}

\bibliographystyle{elsarticle-num} 

\begin{thebibliography}{10}
\expandafter\ifx\csname url\endcsname\relax
  \def\url#1{\texttt{#1}}\fi
\expandafter\ifx\csname urlprefix\endcsname\relax\def\urlprefix{URL }\fi
\expandafter\ifx\csname href\endcsname\relax
  \def\href#1#2{#2} \def\path#1{#1}\fi

\bibitem{Farrar2007}
C.~R. Farrar, K.~Worden, \href{https://royalsocietypublishing.org/doi/10.1098/rsta.2006.1928}{{An introduction to structural health monitoring}}, Philosophical Transactions of the Royal Society A: Mathematical, Physical and Engineering Sciences 365~(1851) (2007) 303--315.
\newblock \href {https://doi.org/10.1098/rsta.2006.1928} {\path{doi:10.1098/rsta.2006.1928}}.
\newline\urlprefix\url{https://royalsocietypublishing.org/doi/10.1098/rsta.2006.1928}

\bibitem{Gelman2020}
L.~Gelman, I.~Petrunin, C.~Parrish, M.~Walters, \href{https://onlinelibrary.wiley.com/doi/10.1002/stc.2479}{{Novel health monitoring technology for in‐service diagnostics of intake separation in aircraft engines}}, Structural Control and Health Monitoring 27~(5) (2020) 1--14.
\newblock \href {https://doi.org/10.1002/stc.2479} {\path{doi:10.1002/stc.2479}}.
\newline\urlprefix\url{https://onlinelibrary.wiley.com/doi/10.1002/stc.2479}

\bibitem{Huang2024}
Y.-T. Huang, D.~Y.-H. Chou, C.-C. Chou, C.-H. Loh, \href{https://onlinelibrary.wiley.com/doi/10.1155/2024/3412305}{{Data‐Driven Structural Health Monitoring on Shaking Table Tests of a 3‐Story Steel Building with Sliding Slabs}}, Structural Control and Health Monitoring 2024~(1) (jan 2024).
\newblock \href {https://doi.org/10.1155/2024/3412305} {\path{doi:10.1155/2024/3412305}}.
\newline\urlprefix\url{https://onlinelibrary.wiley.com/doi/10.1155/2024/3412305}

\bibitem{Zhang2024}
B.~Zhang, Y.~Fu, H.~Liu, Y.~Zhu, W.~Xiong, R.~Ma, \href{https://onlinelibrary.wiley.com/doi/10.1155/2024/2394178}{{A Vibration‐Based Quasi‐Real‐Time Cable Force Identification Method for Cable Replacement Monitoring}}, Structural Control and Health Monitoring 2024~(1) (jan 2024).
\newblock \href {https://doi.org/10.1155/2024/2394178} {\path{doi:10.1155/2024/2394178}}.
\newline\urlprefix\url{https://onlinelibrary.wiley.com/doi/10.1155/2024/2394178}

\bibitem{Rytter1993}
A.~A. Rytter, \href{https://vbn.aau.dk/en/publications/vibrational-based-inspection-of-civil-engineering-structures}{{Vibrational Based inspection of Civil Engineering Structures}}, Phd thesis, Aalborg University (1993).
\newline\urlprefix\url{https://vbn.aau.dk/en/publications/vibrational-based-inspection-of-civil-engineering-structures}

\bibitem{Fan2011}
{W. Fan}, {P. Qiao}, \href{https://journals.sagepub.com/doi/10.1177/1475921710365419}{{Vibration-based Damage Identification Methods: A Review and Comparative Study}}, Structural Health Monitoring 10~(1) (2011) 83--111.
\newblock \href {https://doi.org/10.1177/1475921710365419} {\path{doi:10.1177/1475921710365419}}.
\newline\urlprefix\url{https://journals.sagepub.com/doi/10.1177/1475921710365419}

\bibitem{Civera2021a}
M.~Civera, G.~Calamai, L.~{Zanotti Fragonara}, \href{https://linkinghub.elsevier.com/retrieve/pii/S2352012420307979}{{System identification via Fast Relaxed Vector Fitting for the Structural Health Monitoring of masonry bridges}}, Structures 30~(January) (2021) 277--293.
\newblock \href {https://doi.org/10.1016/j.istruc.2020.12.073} {\path{doi:10.1016/j.istruc.2020.12.073}}.
\newline\urlprefix\url{https://linkinghub.elsevier.com/retrieve/pii/S2352012420307979}

\bibitem{Modesti2024}
M.~Modesti, C.~Gentilini, A.~Palermo, E.~Reynders, G.~Lombaert, \href{https://link.springer.com/10.1007/s13349-024-00839-0}{{A two-step procedure for damage detection in beam structures with incomplete mode shapes}}, Journal of Civil Structural Health Monitoring (aug 2024).
\newblock \href {https://doi.org/10.1007/s13349-024-00839-0} {\path{doi:10.1007/s13349-024-00839-0}}.
\newline\urlprefix\url{https://link.springer.com/10.1007/s13349-024-00839-0}

\bibitem{Reynders2008}
E.~Reynders, R.~Pintelon, G.~{De Roeck}, {Uncertainty bounds on modal parameters obtained from stochastic subspace identification}, Mechanical Systems and Signal Processing 22~(4) (2008) 948--969.
\newblock \href {https://doi.org/10.1016/j.ymssp.2007.10.009} {\path{doi:10.1016/j.ymssp.2007.10.009}}.

\bibitem{Zhang2025}
S.~Zhang, M.~K{\"{a}}ding, M.~Wenner, M.~Cla{\ss}en, S.~Marx, \href{https://linkinghub.elsevier.com/retrieve/pii/S0141029624020078}{{Structural health monitoring (SHM) on a long semi-integral high-speed railway bridge (HSRB) under different traffic loads}}, Engineering Structures 325~(November 2024) (2025) 119445.
\newblock \href {https://doi.org/10.1016/j.engstruct.2024.119445} {\path{doi:10.1016/j.engstruct.2024.119445}}.
\newline\urlprefix\url{https://linkinghub.elsevier.com/retrieve/pii/S0141029624020078}

\bibitem{Vettori2024}
S.~Vettori, E.~{Di Lorenzo}, B.~Peeters, E.~Chatzi, \href{https://linkinghub.elsevier.com/retrieve/pii/S0888327024002012}{{Assessment of alternative covariance functions for joint input-state estimation via Gaussian Process latent force models in structural dynamics}}, Mechanical Systems and Signal Processing 213~(November 2023) (2024) 111303.
\newblock \href {https://doi.org/10.1016/j.ymssp.2024.111303} {\path{doi:10.1016/j.ymssp.2024.111303}}.
\newline\urlprefix\url{https://linkinghub.elsevier.com/retrieve/pii/S0888327024002012}

\bibitem{OConnell2024}
B.~J. O'Connell, T.~J. Rogers, \href{https://linkinghub.elsevier.com/retrieve/pii/S0022460X24001445}{{A robust probabilistic approach to stochastic subspace identification}}, Journal of Sound and Vibration 581~(March) (2024) 118381.
\newblock \href {http://arxiv.org/abs/2305.16836} {\path{arXiv:2305.16836}}, \href {https://doi.org/10.1016/j.jsv.2024.118381} {\path{doi:10.1016/j.jsv.2024.118381}}.
\newline\urlprefix\url{https://linkinghub.elsevier.com/retrieve/pii/S0022460X24001445}

\bibitem{Sibille2023}
L.~Sibille, M.~Civera, L.~{Zanotti Fragonara}, R.~Ceravolo, \href{https://arc.aiaa.org/doi/10.2514/1.J062084}{{Automated Operational Modal Analysis of a Helicopter Blade with a Density-Based Cluster Algorithm}}, AIAA Journal 61~(3) (2023) 1411--1427.
\newblock \href {https://doi.org/10.2514/1.J062084} {\path{doi:10.2514/1.J062084}}.
\newline\urlprefix\url{https://arc.aiaa.org/doi/10.2514/1.J062084}

\bibitem{Wu2025}
C.~Wu, Z.~Yang, S.~He, \href{https://linkinghub.elsevier.com/retrieve/pii/S127096382500183X}{{Efficient modal parameter identification using DMD-DBSCAN and rank stabilization diagrams}}, Aerospace Science and Technology 161~(March) (2025) 110112.
\newblock \href {https://doi.org/10.1016/j.ast.2025.110112} {\path{doi:10.1016/j.ast.2025.110112}}.
\newline\urlprefix\url{https://linkinghub.elsevier.com/retrieve/pii/S127096382500183X}

\bibitem{Li2025}
L.~Li, A.~Br{\"{u}}gger, R.~Betti, Z.~Shen, L.~Gan, H.~Gu, \href{https://linkinghub.elsevier.com/retrieve/pii/S0045794924003213}{{A Cepstrum-Informed neural network for Vibration-Based structural damage assessment}}, Computers {\&} Structures 306~(November 2024) (2025) 107592.
\newblock \href {https://doi.org/10.1016/j.compstruc.2024.107592} {\path{doi:10.1016/j.compstruc.2024.107592}}.
\newline\urlprefix\url{https://linkinghub.elsevier.com/retrieve/pii/S0045794924003213}

\bibitem{Ni2025}
Y.-c. Ni, F.-l. Zhang, \href{https://linkinghub.elsevier.com/retrieve/pii/S0888327024009816}{{Fast Bayesian modal identification based on seismic response considering the ambient effect}}, Mechanical Systems and Signal Processing 224~(September 2024) (2025) 112083.
\newblock \href {https://doi.org/10.1016/j.ymssp.2024.112083} {\path{doi:10.1016/j.ymssp.2024.112083}}.
\newline\urlprefix\url{https://linkinghub.elsevier.com/retrieve/pii/S0888327024009816}

\bibitem{Sanchez-Haro2025}
J.~S{\'{a}}nchez-Haro, M.~Garc{\'{i}}a, G.~Capell{\'{a}}n, A.~da~Costa, P.~Perez, J.~A{\~{n}}{\'{o}}, \href{https://linkinghub.elsevier.com/retrieve/pii/S2352012424020691}{{Digital twin for predictive maintenance on the Espartxo Bridge. Application to early detection of under-foundation scour}}, Structures 71~(February 2024) (2025) 107916.
\newblock \href {https://doi.org/10.1016/j.istruc.2024.107916} {\path{doi:10.1016/j.istruc.2024.107916}}.
\newline\urlprefix\url{https://linkinghub.elsevier.com/retrieve/pii/S2352012424020691}

\bibitem{Cadoret2025}
A.~Cadoret, E.~Denimal-Goy, J.-M. Leroy, J.-L. Pfister, L.~Mevel, \href{https://linkinghub.elsevier.com/retrieve/pii/S088832702400880X}{{Damage detection and localization method for wind turbine rotor based on Operational Modal Analysis and anisotropy tracking}}, Mechanical Systems and Signal Processing 224~(September 2024) (2025) 111982.
\newblock \href {https://doi.org/10.1016/j.ymssp.2024.111982} {\path{doi:10.1016/j.ymssp.2024.111982}}.
\newline\urlprefix\url{https://linkinghub.elsevier.com/retrieve/pii/S088832702400880X}

\bibitem{Dadoulis2025}
G.~Dadoulis, G.~D. Manolis, K.~Katakalos, K.~Dragos, K.~Smarsly, \href{https://doi.org/10.1016/j.engstruct.2024.119216}{{Damage detection in lightweight bridges with traveling masses using machine learning}}, Engineering Structures 322~(PB) (2025) 119216.
\newblock \href {https://doi.org/10.1016/j.engstruct.2024.119216} {\path{doi:10.1016/j.engstruct.2024.119216}}.
\newline\urlprefix\url{https://doi.org/10.1016/j.engstruct.2024.119216}

\bibitem{Ren2026}
Y.~Ren, J.~Li, S.~Yuan, L.~Qiu, L.~Gao, M.~Lu, \href{https://doi.org/10.1016/j.ast.2025.110787}{{An enhanced probability signal construction based on-line damage imaging method for aircraft structures under service environment}}, Aerospace Science and Technology 168~(PA) (2026) 110787.
\newblock \href {https://doi.org/10.1016/j.ast.2025.110787} {\path{doi:10.1016/j.ast.2025.110787}}.
\newline\urlprefix\url{https://doi.org/10.1016/j.ast.2025.110787}

\bibitem{Astorga2025}
A.~Astorga, P.~Gu{\'{e}}guen, M.~Beth, N.~Bessoule, \href{https://linkinghub.elsevier.com/retrieve/pii/S0141029624016092}{{Exploring the effective implementation of population-based SHM in existing buildings. Part I: Structural condition assessment}}, Engineering Structures 322~(January 2024) (2025) 119047.
\newblock \href {https://doi.org/10.1016/j.engstruct.2024.119047} {\path{doi:10.1016/j.engstruct.2024.119047}}.
\newline\urlprefix\url{https://linkinghub.elsevier.com/retrieve/pii/S0141029624016092}

\bibitem{Cawley2018}
P.~Cawley, \href{http://journals.sagepub.com/doi/10.1177/1475921717750047}{{Structural health monitoring: Closing the gap between research and industrial deployment}}, Structural Health Monitoring 17~(5) (2018) 1225--1244.
\newblock \href {https://doi.org/10.1177/1475921717750047} {\path{doi:10.1177/1475921717750047}}.
\newline\urlprefix\url{http://journals.sagepub.com/doi/10.1177/1475921717750047}

\bibitem{Verhulst2022}
T.~Verhulst, D.~Judt, C.~Lawson, Y.~Chung, O.~Al-Tayawe, G.~Ward, \href{https://papers.phmsociety.org/index.php/ijphm/article/view/3134}{{Review for State-of-the-Art Health Monitoring Technologies on Airframe Fuel Pumps}}, International Journal of Prognostics and Health Management 13~(1) (2022) 1--20.
\newblock \href {https://doi.org/10.36001/ijphm.2022.v13i1.3134} {\path{doi:10.36001/ijphm.2022.v13i1.3134}}.
\newline\urlprefix\url{https://papers.phmsociety.org/index.php/ijphm/article/view/3134}

\bibitem{Gao2019}
D.~Gao, Z.~Wu, L.~Yang, Y.~Zheng, W.~Yin, \href{https://linkinghub.elsevier.com/retrieve/pii/S1270963816309439}{{Structural health monitoring for long-term aircraft storage tanks under cryogenic temperature}}, Aerospace Science and Technology 92 (2019) 881--891.
\newblock \href {https://doi.org/10.1016/j.ast.2019.02.045} {\path{doi:10.1016/j.ast.2019.02.045}}.
\newline\urlprefix\url{https://linkinghub.elsevier.com/retrieve/pii/S1270963816309439}

\bibitem{Liu2024}
Y.~Liu, W.~Long, Y.~Chen, H.~Hu, \href{https://doi.org/10.1016/j.ast.2024.109715}{{Nonlinear vibration characteristics and damage detection method of blade with breathing fatigue crack}}, Aerospace Science and Technology 155~(P3) (2024) 109715.
\newblock \href {https://doi.org/10.1016/j.ast.2024.109715} {\path{doi:10.1016/j.ast.2024.109715}}.
\newline\urlprefix\url{https://doi.org/10.1016/j.ast.2024.109715}

\bibitem{Chen2024}
J.~Chen, Y.~Meng, Y.~Xu, \href{https://linkinghub.elsevier.com/retrieve/pii/S1270963823007204}{{A multi-layer ML model evolutionary paradigm for high-accuracy individual aircraft SHM}}, Aerospace Science and Technology 144~(July 2023) (2024) 108824.
\newblock \href {https://doi.org/10.1016/j.ast.2023.108824} {\path{doi:10.1016/j.ast.2023.108824}}.
\newline\urlprefix\url{https://linkinghub.elsevier.com/retrieve/pii/S1270963823007204}

\bibitem{WentaoZhang1}
W.~Zhang, K.~Lu, Y.~Zhang, J.~Chen, D.~Fu, Y.~Yang, C.~Fu, \href{https://linkinghub.elsevier.com/retrieve/pii/S1270963825000823}{{A comprehensive study on dynamic responses of the whole aero-engine system and design of variable stiffness brackets}}, Aerospace Science and Technology 159~(February) (2025) 110010.
\newblock \href {https://doi.org/10.1016/j.ast.2025.110010} {\path{doi:10.1016/j.ast.2025.110010}}.
\newline\urlprefix\url{https://linkinghub.elsevier.com/retrieve/pii/S1270963825000823}

\bibitem{WentaoZhang2}
W.~Zhang, K.~Lu, Y.~Zhang, H.~Cheng, C.~Fu, \href{https://linkinghub.elsevier.com/retrieve/pii/S0263224124002288}{{Nonlinear dynamics analysis of the attachment system and design of variable stiffness connecting bracket based on the complete aero-engine system}}, Measurement 228~(February) (2024) 114344.
\newblock \href {https://doi.org/10.1016/j.measurement.2024.114344} {\path{doi:10.1016/j.measurement.2024.114344}}.
\newline\urlprefix\url{https://linkinghub.elsevier.com/retrieve/pii/S0263224124002288}

\bibitem{WentaoZhang3}
Y.~Zhang, K.~Lu, W.~Zhang, Y.~Yang, H.~Cheng, C.~Fu, \href{https://journals.sagepub.com/doi/10.1177/10775463231202711}{{Design and vibration control of aeroengine bracket with variable stiffness based on shape memory alloy}}, Journal of Vibration and Control 30~(17-18) (2024) 3850--3861.
\newblock \href {https://doi.org/10.1177/10775463231202711} {\path{doi:10.1177/10775463231202711}}.
\newline\urlprefix\url{https://journals.sagepub.com/doi/10.1177/10775463231202711}

\bibitem{Fang2019}
F.~Fang, L.~Qiu, S.~Yuan, Y.~Ren, \href{https://linkinghub.elsevier.com/retrieve/pii/S1270963819322400}{{Dynamic probability modeling-based aircraft structural health monitoring framework under time-varying conditions: Validation in an in-flight test simulated on ground}}, Aerospace Science and Technology 95 (2019) 105467.
\newblock \href {https://doi.org/10.1016/j.ast.2019.105467} {\path{doi:10.1016/j.ast.2019.105467}}.
\newline\urlprefix\url{https://linkinghub.elsevier.com/retrieve/pii/S1270963819322400}

\bibitem{Pappa1998}
R.~S. Pappa, G.~H. James, D.~C. Zimmerman, \href{https://arc.aiaa.org/doi/10.2514/2.3324}{{Autonomous Modal Identification of the Space Shuttle Tail Rudder}}, Journal of Spacecraft and Rockets 35~(2) (1998) 163--169.
\newblock \href {https://doi.org/10.2514/2.3324} {\path{doi:10.2514/2.3324}}.
\newline\urlprefix\url{https://arc.aiaa.org/doi/10.2514/2.3324}

\bibitem{Figueiredo2009}
E.~Figueiredo, G.~Park, J.~Figueiras, C.~Farrar, K.~Worden, \href{https://www.osti.gov/servlets/purl/961604/}{{Structural health monitoring algorithm comparisons using standard data sets}}, Tech. rep., Los Alamos National Laboratory (LANL), Los Alamos, NM (United States) (mar 2009).
\newblock \href {https://doi.org/10.2172/961604} {\path{doi:10.2172/961604}}.
\newline\urlprefix\url{https://www.osti.gov/servlets/purl/961604/}

\bibitem{Maeck2001}
J.~Maeck, B.~Peeters, G.~D. Roeck, \href{https://iopscience.iop.org/article/10.1088/0964-1726/10/3/313}{{Damage identification on the Z24 bridge using vibration monitoring}}, Smart Materials and Structures 10~(3) (2001) 512--517.
\newblock \href {https://doi.org/10.1088/0964-1726/10/3/313} {\path{doi:10.1088/0964-1726/10/3/313}}.
\newline\urlprefix\url{https://iopscience.iop.org/article/10.1088/0964-1726/10/3/313}

\bibitem{Dessena2022}
G.~Dessena, M.~Civera, L.~{Zanotti Fragonara}, D.~I. Ignatyev, J.~F. Whidborne, \href{https://www.hindawi.com/journals/schm/2023/1891062/}{{A Loewner-Based System Identification and Structural Health Monitoring Approach for Mechanical Systems}}, Structural Control and Health Monitoring 2023 (2023) 1--22.
\newblock \href {https://doi.org/10.1155/2023/1891062} {\path{doi:10.1155/2023/1891062}}.
\newline\urlprefix\url{https://www.hindawi.com/journals/schm/2023/1891062/}

\bibitem{Reynders2012a}
E.~Reynders, J.~Houbrechts, G.~{De Roeck}, \href{https://linkinghub.elsevier.com/retrieve/pii/S0888327012000088}{{Fully automated (operational) modal analysis}}, Mechanical Systems and Signal Processing 29 (2012) 228--250.
\newblock \href {https://doi.org/10.1016/j.ymssp.2012.01.007} {\path{doi:10.1016/j.ymssp.2012.01.007}}.
\newline\urlprefix\url{https://linkinghub.elsevier.com/retrieve/pii/S0888327012000088}

\bibitem{Peeters2015}
B.~Peeters, G.~{De Roeck}, \href{https://onlinelibrary.wiley.com/doi/10.1002/1096-9845(200102)30:2{\%}3C149::AID-EQE1{\%}3E3.0.CO;2-Z}{{One-year monitoring of the Z24-Bridge: environmental effects versus damage events}}, Earthquake Engineering {\&} Structural Dynamics 30~(2) (2001) 149--171.
\newblock \href {https://doi.org/10.1002/1096-9845(200102)30:2<149::AID-EQE1>3.0.CO;2-Z} {\path{doi:10.1002/1096-9845(200102)30:2<149::AID-EQE1>3.0.CO;2-Z}}.
\newline\urlprefix\url{https://onlinelibrary.wiley.com/doi/10.1002/1096-9845(200102)30:2%3C149::AID-EQE1%3E3.0.CO;2-Z}

\bibitem{Catbas1998}
F.~N. Catbas, M.~S. Lenett, A.~E. Aktan, D.~L. Brown, {Damage Detection and Condition Assessment of Seymour Bridge}, in: Proceedings of the 16th International Modal Analysis Conference (IMAC), Society for Experimental Mechanics Inc, SPIE-The International Society for Optical Engineering, Santa Barbara, CA, 1998, pp. 1694--1702.

\bibitem{Catbas2004}
F.~N. Catbas, D.~L. Brown, A.~E. Aktan, \href{https://ascelibrary.org/doi/10.1061/{\%}28ASCE{\%}290733-9399{\%}282004{\%}29130{\%}3A8{\%}28921{\%}29}{{Parameter Estimation for Multiple-Input Multiple-Output Modal Analysis of Large Structures}}, Journal of Engineering Mechanics 130~(8) (2004) 921--930.
\newblock \href {https://doi.org/10.1061/(ASCE)0733-9399(2004)130:8(921)} {\path{doi:10.1061/(ASCE)0733-9399(2004)130:8(921)}}.
\newline\urlprefix\url{https://ascelibrary.org/doi/10.1061/%28ASCE%290733-9399%282004%29130%3A8%28921%29}

\bibitem{Dessena2022g}
G.~Dessena, M.~Civera, A.~Pontillo, D.~I. Ignatyev, J.~F. Whidborne, L.~{Zanotti Fragonara}, \href{https://www.emerald.com/insight/content/doi/10.1108/AEAT-06-2024-0178/full/html}{{Noise-robust modal parameter identification and damage assessment for aero-structures}}, Aircraft Engineering and Aerospace Technology 96~(11) (2024) 27--36.
\newblock \href {https://doi.org/10.1108/AEAT-06-2024-0178} {\path{doi:10.1108/AEAT-06-2024-0178}}.
\newline\urlprefix\url{https://www.emerald.com/insight/content/doi/10.1108/AEAT-06-2024-0178/full/html}

\bibitem{Dessena2024b}
G.~Dessena, \href{https://zenodo.org/records/11635814}{{Dataset: Structural Health Monitoring of a Flexible Wing}} (2024).
\newblock \href {https://doi.org/10.5281/zenodo.11635814} {\path{doi:10.5281/zenodo.11635814}}.
\newline\urlprefix\url{https://zenodo.org/records/11635814}

\bibitem{Wilson2024}
J.~Wilson, M.~D. Champneys, M.~Tipuric, R.~Mills, D.~J. Wagg, T.~J. Rogers, \href{https://journals.sagepub.com/doi/10.1177/14759217241297098}{{Multiple-input, multiple-output modal testing of a Hawk T1A aircraft: a new full-scale dataset for structural health monitoring}}, Structural Health Monitoring (2024) 1--23\href {https://doi.org/10.1177/14759217241297098} {\path{doi:10.1177/14759217241297098}}.
\newline\urlprefix\url{https://journals.sagepub.com/doi/10.1177/14759217241297098}

\bibitem{Wilson2024a}
J.~Wilson, M.~D. Champneys, M.~Tipuric, R.~Mills, D.~J. Wagg, T.~J. Rogers, \href{https://orda.shef.ac.uk/articles/dataset/BAE{\_}T1A{\_}Hawk{\_}Full{\_}Structure{\_}Modal{\_}Test/24948549}{{BAE T1A Hawk Full Structure Modal Test}} (2024).
\newblock \href {https://doi.org/10.15131/shef.data.24948549} {\path{doi:10.15131/shef.data.24948549}}.
\newline\urlprefix\url{https://orda.shef.ac.uk/articles/dataset/BAE_T1A_Hawk_Full_Structure_Modal_Test/24948549}

\bibitem{Lju1987}
L.~Ljung, {System Identification: Theory for the User}, 1st Edition, Prentice Hall, Englewood Cliffs, NJ, 1987.

\bibitem{VanOverschee1996}
P.~{Van Overschee}, B.~{De Moor}, \href{http://link.springer.com/10.1007/978-1-4613-0465-4}{{Subspace Identification for Linear Systems}}, no. November 2014, Springer US, Boston, MA, 1996.
\newblock \href {https://doi.org/10.1007/978-1-4613-0465-4} {\path{doi:10.1007/978-1-4613-0465-4}}.
\newline\urlprefix\url{http://link.springer.com/10.1007/978-1-4613-0465-4}

\bibitem{Farrar2025}
C.~R. Farrar, N.~Dervilis, K.~Worden, \href{https://onlinelibrary.wiley.com/doi/10.1111/str.12495}{{The Past, Present and Future of Structural Health Monitoring: An Overview of Three Ages}}, Strain 61~(1) (feb 2025).
\newblock \href {https://doi.org/10.1111/str.12495} {\path{doi:10.1111/str.12495}}.
\newline\urlprefix\url{https://onlinelibrary.wiley.com/doi/10.1111/str.12495}

\bibitem{Civera2021}
M.~Civera, G.~Calamai, L.~{Zanotti Fragonara}, \href{https://onlinelibrary.wiley.com/doi/10.1002/stc.2695}{{Experimental modal analysis of structural systems by using the fast relaxed vector fitting method}}, Structural Control and Health Monitoring 28~(4) (2021) 1--23.
\newblock \href {https://doi.org/10.1002/stc.2695} {\path{doi:10.1002/stc.2695}}.
\newline\urlprefix\url{https://onlinelibrary.wiley.com/doi/10.1002/stc.2695}

\bibitem{Allemang2003}
R.~J. Allemang, \href{http://www.sandv.com/downloads/0308alle.pdf}{{The Modal Assurance Criterion – Twenty Years of Use and Abuse}}, Sound and Vibration (2003) 14--21.
\newline\urlprefix\url{http://www.sandv.com/downloads/0308alle.pdf}

\bibitem{Bull2021}
L.~A. Bull, P.~A. Gardner, J.~Gosliga, T.~J. Rogers, N.~Dervilis, E.~J. Cross, E.~Papatheou, A.~E. Maguire, C.~Campos, K.~Worden, \href{https://linkinghub.elsevier.com/retrieve/pii/S0888327020305276}{{Foundations of population-based SHM, Part I: Homogeneous populations and forms}}, Mechanical Systems and Signal Processing 148 (2021) 107141.
\newblock \href {https://doi.org/10.1016/j.ymssp.2020.107141} {\path{doi:10.1016/j.ymssp.2020.107141}}.
\newline\urlprefix\url{https://linkinghub.elsevier.com/retrieve/pii/S0888327020305276}

\bibitem{Gres2021}
S.~Gre{\'{s}}, M.~D{\"{o}}hler, L.~Mevel, \href{https://linkinghub.elsevier.com/retrieve/pii/S0888327020308438}{{Uncertainty quantification of the Modal Assurance Criterion in operational modal analysis}}, Mechanical Systems and Signal Processing 152 (2021) 107457.
\newblock \href {https://doi.org/10.1016/j.ymssp.2020.107457} {\path{doi:10.1016/j.ymssp.2020.107457}}.
\newline\urlprefix\url{https://linkinghub.elsevier.com/retrieve/pii/S0888327020308438}

\bibitem{Dessena2022d}
G.~Dessena, D.~I. Ignatyev, J.~F. Whidborne, L.~{Zanotti Fragonara}, \href{https://link.springer.com/10.1007/978-3-031-07258-1{\_}26}{{A Kriging Approach to Model Updating for Damage Detection}}, in: P.~Rizzo, A.~Milazzo (Eds.), EWSHM 2022 (LNCE 254), Springer, Singapore, 2023, Ch.~26, pp. 245--255.
\newblock \href {https://doi.org/10.1007/978-3-031-07258-1_26} {\path{doi:10.1007/978-3-031-07258-1_26}}.
\newline\urlprefix\url{https://link.springer.com/10.1007/978-3-031-07258-1_26}

\bibitem{Dessena2022c}
G.~Dessena, D.~I. Ignatyev, J.~F. Whidborne, L.~{Zanotti Fragonara}, \href{https://linkinghub.elsevier.com/retrieve/pii/S0045782523006357}{{A global–local meta-modelling technique for model updating}}, Computer Methods in Applied Mechanics and Engineering 418 (2024) 116511.
\newblock \href {https://doi.org/10.1016/j.cma.2023.116511} {\path{doi:10.1016/j.cma.2023.116511}}.
\newline\urlprefix\url{https://linkinghub.elsevier.com/retrieve/pii/S0045782523006357}

\bibitem{Dessena2024}
G.~Dessena, M.~Civera, \href{https://linkinghub.elsevier.com/retrieve/pii/S0997753824002754}{{Improved tangential interpolation-based multi-input multi-output modal analysis of a full aircraft}}, European Journal of Mechanics - A/Solids 110~(March-April) (2025) 105495.
\newblock \href {https://doi.org/10.1016/j.euromechsol.2024.105495} {\path{doi:10.1016/j.euromechsol.2024.105495}}.
\newline\urlprefix\url{https://linkinghub.elsevier.com/retrieve/pii/S0997753824002754}

\bibitem{Perera2006}
R.~Perera, R.~Torres, \href{https://doi.org/10.1061/(ASCE)0733-9445(2006)132:9(1491)}{{Structural Damage Detection via Modal Data with Genetic Algorithms}}, Journal of Structural Engineering 132~(9) (2006) 1491--1501.
\newblock \href {https://doi.org/10.1061/(ASCE)0733-9445(2006)132:9(1491)} {\path{doi:10.1061/(ASCE)0733-9445(2006)132:9(1491)}}.
\newline\urlprefix\url{https://doi.org/10.1061/(ASCE)0733-9445(2006)132:9(1491)}

\bibitem{Zimmerman2017}
D.~L. Brown, R.~J. Allemang, R.~Zimmerman, M.~Mergeay, \href{https://www.sae.org/content/790221/}{{Parameter estimation techniques for modal analysis}}, in: 1979 Automotive Engineering Congress and Exposition, no. 790221, Univ.of Cincinnati, 1979, p.~19.
\newblock \href {https://doi.org/10.4271/790221} {\path{doi:10.4271/790221}}.
\newline\urlprefix\url{https://www.sae.org/content/790221/}

\bibitem{Georgioudakis2018}
M.~Georgioudakis, V.~Plevris, \href{https://onlinelibrary.wiley.com/doi/10.1155/2018/3183067}{{A Combined Modal Correlation Criterion for Structural Damage Identification with Noisy Modal Data}}, Advances in Civil Engineering 2018~(1) (jan 2018).
\newblock \href {https://doi.org/10.1155/2018/3183067} {\path{doi:10.1155/2018/3183067}}.
\newline\urlprefix\url{https://onlinelibrary.wiley.com/doi/10.1155/2018/3183067}

\bibitem{Lieven1988}
N.~A.~J. Lieven, D.~J. Ewins, {Spatial Correlation of Mode Shapes, the Coordinate Modal Assurance Criterion (COMAC)}, in: Proceedings of the 6th International Modal Analysis Conference, Kissimmee, FL, 1988.

\bibitem{Fayyadh2011}
M.~M. Fayyadh, H.~A. Razak, {Stiffness reduction index for detection of damage location: Analytical study}, International Journal of Physical Sciences 6~(9) (2011) 2194--2204.
\newblock \href {https://doi.org/10.5897/IJPS11.378} {\path{doi:10.5897/IJPS11.378}}.

\bibitem{Perez2014}
M.~A. P{\'{e}}rez, L.~Gil, S.~Oller, \href{https://linkinghub.elsevier.com/retrieve/pii/S0263822313004728}{{Impact damage identification in composite laminates using vibration testing}}, Composite Structures 108~(1) (2014) 267--276.
\newblock \href {https://doi.org/10.1016/j.compstruct.2013.09.025} {\path{doi:10.1016/j.compstruct.2013.09.025}}.
\newline\urlprefix\url{https://linkinghub.elsevier.com/retrieve/pii/S0263822313004728}

\bibitem{Kianfar2024}
E.~Kianfar, K.~Arjomandi, A.~Lloyd, \href{https://link.springer.com/10.1007/978-3-031-61539-9{\_}20}{{Damage Localization of Reinforced Concrete Beams Using Extracted Modal Parameters}}, in: S.~Desjardins, G.~Poitras, A.~{El Damatty}, A.~Elshaer (Eds.), Proceedings of the Canadian Society for Civil Engineering Annual Conference 2023, Volume 13. CSCE 2023. Lecture Notes in Civil Engineering, Vol. 507, Springer Cham, 2024, Ch.~20, pp. 237--251.
\newblock \href {https://doi.org/10.1007/978-3-031-61539-9_20} {\path{doi:10.1007/978-3-031-61539-9_20}}.
\newline\urlprefix\url{https://link.springer.com/10.1007/978-3-031-61539-9_20}

\bibitem{Romero-Carrasco2024}
J.~Romero-Carrasco, F.~Sanhueza-Espinoza, C.~Oyarzo-Vera, \href{https://www.mdpi.com/2075-5309/14/8/2326}{{Variation in the Modal Response of Retrofitted Unreinforced Masonry Walls at Different Levels of Damage}}, Buildings 14~(8) (2024) 2326.
\newblock \href {https://doi.org/10.3390/buildings14082326} {\path{doi:10.3390/buildings14082326}}.
\newline\urlprefix\url{https://www.mdpi.com/2075-5309/14/8/2326}

\bibitem{Shaw1991}
S.~W. Shaw, C.~Pierre, \href{https://linkinghub.elsevier.com/retrieve/pii/0022460X9190412D}{{Non-linear normal modes and invariant manifolds}}, Journal of Sound and Vibration 150~(1) (1991) 170--173.
\newblock \href {https://doi.org/10.1016/0022-460X(91)90412-D} {\path{doi:10.1016/0022-460X(91)90412-D}}.
\newline\urlprefix\url{https://linkinghub.elsevier.com/retrieve/pii/0022460X9190412D}

\bibitem{Kerschen2013}
G.~Kerschen, M.~Peeters, J.~C. Golinval, C.~St{\'{e}}phan, \href{https://arc.aiaa.org/doi/10.2514/1.C031918}{{Nonlinear modal analysis of a full-scale aircraft}}, Journal of Aircraft 50~(5) (2013) 1409--1419.
\newblock \href {https://doi.org/10.2514/1.C031918} {\path{doi:10.2514/1.C031918}}.
\newline\urlprefix\url{https://arc.aiaa.org/doi/10.2514/1.C031918}

\bibitem{Mayo2007}
A.~Mayo, A.~Antoulas, \href{https://linkinghub.elsevier.com/retrieve/pii/S0024379507001280}{{A framework for the solution of the generalized realization problem}}, Linear Algebra and its Applications 425~(2-3) (2007) 634--662.
\newblock \href {https://doi.org/10.1016/j.laa.2007.03.008} {\path{doi:10.1016/j.laa.2007.03.008}}.
\newline\urlprefix\url{https://linkinghub.elsevier.com/retrieve/pii/S0024379507001280}

\bibitem{Antoulas2017}
A.~C. Antoulas, S.~Lefteriu, A.~C. Ionita, \href{http://epubs.siam.org/doi/10.1137/1.9781611974829.ch8}{{A Tutorial Introduction to the Loewner Framework for Model Reduction}}, in: Model Reduction and Approximation, no. May 2011, Society for Industrial and Applied Mathematics, Philadelphia, PA, 2017, Ch.~8, pp. 335--376.
\newblock \href {https://doi.org/10.1137/1.9781611974829.ch8} {\path{doi:10.1137/1.9781611974829.ch8}}.
\newline\urlprefix\url{http://epubs.siam.org/doi/10.1137/1.9781611974829.ch8}

\bibitem{Lefteriu2009}
S.~Lefteriu, A.~C. Antoulas, \href{http://ieeexplore.ieee.org/document/5089847/}{{Modeling multi-port systems from frequency response data via tangential interpolation}}, in: 2009 IEEE Workshop on Signal Propagation on Interconnects, 2009, pp. 1--4.
\newblock \href {https://doi.org/10.1109/SPI.2009.5089847} {\path{doi:10.1109/SPI.2009.5089847}}.
\newline\urlprefix\url{http://ieeexplore.ieee.org/document/5089847/}

\bibitem{Lefteriu2010b}
S.~Lefteriu, A.~C. Antoulas, \href{http://ieeexplore.ieee.org/document/5356286/}{{A new approach to modeling multiport systems from frequency-domain data}}, IEEE Transactions on Computer-Aided Design of Integrated Circuits and Systems 29~(1) (2010) 14--27.
\newblock \href {https://doi.org/10.1109/TCAD.2009.2034500} {\path{doi:10.1109/TCAD.2009.2034500}}.
\newline\urlprefix\url{http://ieeexplore.ieee.org/document/5356286/}

\bibitem{Quero2019}
D.~Quero, P.~Vuillemin, C.~Poussot-Vassal, \href{http://www.mdpi.com/2226-4310/6/1/9}{{A generalized state-space aeroservoelastic model based on tangential interpolation}}, Aerospace 6~(1) (2019) 9.
\newblock \href {https://doi.org/10.3390/aerospace6010009} {\path{doi:10.3390/aerospace6010009}}.
\newline\urlprefix\url{http://www.mdpi.com/2226-4310/6/1/9}

\bibitem{Dessena2022f}
G.~Dessena, M.~Civera, D.~I. Ignatyev, J.~F. Whidborne, L.~{Zanotti Fragonara}, B.~Chiaia, \href{https://www.mdpi.com/2226-4310/10/6/571}{{The Accuracy and Computational Efficiency of the Loewner Framework for the System Identification of Mechanical Systems}}, Aerospace 10~(6) (2023) 571.
\newblock \href {https://doi.org/10.3390/aerospace10060571} {\path{doi:10.3390/aerospace10060571}}.
\newline\urlprefix\url{https://www.mdpi.com/2226-4310/10/6/571}

\bibitem{Dessena2022e}
G.~Dessena, D.~I. Ignatyev, J.~F. Whidborne, A.~Pontillo, L.~{Zanotti Fragonara}, L.~Fragonara, \href{https://www.icas.org/ICAS{\_}ARCHIVE/ICAS2022/data/papers/ICAS2022{\_}0644{\_}paper.pdf}{{Ground vibration testing of a high aspect ratio wing with revolving clamp}}, in: 33rd Congress of the International Council of the Aeronautical Sciences, ICAS 2022, Vol.~6, International Council of the Aeronautical Sciences, Stockholm, Sweden, 2022, pp. 4169--4181.
\newblock \href {https://doi.org/10.17862/cranfield.rd.20486229} {\path{doi:10.17862/cranfield.rd.20486229}}.
\newline\urlprefix\url{https://www.icas.org/ICAS_ARCHIVE/ICAS2022/data/papers/ICAS2022_0644_paper.pdf}

\bibitem{Dessena2024f}
G.~Dessena, M.~Civera, A.~Yousefi, C.~Surace, \href{https://onlinelibrary.wiley.com/doi/10.1002/msd2.70016}{{NExT‐LF: A Novel Operational Modal Analysis Method via Tangential Interpolation}}, International Journal of Mechanical System Dynamics (may 2025).
\newblock \href {https://doi.org/10.1002/msd2.70016} {\path{doi:10.1002/msd2.70016}}.
\newline\urlprefix\url{https://onlinelibrary.wiley.com/doi/10.1002/msd2.70016}

\bibitem{Lowner1934}
K.~L{\"{o}}wner, \href{http://link.springer.com/10.1007/BF01170633}{{{\"{U}}ber monotone matrixfunktionen}}, Mathematische Zeitschrift 38~(1) (1934) 177--216.
\newblock \href {https://doi.org/10.1007/BF01170633} {\path{doi:10.1007/BF01170633}}.
\newline\urlprefix\url{http://link.springer.com/10.1007/BF01170633}

\bibitem{Ionita2014}
A.~C. Ionita, A.~C. Antoulas, \href{http://link.springer.com/10.1007/978-3-319-02090-7{\_}2}{{Case Study: Parametrized Reduction Using Reduced-Basis and the Loewner Framework}}, in: A.~Quarteroni, G.~Rozza (Eds.), Reduced Order Methods for Modeling and Computational Reduction, Springer International Publishing, Cham, 2014, Ch.~2, pp. 51--66.
\newblock \href {https://doi.org/10.1007/978-3-319-02090-7_2} {\path{doi:10.1007/978-3-319-02090-7_2}}.
\newline\urlprefix\url{http://link.springer.com/10.1007/978-3-319-02090-7_2}

\bibitem{Dessena2024e}
G.~Dessena, \href{https://zenodo.org/records/13863292}{{A tutorial for the improved Loewner Framework for modal analysis}} (2024).
\newblock \href {https://doi.org/10.5281/zenodo.13863292} {\path{doi:10.5281/zenodo.13863292}}.
\newline\urlprefix\url{https://zenodo.org/records/13863292}

\bibitem{Allemang1982}
R.~J. Allemang, D.~L. Brown, \href{https://web.archive.org/web/20151018074821/http://sdrl.uc.edu/sdrl/referenceinfo/documents/papers/imac1982-mac.pdf}{{A correlation coefficient for modal vector analysis}}, in: Proceedings of the 1st International Modal Analysis Conference, Union College, Schenectady, NY, 1982, pp. 110--116.
\newline\urlprefix\url{https://web.archive.org/web/20151018074821/http://sdrl.uc.edu/sdrl/referenceinfo/documents/papers/imac1982-mac.pdf}

\bibitem{Haywood-Alexander2024}
M.~Haywood-Alexander, R.~S. Mills, M.~D. Champneys, M.~R. Jones, M.~S. Bonney, D.~Wagg, T.~J. Rogers, \href{https://linkinghub.elsevier.com/retrieve/pii/S0022460X24000592}{{Full-scale modal testing of a Hawk T1A aircraft for benchmarking vibration-based methods}}, Journal of Sound and Vibration 576~(September 2023) (2024) 118295.
\newblock \href {https://doi.org/10.1016/j.jsv.2024.118295} {\path{doi:10.1016/j.jsv.2024.118295}}.
\newline\urlprefix\url{https://linkinghub.elsevier.com/retrieve/pii/S0022460X24000592}

\bibitem{Berghout2025}
T.~Berghout, \href{https://www.mdpi.com/2075-1702/13/3/179}{{Military Training Aircraft Structural Health Monitoring Leveraging an Innovative Biologically Inspired Feedback Mechanism for Neural Networks}}, Machines 13~(3) (2025) 179.
\newblock \href {https://doi.org/10.3390/machines13030179} {\path{doi:10.3390/machines13030179}}.
\newline\urlprefix\url{https://www.mdpi.com/2075-1702/13/3/179}

\bibitem{Noel2013}
J.~P. N{\"{o}}el, L.~Renson, G.~Kerschen, B.~Peeters, S.~Manzato, J.~Debille, {Nonlinear dynamic analysis of an F-16 aircraft using GVT data}, in: IFASD 2013 - International Forum on Aeroelasticity and Structural Dynamics, 2013, pp. 1--13.

\bibitem{Dossogne2015}
T.~Dossogne, J.~P. No{\"{e}}l, C.~Grappasonni, G.~Kerschen, B.~Peeters, J.~Debille, M.~Vaes, J.~Schoukens, {Nonlinear ground vibration identification of an F-16 aircraft - Part II: Understanding nonlinear behaviour in aerospace structures using sine-sweep testing}, International Forum on Aeroelasticity and Structural Dynamics, IFASD 2015 (2015) 1--19.

\bibitem{Worden2007}
K.~Worden, C.~R. Farrar, G.~Manson, G.~Park, \href{https://royalsocietypublishing.org/doi/10.1098/rspa.2007.1834}{{The fundamental axioms of structural health monitoring}}, Proceedings of the Royal Society A: Mathematical, Physical and Engineering Sciences 463~(2082) (2007) 1639--1664.
\newblock \href {https://doi.org/10.1098/rspa.2007.1834} {\path{doi:10.1098/rspa.2007.1834}}.
\newline\urlprefix\url{https://royalsocietypublishing.org/doi/10.1098/rspa.2007.1834}

\end{thebibliography}

\end{document}